\documentclass[a4paper, 11pt, abstract=true, titlepage, oneside]{scrartcl}

\makeatletter

\usepackage[T1]{fontenc}
\usepackage[utf8]{inputenc}
\usepackage[onehalfspacing]{setspace}
\usepackage[headsepline]{scrlayer-scrpage}
\usepackage{layout}
\usepackage{geometry}
\usepackage{adjustbox}
\usepackage{authblk}
\usepackage[titletoc,title]{appendix}
\usepackage[hyphens]{url}
\usepackage{color}

\usepackage{amsmath,amssymb,amsfonts,amsthm, mathtools, bm}
\usepackage{mathrsfs}
\usepackage{accents}
\usepackage{thmtools, thm-restate}
\usepackage{nicefrac}
\usepackage{array}

\usepackage[normal]{threeparttable}
\usepackage{threeparttablex}
\usepackage{array,tabularx}
\usepackage{longtable}
\usepackage{booktabs}
\usepackage{colortbl}
\usepackage{multirow}
\usepackage{makecell}
\usepackage{siunitx}

\usepackage{algorithm}
\usepackage{algpseudocode}  
\usepackage[bottom,hang,multiple]{footmisc}
\usepackage[acronym]{glossaries}

\usepackage{subcaption}
\usepackage{import}
\usepackage{soul}
\usepackage{graphicx}
\usepackage{xcolor} 
\usepackage{placeins}

\usepackage{tikz}
\usetikzlibrary{arrows.meta}
\usetikzlibrary{math}
\usetikzlibrary{shapes}
\usetikzlibrary{patterns,shapes.arrows}
\usepackage{pgfplots}
\pgfplotsset{compat=1.18} 
\usepgfplotslibrary{groupplots}
\usepgfplotslibrary{groupplots,dateplot}
\pgfplotsset{compat=newest}
\usepgfplotslibrary{dateplot}
\usetikzlibrary{backgrounds}
\pgfdeclarelayer{background}
\pgfdeclarelayer{foreground}
\pgfsetlayers{background,main,foreground} 
\usepackage{tikzscale}

\usepackage{svg}
\usepackage{tikz}
\usepackage{pgfplots}
\usepgfplotslibrary{groupplots} 
\pgfplotsset{compat=newest}

\usepackage{enumerate}
\usepackage{multicol}
\usepackage[author=Patrick]{pdfcomment}
\usepackage{xparse}
\usepackage{twoopt}
\usepackage{etoolbox}

\usepackage{eurosym}
\usepackage{ellipsis}
\usepackage{natbib}
\usepackage{paralist}
\usepackage[normalem]{ulem}

\AtBeginDocument{
	\newgeometry{
		textheight=9in,
		textwidth=6.5in,
		top=1in,
		headheight=16.49675pt,
		headsep=25pt,
		footskip=30pt
	}
}
\newsavebox{\abstractbox}
\renewenvironment{abstract}
{\begin{lrbox}{0}\begin{minipage}{\textwidth}
			\begin{center}\normalfont\sectfont\abstractname\end{center}\quotation}
		{\endquotation\end{minipage}\end{lrbox}%
	\global\setbox\abstractbox=\box0 }

\makeatletter
\expandafter\patchcmd\csname\string\maketitle\endcsname
{\vskip\z@\@plus3fill}
{\vskip\z@\@plus2fill\box\abstractbox\vskip\z@\@plus1fill}
{}{}
\makeatother

\allowdisplaybreaks

\addtokomafont{captionlabel}{\footnotesize\bfseries} 
\addtokomafont{caption}{\footnotesize\bfseries} 

\bibpunct[, ]{(}{)}{,}{a}{}{,}%
\def\newblock{\ }%
\newtheoremstyle{upright} 
  {3pt}                    
  {3pt}                    
  {\upshape}               
  {}                       
  {\bfseries}              
  {.}                      
  {.5em}                   
  {}                       

\theoremstyle{upright}

\let\theoremstyle\relax

\counterwithin*{equation}{section}

\newenvironment{myfont}{\fontfamily{ccr}\selectfont}{\par}
\DeclareTextFontCommand{\textmyfont}{\myfont}

\AtBeginDocument{%
	\abovedisplayskip=7.5pt plus 0pt minus 0pt
	\abovedisplayshortskip=7.5pt plus 0pt minus 0pt
	\belowdisplayskip=7.5pt plus 0pt minus 0pt
	\belowdisplayshortskip=7.5pt plus 0pt minus 0pt
}

\newcolumntype{L}[1]{>{\raggedright\let\newline\\\arraybackslash\hspace{0pt}}p{#1}}
\newcolumntype{C}[1]{>{\centering\let\newline\\\arraybackslash\hspace{0pt}}p{#1}}
\newcolumntype{R}[1]{>{\raggedleft\let\newline\\\arraybackslash\hspace{0pt}}p{#1}}
\renewcommand{\emph}[1]{\textit{#1}}

\algtext*{EndIf} 
\algtext*{EndWhile} 
\algtext*{EndFor} 

\usepackage{comment}
\begin{document}
\emergencystretch 3em
\newacronym{acr:mdp}{MDP}{Markov decision process}

\newacronym{acr:ai}{AI}{Artificial Intelligence}

\newacronym{acr:ftc}{FTC}{Federal Trade Commission}

\newacronym{acr:ec}{EC}{European Commission}

\newacronym{acr:rl}{RL}{Reinforcement Learning}

\newacronym{acr:drl}{DRL}{Deep Reinforcement Learning}

\newacronym{acr:ppo}{PPO}{Proximal Policy Optimization}

\newacronym{acr:ppoc}{PPO-C}{\gls{acr:ppo} with continuous action space}

\newacronym{acr:ppod}{PPO-D}{\gls{acr:ppo} with discrete action space}

\newacronym{acr:dqn}{DQN}{Deep Q-Networks}

\newacronym{acr:tql}{TQL}{Tabular Q-learning}

\newacronym{acr:sac}{SAC}{Soft Actor-Critic}

\newacronym{acr:b2c}{B2C}{Business-to-Consumer}

\newacronym{acr:rpdi}{RPDI}{Relative Price Deviation Index}

\newacronym{acr:dnn}{DNN}{Deep Neural Network}


\title{\large Exploring Competitive and Collusive Behaviors in Algorithmic Pricing with Deep Reinforcement Learning}

\author[1]{\normalsize Shidi Deng}
\author[2]{\normalsize Maximilian Schiffer}
\author[3]{\normalsize Martin Bichler}
\affil{\small 
	School of Management, Technical University of Munich, Germany	
 
	\scriptsize shidi.deng@tum.de

        \small
	\textsuperscript{2}School of Management \& Munich Data Science Institute,
	
	Technical University of Munich, Germany
	
	\scriptsize schiffer@tum.de

         \small
	\textsuperscript{3}School of Computation, Information and Technology, 
      Technical University of Munich, Germany
	
	\scriptsize bichler@cit.tum.de

 }

\date{}

\lehead{\pagemark}
\rohead{\pagemark}

\begin{abstract}
\begin{singlespace}
{\small\noindent Nowadays, a significant share of the business-to-consumer sector is based on online platforms like Amazon and Alibaba and uses \gls{acr:ai} for pricing strategies. This has sparked debate on whether pricing algorithms may tacitly collude to set supra-competitive prices without being explicitly designed to do so. Our study addresses these concerns by examining the risk of collusion when \gls{acr:rl} algorithms are used to decide on pricing strategies in competitive markets. Prior research in this field focused on \gls{acr:tql} and led to opposing views on whether learning-based algorithms can result in supra-competitive prices. Building on this, our work contributes to this ongoing discussion by providing a more nuanced numerical study that goes beyond \gls{acr:tql}, additionally capturing off- and on-policy \gls{acr:drl} algorithms, two distinct families of \gls{acr:drl} algorithms that recently gained attention for algorithmic pricing. We study multiple Bertrand oligopoly variants and show that algorithmic collusion depends on the algorithm used. In our experiments, we observed that \gls{acr:tql} tends to exhibit higher collusion and price dispersion. Moreover, it suffers from instability and disparity, as agents with higher learning rates consistently achieve higher profits, and it lacks robustness in state representation, with pricing dynamics varying significantly based on information access. In contrast, \gls{acr:drl} algorithms, such as \gls{acr:ppo} and \gls{acr:dqn}, generally converge to lower prices closer to the Nash equilibrium. Additionally, we show that when pre-trained \gls{acr:tql} agents interact with \gls{acr:drl} agents, the latter quickly outperforms the former, highlighting the advantages of \gls{acr:drl} in pricing competition. Lastly, we find that competition between heterogeneous \gls{acr:drl} algorithms, such as \gls{acr:ppo} and \gls{acr:dqn}, tends to reduce the likelihood of supra-competitive pricing.

\smallskip}
\end{singlespace}
\end{abstract}

\maketitle
\glsresetall 
\section{Introduction}
\label{sec:introduction}

During the past two decades, a significant share of the B2C business has transitioned towards leading online retailers such as Amazon and Alibaba. With the rise of \gls{acr:ai} and big data, sellers on these platforms increasingly rely on pricing algorithms to explore market dynamics and demand elasticity. In practice, algorithms help businesses adapt to market changes, find optimal strategies, and improve decision-making efficiency. However, some real-world cases have provided evidence that algorithmic decision-making can lead to collusive outcomes, resulting in supra-competitive pricing and sustained collusion \citep{assad2020algorithmic, brown2021competition}. In this context, algorithmic collusion refers to a phenomenon where independently operating pricing algorithms learn to agree on setting prices above the Nash equilibrium, leading to supra-competitive results, even though the algorithms were not explicitly designed for collusion.

While such tacitly collusive outcomes remain problematic from a market perspective, independently operating pricing algorithms cannot be held responsible for collusion due to their lack of communication and mutual understanding capabilities \citep{harrington2018developing}. This creates significant challenges for regulatory bodies in enforcing competition law and antitrust policies \citep{mehra2015antitrust}. Accordingly, automated pricing algorithms pose a potential risk of tacit collusion in modern markets, drawing widespread attention and concern from scientists, antitrust authorities, and practitioners \citep{ezrachi2017two, varian2018artificial, agrawal2019economics}.
In this paper, we focus on supra-competitive outcomes higher than the Nash equilibrium and refer to the phenomenon as algorithmic collusion or collusive prices.

These concerns go beyond deliberately designed price-manipulating algorithms, and hold particularly for hypothetical or experimental
self-learning algorithms that aim to maximize profits. Among such algorithms, \gls{acr:rl} serves as a widely studied example, enabling autonomous learning of pricing strategies through self-play without prior knowledge or explicit guidance from developers. 
Based on the observed pricing history, algorithms can communicate and coordinate collusive behaviors to maintain an off-equilibrium state. However, in most practical settings, the length of pricing history considered in the state representation is limited, as including longer histories significantly increases the state space, making it computationally challenging. Nevertheless, we analyze this aspect in our study. If such pricing dynamics exist, they pose a serious threat to markets, potentially leading to inefficiencies that benefit companies at the expense of consumers. Although many prices today are still established by human salespeople and subjective judgments, these pricing algorithms may become widespread in retail markets in the near future \citep{oecd2017algorithms}. Detecting these issues early is crucial to preventing their widespread adoption, which may lead to unintended consequences such as market distortions and reduced competition regarding algorithmic decision-making.

So far, there is no consensus on whether algorithmic collusion exists or whether it leads to supra-competitive prices that negatively impact social welfare. This lack of consensus highlights the theoretical and practical challenges of analyzing tacit collusion between algorithms. On the one hand, there are no clear standards for detecting algorithmic collusion in real markets, and data or information gaps, often exacerbated by confidentiality, complicate rigorous analyses even further. On the other hand, studying algorithmic collusion through stylized market models can detect tacit collusion by comparing results to analytically derived equilibria, but existing analyses {have often been limited to single, simple algorithms, which has led to concerns about their generalizability. 
In fact, most existing studies focus on \gls{acr:tql}  and argue that if simple algorithms can learn to collude, more complex algorithms, including those that adopt deep learning, will pose even greater risks of tacit collusion \citep{calvano2020artificial, klein2021autonomous}. Several studies have explored the use of \gls{acr:drl} in algorithmic pricing \citep{liu2019dynamic, chen2021spatial}, highlighting its potential for strategic pricing decisions. While most existing analyses remain academic, the increasing interest in \gls{acr:drl} makes these algorithms strong candidates for real-world pricing applications. However, details of such implementations are rarely disclosed, as firms withhold information about proprietary algorithms to prevent exploitation. In this context, our study aims to provide more nuanced numerical analyses, capturing \gls{acr:drl} algorithms beyond \gls{acr:rl} to reveal the risks of collusion in algorithmic pricing.\medskip

\noindent\textbf{Contribution}\quad This paper advances the study of algorithmic collusion by systematically analyzing the pricing behavior of \gls{acr:drl} algorithms in competitive settings. Building upon prior research focused on \gls{acr:tql}, we extend the analysis to more sophisticated on-policy and off-policy \gls{acr:drl} methods, including \gls{acr:ppo} and \gls{acr:dqn}. Using a numerical simulation framework based on Bertrand competition models with various demand specifications, we compare pricing strategies, the emergence of supra-competitive prices, and market stability across these learning paradigms.

Our results provide several key insights. First, we demonstrate that \gls{acr:tql} exhibits a higher propensity for collusion, leading to supra-competitive pricing and instability in competitive environments. Agents with higher learning rates systematically outperform slower-learning counterparts, creating unfair market dynamics. Second, \gls{acr:drl} agents, particularly \gls{acr:ppo} and \gls{acr:dqn}, exhibit stronger competitive behavior, consistently converging to prices closer to the Nash equilibrium. When competing against pre-trained \gls{acr:tql} agents, \gls{acr:drl} agents rapidly outperform them, suggesting that advanced \gls{acr:rl} approaches mitigate collusion risks. Lastly, we find that interactions between heterogeneous \gls{acr:drl} agents, such as \gls{acr:ppo} and \gls{acr:dqn}, further reduce the likelihood of collusive behavior, reinforcing the notion that algorithmic diversity fosters competition. Our study provides a in-depth evaluation of algorithmic collusion risks in \gls{acr:drl}-based pricing and highlights the potential for competitive pricing dynamics in algorithmically governed markets. By advancing our understanding of \gls{acr:rl}  in pricing competition, we contribute to the ongoing efforts to design fairer and more competitive digital marketplaces.

The following sections provide a detailed exploration of our approach and findings.  Section 2 reviews the state of the art. Section 3 defines the problem setting, outlining key assumptions and market dynamics. Section 4 presents the algorithmic framework, describing the \gls{acr:rl} algorithms used. Section 5 details the experimental design, specifying key parameters and settings. In Section 6, we present the results, followed by a discussion of their significance. Finally, Section 7 concludes with a summary of contributions and suggestions for future research. 
\raggedbottom

\section{State of the art}
Our work is closely related to two key research directions: the application of \gls{acr:rl} algorithms in pricing markets and the discussion surrounding whether automated pricing algorithms may lead to collusion. In the following, we review both research streams concisely.\medskip  

\noindent\textbf{Application of \gls{acr:rl} Algorithms in Pricing Markets}\quad
Research on the application of \gls{acr:rl} algorithms in pricing markets has evolved significantly over time. Early studies primarily focused on basic \gls{acr:rl} algorithms, particularly the application of \gls{acr:tql} in pricing problems. \citet{kephart2000pseudo} demonstrated that \gls{acr:tql} could converge to equilibrium in a simulated market environment where two competing pricebots repeatedly interacted, while \citet{kutschinski2003learning} showed its ability to find near-optimal strategies in multi-agent markets, though constrained by a zero discount factor. \citet{rana2014real} proposed using \gls{acr:tql} for pricing multiple products with interdependent demand, addressing scenarios where users lack precise knowledge of product relationships. Similarly, \citet{kim2015dynamic} applied \gls{acr:tql} in energy markets to optimize pricing strategies through specific model adjustments. Collectively, these studies demonstrate the flexibility of \gls{acr:tql} across diverse market settings but also reveal inherent limitations in handling complex and dynamic environments.

Building on the foundations laid by basic \gls{acr:rl} algorithms, recent advancements in \gls{acr:ai} and big data have shifted the focus toward \gls{acr:drl} algorithms in pricing research. These advanced methods address many of the scalability and adaptability challenges faced by \gls{acr:tql}, enabling applications in increasingly complex and dynamic market environments. Notably, their potential has been validated through practical implementations in real-world settings. For instance, \citet{liu2019dynamic} pioneered the application of \gls{acr:drl} for discrete and continuous dynamic pricing on real-world e-commerce platforms, demonstrating its significant superiority over manual pricing through large-scale field experiments at Alibaba. Beyond e-commerce, \citet{qiu2020deep} proposed a deep deterministic policy gradient method for pricing electric vehicles with discrete charging levels. Similarly, \citet{chen2021spatial} explored spatial-temporal pricing for ride-hailing platforms using \gls{acr:ppo}. Meanwhile, 
\citet{yan2022hierarchical} developed a hierarchical \gls{acr:rl}-based community energy trading scheme using historical data.  Overall, these studies highlight the versatility of \gls{acr:drl} algorithms in addressing diverse pricing challenges, underscoring their potential to enhance decision-making efficiency and adaptability across various industries.\medskip

\noindent\textbf{Automated Pricing Algorithms and Collusion}\quad The second research stream examines how automated pricing algorithms may facilitate collusion, resulting in supra-competitive pricing and reduced consumer welfare. This phenomenon has drawn significant attention to antitrust laws, with numerous studies emphasizing the urgent need for policies to address this growing challenge \citep{werner2023algorithmic, constantine2018oecd, capobianco2020competition}. Existing research on algorithm-driven collusion primarily focuses on two directions: empirical studies and simulation-based approaches.  

Empirical studies on algorithmic collusion are limited but provide valuable insights into its real-world risks. For instance, \citet{assad2020algorithmic} investigated the German retail gasoline market and found that since 2017, algorithmic pricing has increased average station profit margins by 9\%. Similarly, \citet{brown2021competition} analyzed the U.S. retail market, showing that algorithmic pricing increased price dispersion and resulted in supra-competitive pricing. However, no direct evidence has proven that algorithms autonomously collude, prompting researchers to explore simulation-based analysis in synthetic environments. Claims about the ability of \gls{acr:tql} to sustain supra-competitive prices are often grounded in academic studies \citep{azzutti2021machine, buckmann2021comparing}, and the most influential simulations supporting algorithmic collusion thereby primarily focus on \gls{acr:tql}. For example, \citet{klein2021autonomous} simulated sequential pricing scenarios using \gls{acr:tql} and demonstrated that algorithms can autonomously learn collusive behaviors even without explicit human intervention. \citet{calvano2020artificial} applied \gls{acr:tql} to the Bertrand competition with logit demand and showed that the algorithm could sustain supra-competitive prices through reward-punishment strategies. Similarly, \citet{abada2023artificial} simulated the electricity market using a Cournot framework and observed comparable collusive behaviors.  However, as \citet{calvano2020artificial} highlighted, a key limitation of \gls{acr:tql} is its slow learning speed, often requiring hundreds of thousands of cycles to form collusion. 

Although \gls{acr:tql} has shown potential for collusion in simulations, its limitations have sparked academic debate regarding the validity of these findings. In particular, some studies question whether the observed collusive behaviors truly reflect the algorithm's inherent properties or are artifacts of specific experimental setups. For example, \citet{abada2024collusion} criticized the single-period price reduction reward-punishment scheme used by \citet{calvano2020artificial}, arguing that insufficient exploration was a key driver of their collusion results. \citet{asker2022artificial} emphasized that supra-competitive pricing heavily depends on the learning protocol, and synchronous learning where algorithms account for competitors' behavior may lead to more competitive pricing outcomes. At the same time, \citet{meylahn2022learning} and \citet{epivent2022algorithmic} stressed the need for further research and comprehensive analysis to validate these collusive findings. To date, most existing studies focus on basic \gls{acr:rl} algorithms (e.g., \gls{acr:tql})\citep{brero2022learning,dolgopolov2022reinforcement, sanchez2022artificial, bichler2024online}, while research on \gls{acr:drl} in pricing collusion has gained increasing attention. Recent studies, in particular, have more frequently explored \gls{acr:dqn} and \gls{acr:ppo} \citep{friedrich2024learning, kastius2022dynamic, schlechtinger2023price}, supporting our algorithm choices. However, these works primarily focus on specific or single-industry models. In contrast, our study examines \gls{acr:drl} across multiple Bertrand oligopoly variants, providing a broader perspective on algorithmic collusion. 

\section{Problem Setting}
We study the Bertrand model \citep{bertrand1883review}, which describes oligopolistic price competition among firms producing homogeneous products. In this context, each firm employs an independent pricing algorithm.  
Our analysis focuses on duopoly markets, consisting of two firms, indexed as \(i = 0, 1\). Each firm \(i\) produces a product with a given quality \(g\) and incurs a marginal cost \(c\). The Bertrand model primarily examines the interaction between firms, which set product prices \(p_i\), and consumers, who choose products based on these prices. This interaction directly determines the demand \(d_i\) for each firm's product, and the profit for firm \(i\) is given by \(r_i = (p_i - c) \times d_i\). Firms set prices simultaneously, aiming to maximize their profits.  This interaction between firms and consumers creates a dynamic relationship between prices and demand, repeated over multiple rounds. At the start of each round, the two firms simultaneously decide on the prices for their products. Based on these prices, consumers determine the demand for each product, and the interaction proceeds to the next round.  

We explore various extensions of the Bertrand model, differing in assumptions about product capacity, product quality, and the form of the demand function. In the remainder of this section, we first discuss the characteristics of these market models (Section~\ref{subsec:marketModels}), before we elaborate on the respective Nash and monopoly prices (Section~\ref{subsec:NMPprices}).

\subsection{Market Models}\label{subsec:marketModels}
We consider three market models: the Standard Bertrand model, the Bertrand-Edgeworth model in which firms face production capacity constraints, and the Logit Bertrand model.

In the \textbf{Standard Bertrand Model}, products are homogeneous, and consumers choose the product with the lower price. The demand function is given by \( d(p) = 1 - p \). For two firms with prices \( p_i \) and \( p_{-i} \), demand is allocated as follows:  
\begin{equation}
d_i(p_i, p_{-i}) = 
\begin{cases} 
d(p_i) & \text{if } p_i < p_{-i}, \\ 
\frac{1}{2}d(p_i) & \text{if } p_i = p_{-i}, \\ 
0 & \text{if } p_i > p_{-i}.
\end{cases}
\label{eq:standardDemand}
\end{equation}
This implies that the firm with the lower price captures all demand, while equal prices result in an equal demand split, assuming unlimited production capacity for both firms.  

In the \textbf{Bertrand-Edgeworth Model}, firms face production capacity constraints \citep{edgeworth1925papers}, which can lead to market prices above marginal cost if their capacity is fully utilized. While this extension adds a realistic constraint to the standard Bertrand model, it complicates the closed-form equilibrium analysis. To ensure well-defined and unique Nash and monopoly prices, we focus on a specific case where two competing firms have identical production capacities \( k > 0.5 \), restrict their pricing strategies to the interval \([0, 1]\), and face a demand function of the form \( d(p) = 1 - p \). In this scenario, the demand allocation results in:
\begin{equation}
d_i(p_i, p_{-i}) = 
\begin{cases} 
\min\{k, d(p_i)\} & \text{if } p_i < p_{-i}, \\ 
\frac{1}{2}d(p_i) & \text{if } p_i = p_{-i}, \\ 
0 & \text{if } p_i > p_{-i}.
\end{cases}
\label{eq:edgeqorthDemand}
\end{equation}

In the \textbf{Logit Bertrand Model}, we extend the Standard Bertrand model by relaxing the assumption that consumers always purchase from the firm offering the lowest price. Instead, we incorporate demand variation, product substitutability, and the allocation of consumer expenditure among firms, reflecting a more complex market behavior. Following \cite{calvano2020artificial}, we consider a logit demand model where marginal costs remain constant, and each firm sells a single product. In this model, consumers allocate their expenditure across firms based on pricing rather than solely choosing the lowest-priced product, allowing for a more nuanced modeling of substitution effects and competitive dynamics.  In this case, the demand for firm \( i \)'s product is given by:
\begin{equation}
d_i(p_i, p_{-i}) = \frac{e^{\frac{g - p_i}{\mu}}}{\sum_{j=0}^{1} e^{\frac{g - p_j}{\mu}} + 1},
\label{eq:logitDemand}
\end{equation}
where the term \( g - p_i \) reflects the utility consumers derive from purchasing product \( i \). The parameter \( \mu \) captures the degree of horizontal differentiation or substitutability between products. As \( \mu \to 0 \), products become perfect substitutes, while larger \( \mu \) values correspond to increasing product differentiation and lower substitutability.

Figure~\ref{fig:three_rewards} visualizes the profit functions \(r\) for all three market models, assuming the demand to behave as described in Equations~\eqref{eq:standardDemand}--\eqref{eq:logitDemand}. In the Standard Bertrand model, the profit surface exhibits a symmetric and concave shape, reflecting the homogeneous nature of the products and the fierce characterization of the resulting competition, which leads to an all-or-nothing outcome for the players. In the Bertrand-Edgeworth model, the profit function retains a similar structure but shows the influence of capacity constraints, 
\begin{figure}[!hb]
    \centering
    \includegraphics[width=1.0\linewidth]{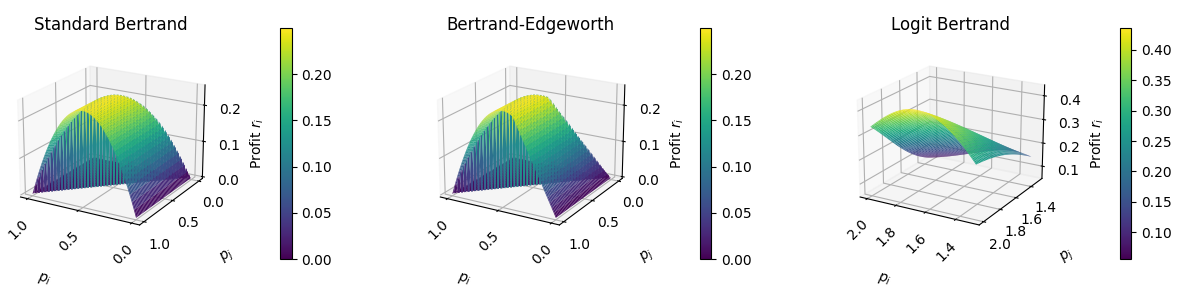}
    \caption{Profit functions for one firm under Standard Bertrand, Bertrand-Edgeworth, and Logit Bertrand models in a homogeneous duopoly. The Logit Bertrand model shows a smooth and convex reward surface, followed by the Standard Bertrand model, while the Bertrand-Edgeworth model shows greater irregularities.}
    \label{fig:three_rewards}
\end{figure}
particularly when demand exceeds the capacity limit. Lastly, the Logit Bertrand model introduces greater asymmetry and smoother transitions in the profit function, capturing the effects of product differentiation and consumer preference heterogeneity. 

\subsection{Nash Price \& Monopoly Price}\label{subsec:NMPprices}
In the previous subsection, we introduced three distinct demand models, each capturing the effects of product homogeneity, capacity constraints, and differences in consumer demand distribution on firm competition. These models not only differ in terms of demand and competitive behavior but also exhibit significant variations in key pricing strategies, which leads to different Nash prices \( p^N \) and monopoly \( p^M \)  prices. To characterize the competitive dynamics of the market models introduced in Section~\ref{subsec:marketModels}, we provide a detailed discussion of the definitions and economic implications of the Nash prices and monopoly prices in the following. These analyses lay the theoretical foundation for studying the evolution of pricing behavior and the potential for collusion.

The \textbf{Nash  price \( p^N \)} represents a price at which all firms, while seeking to maximize their profits, have no incentive to unilaterally deviate from their pricing strategies. A key characteristic of this price is its stability, as all participants' strategies are balanced at the Nash equilibrium point, such that deviating from this price yields no additional benefit to any firm. The computation of \( p^N \) varies across different demand models.

In the \textit{Standard Bertrand model}, where firms produce homogeneous products and consumers always choose the cheaper option, firms continuously undercut each other's prices to capture the market until prices fall to the marginal cost \( c \). At this point, any further price reduction becomes unprofitable, making \( p^N = c \) the unique Nash price \citep{bertrand1883review}.

In the \textit{Bertrand-Edgeworth model}, considering limited production capacities \( k > 0.5 \), the price-cutting competition between firms still drives prices toward the marginal cost \( c \). As a result, the Nash price remains \( p^N = c \), similar to the Standard Bertrand model \citep{levitan1972price}.

In the \textit{Logit Bertrand Model}, products are no longer perfectly homogeneous, and consumers choose with a probabilistic preference, leading to a Nash price \( p^N \) that no longer equals the marginal cost. In this case, firms determine \( p^N \) by optimizing their profit functions with respect to price, solving the following equation:
\begin{equation}
\frac{d}{dp_{i}} r_{i} = \frac{d}{dp_{i}} \left( (p_{i} - c_i) d_{i} \right) = 0.
\label{eq:nashLogit}
\end{equation}
Accordingly, we obtain the Nash price by solving Equation~\ref{eq:nashLogit} simultaneously for all firms. In practice, iterative search methods can be employed, such as scanning over a range of prices and verifying whether any firm can increase its profit by deviating. The price at which no firm benefits from deviation is identified as the Nash price \( p^N \).

The \textbf{monopoly price \( p^M \)} is the price that maximizes collective profits when firms coordinate their pricing strategies as if they were a single entity. By reducing output and increasing prices, firms exploit market demand to achieve higher joint profits. This price is typically higher than the Nash price \( p^N \), as it reflects the absence of competition and the pursuit of monopolistic gains.

In the \textit{Standard Bertrand Model}, given a total demand of \( d(p) = 1 - p \) and a marginal cost of \( c = 0 \), the profit function is \( r(p) = p(1 - p) \). Taking the derivative, \( \frac{dr(p)}{dp} = 1 - 2p \), and setting it to zero, the profit-maximizing price is \( p^M = 0.5 \). At this price, each firm captures an equal share of the market demand, \( d_i(p^M) = 0.25 \), and earns a profit of \( r_i(p^M) =  0.125\).

In the \textit{Bertrand-Edgeworth Model}, the introduction of capacity constraints significantly affects how demand is allocated between firms. When the production capacity \( k \) is limited, neither firm can fully satisfy the total market demand on its own. This creates a setting where firms may benefit from tacit or explicit collusion to coordinate prices and avoid aggressive competition that would otherwise drive prices to marginal cost. When \( k > 0.5 \), the monopoly price \( p^M \) is derived as in the standard Bertrand case. Given \( d(p) = 1 - p \) and \( c = 0 \), the profit function \( r(p) = p(1 - p) \) is maximized at \( p^M = 0.5 \). 

In the \textit{Logit Bertrand Model}, the non-linear nature of demand makes calculating the monopoly price more complex. The total profit is given by \( r(p_0, p_1) = \sum_{i=0}^1 (p_i - c_i)d_i(p_i, p_{-i}) \). To determine the monopoly price, partial derivatives of \( r(p_0, p_1) \) with respect to \( p_0 \) and \( p_1 \) are set to zero. Numerical methods or iterative algorithms approximate the solution by evaluating \( \frac{\partial r}{\partial p_0} = 0 \) and \( \frac{\partial r}{\partial p_1} = 0 \) over a discretized range of prices, iteratively refining estimates until convergence criteria are met.

The monopoly price \( p^M \) is generally higher than the Nash price \( p^N \), as firms, acting as monopolists, exploit market demand to maximize collective profits. Comparing the Nash price with the monopoly price provides insights into whether and to which extent firms engage in collusive behavior.
\section{Algorithmic Framework}
\label{sec: algorithms}
This section formalizes the algorithmic framework underpinning our study. We first define the agent's learning behavior within an \gls{acr:mdp} framework. Following this, we describe \gls{acr:rl} algorithms in our study.

\subsection{Markov Decision Process}
We analyze the dynamics of algorithmic collusion in pricing decisions involving \gls{acr:rl} agents. Each agent represents a firm engaging in pricing competitions to maximize profits. The decision-making process is modeled as an infinite-horizon \gls{acr:mdp}, which captures the ongoing interaction and competition among firms. An \gls{acr:mdp} provides a mathematical framework for decision-making in stochastic environments, defined by states, actions, transition dynamics, and rewards. In this context, the market serves as the system, the firms act as agents, and demand response functions driven by pricing decisions determine the transition dynamics. Below, we detail the components of the \gls{acr:mdp}.

\textbf{State space:} At time $t$, the state $s_t$ for each agent consists of a historical set of pricing decisions with a memory length $l$, capturing the agent's and its competitor's pricing decisions. Formally, the state space is $\mathcal{S} = \{s_t \mid s_t = (p_{i,t-k})_{i=0,1, k=1,\dots,l}\}$, where $l$ determines the history length. For simplicity, when $l = 1$, the state $s_t$ reduces to $s_t = (p_{i,t-1})_{i=0,1}$, relying only on the most recent pricing decisions.

\textbf{Action space:} At each time step~$t$, an agent chooses a price $p_t \in \mathcal{A}$ from the action space $\mathcal{A}$, which is constrained by a price range $[\underline{p}, \bar{p}]$. For Standard Bertrand and Bertrand-Edgeworth models, this range is $[0, 1]$. For the Logit Bertrand model, we follow \citet{calvano2020artificial} and set these bounds to $\underline{p} = p^N - \zeta(p^M - p^N)$ and $\bar{p} = p^M + \zeta(p^M - p^N)$, where $\zeta$ controls pricing flexibility. When considering a discretized action space, we divide this range into $m$ equidistant values, and when $l=1$, the state space $\mathcal{S} = \mathcal{A} \times \mathcal{A}$ grows quadratically with $m$.

\textbf{Transition dynamics:} Agents' pricing decisions $(p_{i,t})_{i=0,1}$ determine the transition to the next state $S_{t+1}$. Each agent $i$ receives a reward $R_{i,t}$, calculated as $R_{i,t} = (p_{i,t} - c)d_i(p_{i,t}, p_{-i,t})$, where $c$ is the marginal cost and $d_i(p_{i,t}, p_{-i,t})$ represents the demand for agent $i$'s product based on its own price $p_{i,t}$ and the competitor's pricing decisions $p_{-i,t}$.

\textbf{Objective:} Each agent maximizes discounted future rewards, given by 
\begin{equation}
   G_t = \sum_{k=0}^{\infty}\gamma^kR_{t+k+1},    
\end{equation}
where $\gamma \in [0, 1)$ is the discount factor prioritizing immediate rewards.

\subsection{Reinforcement Learning Approaches and Algorithm Selection}
\label{subsec: rl_approaches}

Building on the introduced \gls{acr:mdp} framework, we study \gls{acr:rl} algorithms that enable agents to learn optimal policies through interactions with the environment. \gls{acr:rl} methods fall into two categories: value-based and policy-based approaches.

\textbf{Value-based methods} estimate value functions, such as the state-value function $V(s)$ or the action-value function $Q(s, a)$, and guide decision-making by selecting actions that maximize the expected return based on these value functions. The Bellman optimality equation for value functions underpins these methods:
\begin{equation}
    V^{\pi^*}(s_t) = \max_{a_t} Q^{\pi^*}(s_t, a_t) = \max_{a_t} \sum_{s_{t+1}} p(s_t, a_t, s_{t+1}) \left[r(s_t, a_t, s_{t+1}) + \gamma V^{\pi^*}(s_{t+1}) \right].
\end{equation}
Here, \(V^{\pi^*}(s)\) represents the expected utility of starting in state \(s\) and following the optimal policy \(\pi^*\), while \(Q^{\pi^*}(s, a)\) is the expected utility of taking action \(a\) in state \(s\) and then following the optimal policy. The discount factor \(\gamma\) balances immediate and future rewards. The essence of value-based methods lies in selecting actions that maximize the \(Q\) value, guiding decision-making towards maximizing long-term returns.

\textbf{Policy-based methods} directly optimize a (stochastic) policy $\pi_\theta(a_t | s_t)$, representing the probability of taking action $a_t$ in state $s_t$. These methods use gradient ascent to adjust policy parameters $\theta$, maximizing the expected return:
\begin{equation}
    \eta(\theta) = \mathbb{E}_{\tau \sim \pi_\theta(\tau)} \left[ \sum_t r(a_t, s_t) \right],
\end{equation}
where \(\tau\) represents a trajectory generated by the policy \(\pi_\theta\). By performing gradient ascent on the policy parameters, the agent progressively improves its policy, increasing the probability of selecting better actions, maximizing long-term total returns by converging to an optimal policy.

Value-based \gls{acr:rl} algorithms can be implemented in a traditional tabular setting, while both value-based and policy-based methods can be realized in a \gls{acr:drl} fashion. Traditional tabular \gls{acr:rl} often struggles with high-dimensional state spaces and complex tasks. \gls{acr:drl} overcomes these limitations by leveraging deep neural networks to enhance representation learning, significantly broadening the applicability of the \gls{acr:rl} framework. Unlike tabular methods or simple function approximations, \gls{acr:drl} employs neural networks to approximate value functions, policies, or state representations, making it well-suited for continuous, high-dimensional, and nonlinear problems. Despite challenges such as training instability and sample inefficiency, its strong generalization capabilities establish \gls{acr:drl} as a powerful tool for solving complex decision-making problems.

In this study, we aim for a comprehensive analysis of algorithm dynamics, which is why we consider both tabular \gls{acr:rl} and \gls{acr:drl} algorithms in our experiments. Alongside the basic \gls{acr:tql} algorithm, we selected \gls{acr:dqn} and \gls{acr:ppo} as representative value-based and policy-based \gls{acr:drl} methods, respectively. \gls{acr:dqn} is well-suited for discrete action spaces, that arise in competitive pricing scenarios. By approximating Q-values with deep neural networks, it effectively handles large state spaces and mitigates the curse of dimensionality. Its extensive use in discrete-action tasks, such as games and market bidding \citep{mnih2015human, ye2019deep}, further supports its selection. \gls{acr:ppo} was chosen for its flexibility in handling both continuous and discrete action spaces, making it suitable for a wide range of pricing models. Unlike value-based approaches, policy gradient methods directly optimize policies, improving efficiency in large state-action spaces. \gls{acr:ppo} enhances stability by constraining policy updates, leading to faster convergence and strong performance in practical applications \citep{schulman2017proximal, berner2019dota}. To ensure a fair comparison with \gls{acr:dqn} and \gls{acr:tql}, we focus our studies on its discrete-action space variant.

\textbf{\gls{acr:tql}:} Tabular Q-Learning (\gls{acr:tql}), introduced by \citet{watkins1989learning}, uses a tabular representation for the Q-function $Q(s, a)$ with dimensions $|S| \times |A|$, where $|S|$ and $|A|$ are the sizes of the state and action spaces. Agents update Q-values at each step using the Bellman equation:
\begin{equation}
    Q_{t+1}(s, a) =(1-\alpha)Q_t(s,a)+\alpha[R_{t+1} + \gamma \max_{a'} Q_t(s', a')],
\end{equation}
where $\alpha$ is the learning rate, $s'$ is the next state, and $\gamma$ is the discount factor. This method works well in small state-action spaces but struggles to scale due to its reliance on explicit storage of Q-tables.

\textbf{\gls{acr:dqn}:} Deep Q-Networks (\gls{acr:dqn}), proposed by \citet{mnih2015human}, extend \gls{acr:tql} by using deep neural networks to approximate the Q-function $Q(s, a; \theta)$. This approach overcomes scalability issues by generalizing across state-action pairs. Key techniques include experience replay, which randomizes training samples to reduce correlations, and target networks, which stabilize training by decoupling target Q-value computation from network updates. \gls{acr:dqn} performs well in discrete action spaces and high-dimensional state representations.

\textbf{\gls{acr:ppo}:} Proximal Policy Optimization (\gls{acr:ppo}), developed by \citet{schulman2017proximal}, directly optimizes the policy $\pi(s, a; \theta)$ using neural networks. Unlike value-based methods, \gls{acr:ppo} explicitly models the policy and updates it with a clipped surrogate objective:
\begin{equation}
    L^{\text{CLIP}}(\theta) = \mathbb{E}_t \left[ \min \left( r_t(\theta) \hat{A}_t, \operatorname{clip} \left( r_t(\theta), 1 - \epsilon, 1 + \epsilon \right) \hat{A}_t \right) \right]
\end{equation}
where $r_t(\theta)$ is the probability ratio between new and old policies, $\hat{A}_t$ is the advantage estimate, and $\epsilon$ controls the clipping range. \gls{acr:ppo} excels in both discrete and continuous action spaces. Gradient clipping ensures incremental updates that prevent performance degradation.

For all three algorithms used, we provide a pseudocode, accompanied by a more detailed explanation in Appendix~\ref{appendix: pseudocodes_drl}.
\section{Experimental Design}
\label{subsec:experimental_design}
We investigate three variants of the Bertrand competition model to understand how different market structures affect competitive behavior. In the Standard Bertrand and Bertrand-Edgeworth models, all competitive agents have a marginal cost $c$ set to zero. In the Logit Bertrand model, we set $c$ to one to align with the data from \citet{calvano2020artificial}. For the Bertrand-Edgeworth model, we apply a capacity constraint of $k=0.6$ to both agents.

In the Logit Bertrand model, we set the product quality to $g=2$ and the inter-product substitutability to $\mu= 0.25$. For the Standard Bertrand and Bertrand-Edgeworth models, the Nash price $p^{N}$ and the monopoly price $p^{M}$ are 0 and 0.5, respectively, while the Nash profit $\pi^{N}$ and monopoly profit $\pi^{M}$ are 0 and 0.125, respectively. In the Logit Bertrand model, $p^{N}$ increases to 1.473, $p^{M}$ to 1.925, and $\pi^{N}$ and $\pi^{M}$ increase to 0.223 and 0.337, respectively.\medskip

\noindent\textbf{Algorithm and Parameter Consistency}\quad We use consistent neural network architectures for all \gls{acr:drl} algorithms: a fully connected feedforward network with two hidden layers of 64 nodes each. Each layer employs ReLU activation functions, and we trained the network using the Adam optimizer. Although tuning architectures for specific algorithms could enhance performance, we prioritize uniformity to ensure that performance differences reflect intrinsic algorithm characteristics. To standardize action spaces, we configure \gls{acr:ppo} to operate within a discrete action space, aligning it with \gls{acr:tql} and \gls{acr:dqn}. The action space includes $m = 15$ price options, evenly divided. In the Logit Bertrand model, we set the relaxation parameter $\eta = 0.1$. This ensures that observed differences arise from algorithm characteristics rather than action space discrepancies.\medskip

\noindent\textbf{Experimental Configuration \& Convergence}\quad Our design involves two \gls{acr:rl} agents simulating pricing decisions for two firms, focusing on \gls{acr:tql}, \gls{acr:dqn}, and \gls{acr:ppo}. 

Note that we do not define a unified convergence criterion across the three \gls{acr:rl} algorithms. Instead, we execute each algorithm over a sufficiently long time horizon $T$, based on empirical observations, to allow for price stabilization. Two factors motivate this choice. First, \gls{acr:drl} algorithms like \gls{acr:ppo} retain an intrinsic level of exploration during training, complicating the identification of stable policies. Unlike tabular methods, which gradually reduce exploration, \gls{acr:ppo} continues exploring, making convergence to a fixed or low-variance policy unlikely. Second, algorithms using neural networks and replay buffers, such as \gls{acr:dqn}, are prone to catastrophic forgetting, especially without a learning rate decay. Even with stable rewards, policy adjustments may occur due to replay buffer dynamics, causing shifts from seemingly stable policies. Thus, we select a sufficiently extended time horizon $T$ based on observed empirical stability in prices and profits, aligning with the adaptive dynamics of pricing problems and \gls{acr:rl} algorithms.

Our experiments were run on high-performance hardware featuring an AMD Ryzen 9 7950X CPU (32 cores @ 4.5 GHz), 128 GB of RAM, and an NVIDIA RTX 4090 GPU (24 GB). Each experiment was independently repeated 20 times, with the mean and standard deviation of results calculated to assess stability and robustness.\medskip

\noindent\textbf{Evaluation Criteria}\quad We analyze individual pricing trends and profit trajectories using two indicators: the \gls{acr:rpdi}, and the profit metric $\Delta$, which was proposed by \citet{calvano2020artificial}. The \gls{acr:rpdi} measures an agent's pricing relative to Nash equilibrium and monopoly pricing, indicating the extent of supra-competitive pricing. The profit metric $\Delta$ assesses an agent's average profit over time, normalized relative to Nash and monopoly profits. Both indicators use the average values from the last 10,000 timesteps to ensure robustness.

The \gls{acr:rpdi} is defined as:
\begin{equation}
     \hat{p_i} = \frac{\Bar{p_i} - p^{\text{N}}}{p^{\text{M}} - p^{\text{N}}},
\end{equation}
measuring an agent's pricing relative to Nash and monopoly levels. The profit metric from \citet{calvano2020artificial} is given by:
\begin{equation}
     \Delta_i = \frac{\Bar{\pi_i} - \pi^N}{\pi^M - \pi^N},
\end{equation}
indicating changes in each firm $i$'s profits relative to Nash and monopoly profits. Together, these metrics provide a comprehensive view: when \gls{acr:rpdi} and $\Delta_i$ values approach 0, behavior resembles perfect competition; values near 1 suggest a higher propensity for collusion.
\section{Results}

To introduce the findings presented in this section, we briefly outline five core results examining different aspects of \gls{acr:rl} behavior in pricing. These results first investigate how variations in learning rates with \gls{acr:tql} can influence market dynamics. We then turn to the dynamics within a homogeneous \gls{acr:rl} agents' setting, evaluating how homogeneous agents interact in competitive environments. Subsequently, we analyze the comparative effectiveness of \gls{acr:drl} algorithms and \gls{acr:tql} in pricing strategies. Next, we explore competition among heterogeneous \gls{acr:drl} algorithms, highlighting the distinct influences that each type of algorithm brings to market behavior. We then examine how different state definitions impact pricing strategies. Building on this, we further analyze the underlying factors driving differences in pricing behavior among the \gls{acr:rl} algorithms studied.

\subsection{Impact of Learning Rate on Market Stability in \gls{acr:tql}}

In the following, we analyze how learning rates affect pricing strategies, focusing on competition between two \gls{acr:tql} agents. Each agent adjusts its pricing strategy based on its assigned learning rate \(\alpha\), revealing how variations in \(\alpha\) influence decision-making and profitability. We assign Agent 0 a higher learning rate than Agent 1 to evaluate differences in market adaptation and profit outcomes. Both agents use an exploration strategy with an exponentially decaying rate \(\epsilon(t) = e^{-\beta \cdot t}\), where \(\beta\) controls decay speed. This strategy encourages early exploration and gradually emphasizes profit maximization as \(\epsilon\) decreases.

Specifically, we select learning rates \(\alpha\) from \([0.01, 0.05, 0.1, 0.5]\), assigning Agent 0 consistently higher rates than Agent 1. This setup creates six unique learning rate combinations. To evaluate performance, we calculate differences in the \gls{acr:rpdi} and normalized profit indices \(\Delta\), using Agent~1 as reference point. Boxplots summarize these results for 20 independent runs, capturing variability across combinations.\medskip

\noindent\textbf{Main Findings}\quad The experimental results for the Logit Bertrand competition show that learning rate variations significantly affect pricing strategies and profitability. Figure~\ref{fig:normalized_price_profit_diff_boxplot_logit} demonstrates that agents with higher \(\alpha\) adjust pricing strategies more aggressively and often outperform competitors. For instance, when Agent 0's \(\alpha\) is much higher than Agent 1's (e.g., in 0.5\_0.05 and 0.5\_0.01 combinations), Agent 0 sets lower prices, capturing more market share. Agents with aggressive pricing strategies frequently achieve higher profitability, especially with significant learning rate disparities. In the 0.5\_0.01 combination, for example, Agent 0's profit surpasses Agent 1's by a wide margin. Faster adaptation enables Agent 0 to leverage lower prices for sustained profitability. This trend persists across combinations, showing that faster-learning agents dominate by setting lower prices and achieving higher profit margins.

When learning rates are similar, pricing and profitability differences diminish. For example, in the 0.5\_0.1 combination, similar learning rates produce minimal differences, 
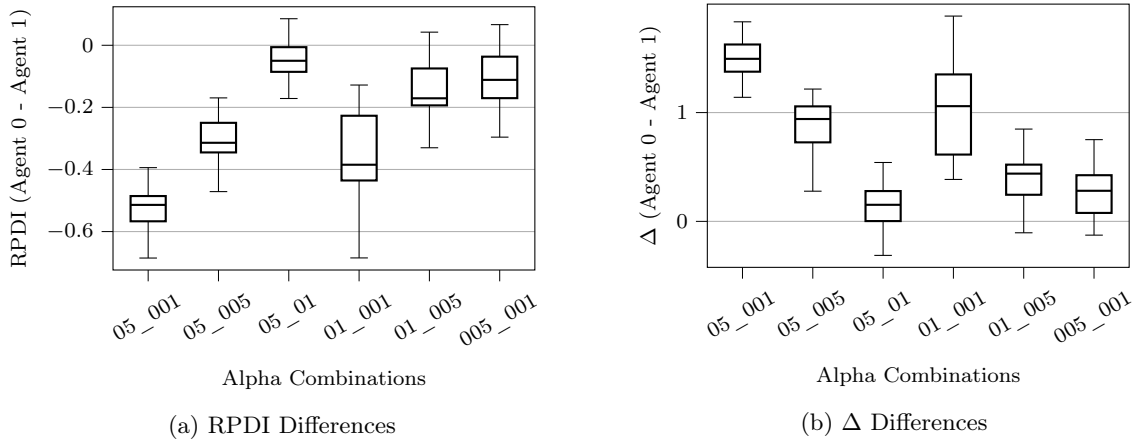
\begin{figure}[!hb]
	\centering
	\begin{minipage}{0.48\linewidth}
		\centering       
\begin{tikzpicture}

\definecolor{darkgray176}{RGB}{176,176,176}
\definecolor{skyblue}{RGB}{135,206,235}
\scriptsize

\begin{axis}[
width=0.9\textwidth, 
height=0.64\textwidth, 
tick align=outside,
tick pos=left,
x grid style={darkgray176},
xlabel={Alpha Combinations},
xmin=0.5, xmax=6.5,
xtick style={color=black},
xtick={1,2,3,4,5,6},
xticklabel style={rotate=30.0},
xticklabels={05\_001,05\_005,05\_01,01\_001,01\_005,005\_001},
y grid style={darkgray176},
ylabel={RPDI (Agent 0 - Agent 1)},
ymajorgrids,
ymin=-0.723751147524132, ymax=0.124019896448791,
ytick style={color=black}
]
\path [draw=black, fill=white, thick]
(axis cs:0.75,-0.566860046354815)
--(axis cs:1.25,-0.566860046354815)
--(axis cs:1.25,-0.485569287047571)
--(axis cs:0.75,-0.485569287047571)
--(axis cs:0.75,-0.566860046354815)
--cycle;
\addplot [black]
table {%
1 -0.566860046354815
1 -0.685216100070818
};
\addplot [black]
table {%
1 -0.485569287047571
1 -0.394068694692433
};
\addplot [black]
table {%
0.875 -0.685216100070818
1.125 -0.685216100070818
};
\addplot [black]
table {%
0.875 -0.394068694692433
1.125 -0.394068694692433
};
\path [draw=black, fill=white, thick]
(axis cs:1.75,-0.345076926540693)
--(axis cs:2.25,-0.345076926540693)
--(axis cs:2.25,-0.249793723914873)
--(axis cs:1.75,-0.249793723914873)
--(axis cs:1.75,-0.345076926540693)
--cycle;
\addplot [black]
table {%
2 -0.345076926540693
2 -0.471272520409765
};
\addplot [black]
table {%
2 -0.249793723914873
2 -0.169443795150877
};
\addplot [black]
table {%
1.875 -0.471272520409765
2.125 -0.471272520409765
};
\addplot [black]
table {%
1.875 -0.169443795150877
2.125 -0.169443795150877
};
\path [draw=black, fill=white, thick]
(axis cs:2.75,-0.0855941482157605)
--(axis cs:3.25,-0.0855941482157605)
--(axis cs:3.25,-0.0060886095065213)
--(axis cs:2.75,-0.0060886095065213)
--(axis cs:2.75,-0.0855941482157605)
--cycle;
\addplot [black]
table {%
3 -0.0855941482157605
3 -0.171406894872095
};
\addplot [black]
table {%
3 -0.0060886095065213
3 0.0854848489954762
};
\addplot [black]
table {%
2.875 -0.171406894872095
3.125 -0.171406894872095
};
\addplot [black]
table {%
2.875 0.0854848489954762
3.125 0.0854848489954762
};
\path [draw=black, fill=white, thick]
(axis cs:3.75,-0.435428805521846)
--(axis cs:4.25,-0.435428805521846)
--(axis cs:4.25,-0.226967331850824)
--(axis cs:3.75,-0.226967331850824)
--(axis cs:3.75,-0.435428805521846)
--cycle;
\addplot [black]
table {%
4 -0.435428805521846
4 -0.684821765629005
};
\addplot [black]
table {%
4 -0.226967331850824
4 -0.128047251248033
};
\addplot [black]
table {%
3.875 -0.684821765629005
4.125 -0.684821765629005
};
\addplot [black]
table {%
3.875 -0.128047251248033
4.125 -0.128047251248033
};
\path [draw=black, fill=white, thick]
(axis cs:4.75,-0.193577491614425)
--(axis cs:5.25,-0.193577491614425)
--(axis cs:5.25,-0.074687800528786)
--(axis cs:4.75,-0.074687800528786)
--(axis cs:4.75,-0.193577491614425)
--cycle;
\addplot [black]
table {%
5 -0.193577491614425
5 -0.330066500288078
};
\addplot [black]
table {%
5 -0.074687800528786
5 0.0422623651772654
};
\addplot [black]
table {%
4.875 -0.330066500288078
5.125 -0.330066500288078
};
\addplot [black]
table {%
4.875 0.0422623651772654
5.125 0.0422623651772654
};
\path [draw=black, fill=white, thick]
(axis cs:5.75,-0.170281755839735)
--(axis cs:6.25,-0.170281755839735)
--(axis cs:6.25,-0.0366945343085765)
--(axis cs:5.75,-0.0366945343085765)
--(axis cs:5.75,-0.170281755839735)
--cycle;
\addplot [black]
table {%
6 -0.170281755839735
6 -0.295870846192317
};
\addplot [black]
table {%
6 -0.0366945343085765
6 0.0664710709096377
};
\addplot [black]
table {%
5.875 -0.295870846192317
6.125 -0.295870846192317
};
\addplot [black]
table {%
5.875 0.0664710709096377
6.125 0.0664710709096377
};
\addplot [thick, black]
table {%
0.75 -0.514049234858963
1.25 -0.514049234858963
};
\addplot [thick, black]
table {%
1.75 -0.314053092955193
2.25 -0.314053092955193
};
\addplot [thick, black]
table {%
2.75 -0.0499390280608836
3.25 -0.0499390280608836
};
\addplot [thick, black]
table {%
3.75 -0.384720396675028
4.25 -0.384720396675028
};
\addplot [thick, black]
table {%
4.75 -0.170755385794308
5.25 -0.170755385794308
};
\addplot [thick, black]
table {%
5.75 -0.111412338544892
6.25 -0.111412338544892
};
\end{axis}

\end{tikzpicture}
        \subcaption*{(a) RPDI Differences}
        \label{fig:normalized_price_diff_boxplot_logit}
	\end{minipage}
	\begin{minipage}{0.48\linewidth}
		\centering       
\begin{tikzpicture}

\definecolor{darkgray176}{RGB}{176,176,176}
\definecolor{skyblue}{RGB}{135,206,235}
\scriptsize

\begin{axis}[
width=0.9\textwidth, 
height=0.64\textwidth, 
tick align=outside,
tick pos=left,
x grid style={darkgray176},
xlabel={Alpha Combinations},
xmin=0.5, xmax=6.5,
xtick style={color=black},
xtick={1,2,3,4,5,6},
xticklabel style={rotate=30.0},
xticklabels={05\_001,05\_005,05\_01,01\_001,01\_005,005\_001},
y grid style={darkgray176},
ylabel={\(\displaystyle \Delta\) (Agent 0 - Agent 1)},
ymajorgrids,
ymin=-0.4227159425487, ymax=1.9970299793081,
ytick style={color=black}
]
\path [draw=black, fill=white, thick]
(axis cs:0.75,1.37591344470336)
--(axis cs:1.25,1.37591344470336)
--(axis cs:1.25,1.625893564775)
--(axis cs:0.75,1.625893564775)
--(axis cs:0.75,1.37591344470336)
--cycle;
\addplot [black]
table {%
1 1.37591344470336
1 1.14028777364025
};
\addplot [black]
table {%
1 1.625893564775
1 1.83433192127645
};
\addplot [black]
table {%
0.875 1.14028777364025
1.125 1.14028777364025
};
\addplot [black]
table {%
0.875 1.83433192127645
1.125 1.83433192127645
};
\path [draw=black, fill=white, thick]
(axis cs:1.75,0.726322572081175)
--(axis cs:2.25,0.726322572081175)
--(axis cs:2.25,1.05710959762201)
--(axis cs:1.75,1.05710959762201)
--(axis cs:1.75,0.726322572081175)
--cycle;
\addplot [black]
table {%
2 0.726322572081175
2 0.277380005464325
};
\addplot [black]
table {%
2 1.05710959762201
2 1.2162209613964
};
\addplot [black]
table {%
1.875 0.277380005464325
2.125 0.277380005464325
};
\addplot [black]
table {%
1.875 1.2162209613964
2.125 1.2162209613964
};
\path [draw=black, fill=white, thick]
(axis cs:2.75,0.00267481249133206)
--(axis cs:3.25,0.00267481249133206)
--(axis cs:3.25,0.278648140370911)
--(axis cs:2.75,0.278648140370911)
--(axis cs:2.75,0.00267481249133206)
--cycle;
\addplot [black]
table {%
3 0.00267481249133206
3 -0.312727491555209
};
\addplot [black]
table {%
3 0.278648140370911
3 0.541273783970066
};
\addplot [black]
table {%
2.875 -0.312727491555209
3.125 -0.312727491555209
};
\addplot [black]
table {%
2.875 0.541273783970066
3.125 0.541273783970066
};
\path [draw=black, fill=white, thick]
(axis cs:3.75,0.614049326104641)
--(axis cs:4.25,0.614049326104641)
--(axis cs:4.25,1.35074149038526)
--(axis cs:3.75,1.35074149038526)
--(axis cs:3.75,0.614049326104641)
--cycle;
\addplot [black]
table {%
4 0.614049326104641
4 0.385459736616627
};
\addplot [black]
table {%
4 1.35074149038526
4 1.88704152831461
};
\addplot [black]
table {%
3.875 0.385459736616627
4.125 0.385459736616627
};
\addplot [black]
table {%
3.875 1.88704152831461
4.125 1.88704152831461
};
\path [draw=black, fill=white, thick]
(axis cs:4.75,0.244522767637906)
--(axis cs:5.25,0.244522767637906)
--(axis cs:5.25,0.521549813536724)
--(axis cs:4.75,0.521549813536724)
--(axis cs:4.75,0.244522767637906)
--cycle;
\addplot [black]
table {%
5 0.244522767637906
5 -0.104932889059974
};
\addplot [black]
table {%
5 0.521549813536724
5 0.848336536039317
};
\addplot [black]
table {%
4.875 -0.104932889059974
5.125 -0.104932889059974
};
\addplot [black]
table {%
4.875 0.848336536039317
5.125 0.848336536039317
};
\path [draw=black, fill=white, thick]
(axis cs:5.75,0.0776189432854876)
--(axis cs:6.25,0.0776189432854876)
--(axis cs:6.25,0.424349535186885)
--(axis cs:5.75,0.424349535186885)
--(axis cs:5.75,0.0776189432854876)
--cycle;
\addplot [black]
table {%
6 0.0776189432854876
6 -0.126592420264579
};
\addplot [black]
table {%
6 0.424349535186885
6 0.751204266878471
};
\addplot [black]
table {%
5.875 -0.126592420264579
6.125 -0.126592420264579
};
\addplot [black]
table {%
5.875 0.751204266878471
6.125 0.751204266878471
};
\addplot [thick, black]
table {%
0.75 1.49415732682755
1.25 1.49415732682755
};
\addplot [thick, black]
table {%
1.75 0.940852268944288
2.25 0.940852268944288
};
\addplot [thick, black]
table {%
2.75 0.152274617825748
3.25 0.152274617825748
};
\addplot [thick, black]
table {%
3.75 1.05898149769883
4.25 1.05898149769883
};
\addplot [thick, black]
table {%
4.75 0.438903537865013
5.25 0.438903537865013
};
\addplot [thick, black]
table {%
5.75 0.281130929377356
6.25 0.281130929377356
};
\end{axis}

\end{tikzpicture}  
        \subcaption*{(b) \(\displaystyle \Delta\) Differences}
        \label{fig:normalized_profit_diff_boxplot_logit}
	\end{minipage}
        \caption{Logit Bertrand: RPDI and $\Delta$ differences between Agent 0 and Agent 1. A negative RPDI difference indicates that Agent 0 sets lower prices , while a positive $\Delta$ difference indicates that Agent 0 gets a higher profit compared to Agent 1.}
        \label{fig:normalized_price_profit_diff_boxplot_logit}
\end{figure}
reflecting a balanced competitive environment. Accordingly, similar learning rates encourage convergence in strategies and outcomes, fostering fairer competition.

In the Standard Bertrand and Edgeworth Bertrand models, results align with these trends. As shown in Appendix~\ref{appendix:supplementary_results}, agents with higher learning rates adopt lower pricing strategies in both models.
However, this effect is less pronounced compared to the Logit Bertrand model. Pricing and profitability differences exhibit greater volatility in the Edgeworth Bertrand model due to its dynamic nature. Nevertheless, agents with higher learning rates consistently achieve higher profitability, particularly in cases with significant disparities (e.g., 0.5\_0.01 and 0.1\_0.01).\medskip

\noindent\textbf{Implications}\quad These findings highlight critical considerations for companies using \gls{acr:tql} as a pricing algorithm. Higher learning rates allow companies to adapt quickly to market dynamics, gaining advantages in pricing agility and profitability. However, these advantages create structural biases that disadvantage companies with lower learning rates, regardless of other strategic capabilities. Companies face challenges in mitigating these biases due to \gls{acr:tql}'s reliance on fixed learning rates. This rigidity creates risks of volatile market outcomes when firms compete with optimized learning rates. Over time, such imbalances may destabilize markets, enabling high-learning-rate firms to dominate by undercutting competitors, reducing market diversity, and fostering monopolistic behavior. These biases make \gls{acr:tql} unsuitable for companies seeking equitable and stable market positions.

\textit{\textbf{Insight 1.} Higher learning rates in \gls{acr:tql} lead to aggressive pricing, giving faster-learning agents significant advantages in achieving lower prices and higher profits, particularly in scenarios with large learning rate disparities. This creates structural biases favoring faster-learning agents and risks destabilizing the market, making \gls{acr:tql} unsuitable for fair competition.}

\subsection{Pricing behavior of Identical RL Agents}
Previous works claim that agents employing identical strategies often exhibit predictable pricing behaviors, which can increase the likelihood of collusion \citep{ezrachi2017artificial, beneke2019artificial}. 
 
A real-world manifestation of this phenomenon can be observed in hub-and-spoke scenarios, where multiple firms utilize similar algorithmic pricing tools provided by third-party platforms.  
For example, many sellers on Amazon and eBay adopt third-party pricing software instead of developing their own \citep{chen2016empirical}. Likewise, online travel agencies (OTAs) such as Expedia and Booking.com provide hotels with algorithmic pricing solutions \citep{wang2023algorithms}.   

To explore this hypothesis, we analyze interactions between two homogeneous \gls{acr:rl} agents. Homogeneous agents share identical algorithmic parameters and training configurations, operate in the same market environment, and follow identical training protocols. Despite these shared settings, each agent updates its strategies independently, with learning trajectories diverging due to variations in data sampling paths. This setup reflects real-world scenarios where competing firms use standardized pricing software configurations.

Specifically, we perform experiments with two homogeneous \gls{acr:tql} agents to establish a baseline, even though earlier results suggest \gls{acr:tql} offers limited advantages compared to \gls{acr:drl}. This baseline allows us to quantify the performance of identical \gls{acr:tql} agents and compare it with \gls{acr:drl} algorithms. Simulations with \gls{acr:drl} agents run for 100,000 timesteps due to their faster convergence in dynamic environments. \gls{acr:tql} agents interact over 1,000,000 timesteps to achieve stable pricing strategies. We conduct each experiment over 20 independent random seeds to ensure robust and reproducible results.\medskip

\noindent\textbf{Main Findings}\quad In the Logit Bertrand model, pricing behaviors differ significantly across \gls{acr:dqn}, \gls{acr:ppo}, and \gls{acr:tql} agents, as shown in  Figure~\ref{fig:three_models_heatmap}(a). 
\begin{figure}[!hb]
    \centering
    \begin{minipage}{0.8\linewidth}
        \centering
        \includegraphics[width=0.9\linewidth]{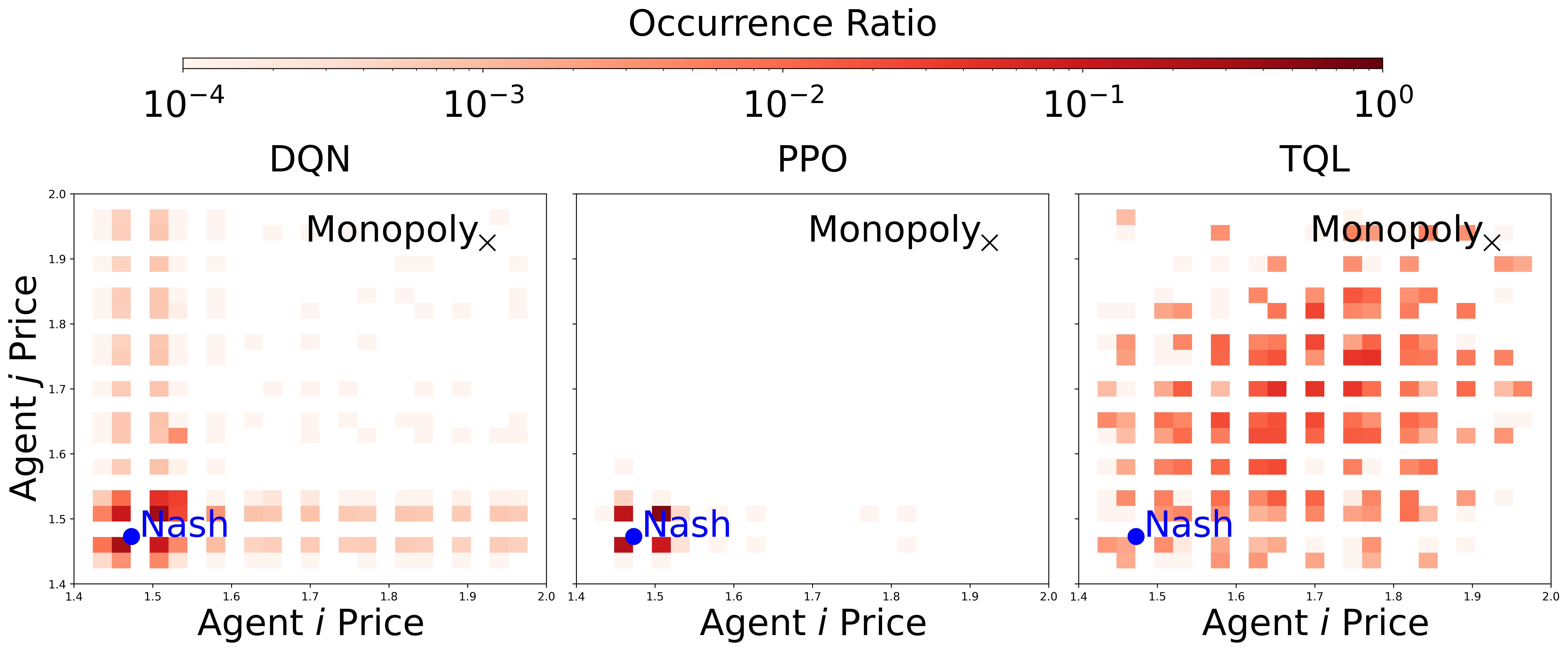}
        \subcaption*{(a) Logit Bertrand}
        \label{Delta_tql_ppo_logit}
        \label{logit_price_selection}
    \end{minipage}
    \vspace{0.1cm} 
    \begin{minipage}{0.8\linewidth}
        \centering
        \includegraphics[width=0.9\linewidth]{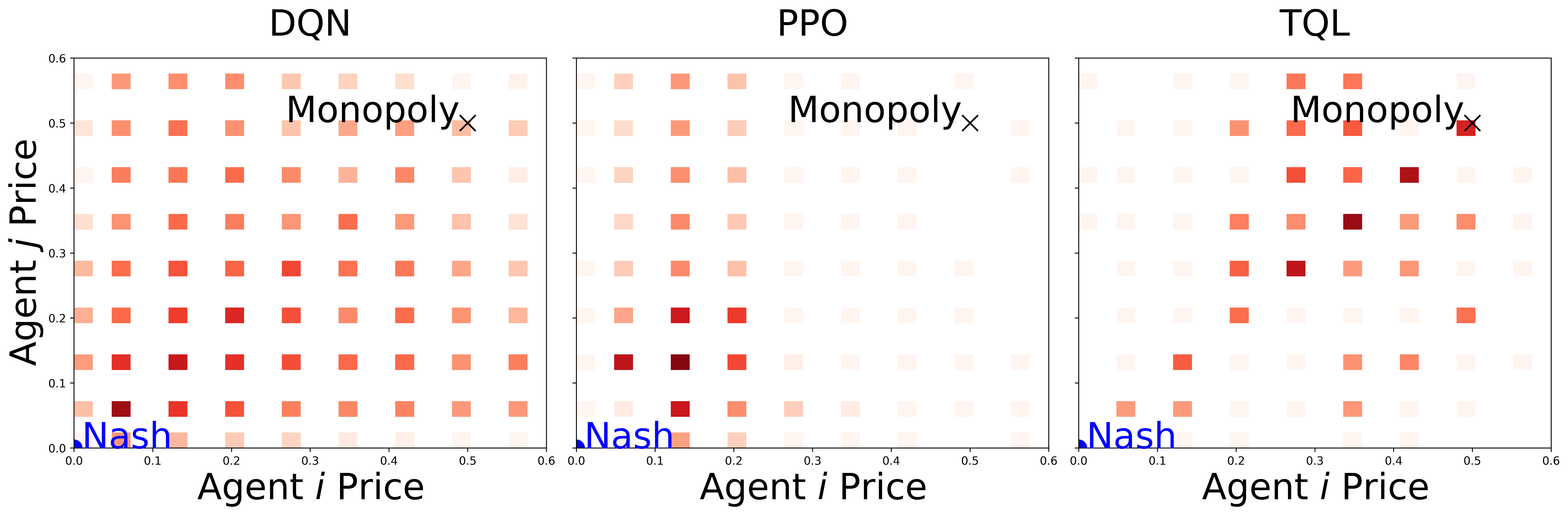}
        \subcaption*{(b) Standard Bertrand}
        \label{standard_price_selection}
    \end{minipage}
    \vspace{0.1cm} 
    \begin{minipage}{0.8\linewidth}
        \centering
        \includegraphics[width=0.9\linewidth]{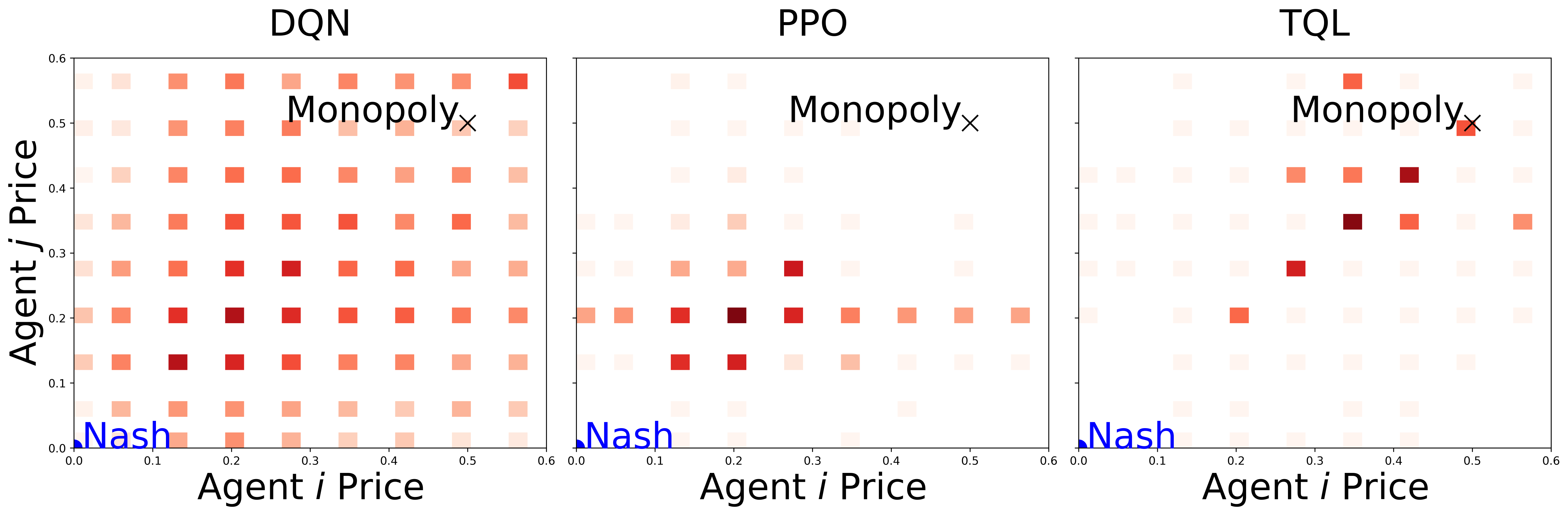} 
        \subcaption*{(c) Edgeworth Bertrand}
        \label{edgeworth_price_selection}
    \end{minipage}
    \caption{Price distribution heatmaps illustrating price selection behaviors for homogeneous DQN, PPO, and TQL agents during the last 10,000 timesteps. Results are based on 20 independent runs. Price clustering near the Nash price reflects competitive strategies, while clustering toward the monopoly price indicates collusive tendencies.}
    \label{fig:three_models_heatmap}
\end{figure}
\gls{acr:ppo} agents tightly cluster their pricing distributions around the Nash price, demonstrating strong competitiveness and minimal dispersion. Similarly, \gls{acr:dqn} agents primarily select prices near the Nash price, with symmetric distributions reflecting coordinated and fair competition. In contrast, \gls{acr:tql} agents exhibit broader and more asymmetric price distributions, with some prices approaching the Nash price but many skewing toward higher levels near monopoly pricing. These results highlight \gls{acr:tql}'s tendency to favor high-price strategies without achieving pricing consistency.

Normalized profit values \(\Delta\), as shown in Figure\ref{fig:bertrand_models_delta_boxplot}(a), further emphasize these trends. \gls{acr:dqn} agents achieve \(\Delta\) values centered around 0.1, reflecting stable pricing near Nash equilibrium. \gls{acr:ppo} agents maintain \(\Delta\) values near 0, indicating negligible collusion and strong competitiveness. By contrast, \gls{acr:tql} agents achieve significantly higher \(\Delta\) values around 0.6, with greater variability, indicating higher profits but reduced stability.

In the Standard and Edgeworth Bertrand models, \gls{acr:dqn} and \gls{acr:ppo} agents show more dispersed pricing behaviors, as depicted in Figure~\ref{fig:three_models_heatmap}(b) and Figure~\ref{fig:three_models_heatmap}(c), though their prices still cluster around the Nash price. This dispersion reflects reduced pricing stability in these environments compared to Logit Bertrand. \gls{acr:tql} agents, in contrast, display concentrated pricing distributions near monopoly prices, with lower variability, highlighting their consistent preference for high-price strategies.

Figure~\ref{fig:bertrand_models_delta_boxplot}
show \(\Delta\) results across the models. In the Standard and Edgeworth Bertrand models, \gls{acr:dqn} agents achieve \(\Delta\) values around 0.3, reflecting a higher tendency toward supra-competitive pricing compared to Logit Bertrand. \gls{acr:ppo} agents maintain \(\Delta\) values near 0 in the Standard model and around 0.34 in the Edgeworth model, demonstrating strong competitive behavior and stability. \gls{acr:tql} agents achieve \(\Delta\) values near 0.5 in both models, slightly lower than their 0.6 \(\Delta\) in Logit Bertrand, suggesting marginally reduced collusion and variability.\medskip

\noindent\textbf{Implications}\quad
These findings highlight the adaptability of \gls{acr:drl} agents, which consistently converge to competitive pricing strategies near the Nash price, especially in the Logit Bertrand model.
However, \gls{acr:drl} agents show reduced pricing stability in the Standard and Edgeworth Bertrand models, with greater dispersion in their pricing strategies. In contrast, \gls{acr:tql} agents maintain consistent high-price strategies across all models, favoring collusion but sacrificing stability and competitiveness.
\begin{figure}[!hb]
    \centering
    \begin{subfigure}[t]{0.31\linewidth}
        \centering
        \resizebox{\linewidth}{!}{
\begin{tikzpicture}

\definecolor{darkgray176}{RGB}{176,176,176}
\definecolor{darkorange25512714}{RGB}{255,127,14}
\definecolor{forestgreen4416044}{RGB}{44,160,44}
\definecolor{steelblue31119180}{RGB}{31,119,180}
\scriptsize

\begin{axis}[
width=0.9\textwidth, 
height=0.64\textwidth, 
tick align=outside,
tick pos=left,
x grid style={darkgray176},
xmin=0.5, xmax=3.5,
xtick style={color=black},
xtick={1,2,3},
xticklabels={DQN,PPO,TQL},
y grid style={darkgray176},
ylabel={\(\displaystyle \Delta\)},
ymajorgrids,
ymin=-0.0820634489027114, ymax=0.994413891494432,
ytick style={color=black}
]
\path [draw=black, fill=white, opacity=0.7]
(axis cs:0.7,0.053988935039329)
--(axis cs:1.3,0.053988935039329)
--(axis cs:1.3,0.0983223743668865)
--(axis cs:0.7,0.0983223743668865)
--(axis cs:0.7,0.053988935039329)
--cycle;
\addplot [black]
table {%
1 0.053988935039329
1 0.00396320331744522
};
\addplot [black]
table {%
1 0.0983223743668865
1 0.130471846606564
};
\addplot [black]
table {%
0.85 0.00396320331744522
1.15 0.00396320331744522
};
\addplot [black]
table {%
0.85 0.130471846606564
1.15 0.130471846606564
};
\path [draw=black, fill=white, opacity=0.7]
(axis cs:1.7,-0.0234424839227431)
--(axis cs:2.3,-0.0234424839227431)
--(axis cs:2.3,0.116774817058084)
--(axis cs:1.7,0.116774817058084)
--(axis cs:1.7,-0.0234424839227431)
--cycle;
\addplot [black]
table {%
2 -0.0234424839227431
2 -0.0331326607028413
};
\addplot [black]
table {%
2 0.116774817058084
2 0.11705410420102
};
\addplot [black]
table {%
1.85 -0.0331326607028413
2.15 -0.0331326607028413
};
\addplot [black]
table {%
1.85 0.11705410420102
2.15 0.11705410420102
};
\path [draw=black, fill=white, opacity=0.7]
(axis cs:2.7,0.518343467263942)
--(axis cs:3.3,0.518343467263942)
--(axis cs:3.3,0.697089797960653)
--(axis cs:2.7,0.697089797960653)
--(axis cs:2.7,0.518343467263942)
--cycle;
\addplot [black]
table {%
3 0.518343467263942
3 0.333363453670027
};
\addplot [black]
table {%
3 0.697089797960653
3 0.945483103294562
};
\addplot [black]
table {%
2.85 0.333363453670027
3.15 0.333363453670027
};
\addplot [black]
table {%
2.85 0.945483103294562
3.15 0.945483103294562
};
\addplot [thick, black]
table {%
0.7 0.0798976910262989
1.3 0.0798976910262989
};
\addplot [thick, black]
table {%
1.7 0.116744850996605
2.3 0.116744850996605
};
\addplot [thick, black]
table {%
2.7 0.601911361124162
3.3 0.601911361124162
};
\end{axis}

\end{tikzpicture}}
        \subcaption*{(a) Logit Bertrand}
        \label{fig:logit_delta_boxplot}
    \end{subfigure}
    \hspace{0.01\linewidth}
    \begin{subfigure}[t]{0.31\linewidth}
        \centering
        \resizebox{\linewidth}{!}{
\begin{tikzpicture}

\definecolor{darkgray176}{RGB}{176,176,176}
\definecolor{darkorange25512714}{RGB}{255,127,14}
\definecolor{forestgreen4416044}{RGB}{44,160,44}
\definecolor{steelblue31119180}{RGB}{31,119,180}
\scriptsize

\begin{axis}[
width=0.9\textwidth, 
height=0.64\textwidth, 
tick align=outside,
tick pos=left,
x grid style={darkgray176},
xmin=0.5, xmax=3.5,
xtick style={color=black},
xtick={1,2,3},
xticklabels={DQN,PPO,TQL},
y grid style={darkgray176},
ylabel={\(\displaystyle \Delta\)},
ymajorgrids,
ymin=0.11091831632653, ymax=0.573727602040816,
ytick style={color=black}
]
\path [draw=black, fill=white, opacity=0.7]
(axis cs:0.7,0.141245918367347)
--(axis cs:1.3,0.141245918367347)
--(axis cs:1.3,0.268627551020408)
--(axis cs:0.7,0.268627551020408)
--(axis cs:0.7,0.141245918367347)
--cycle;
\addplot [black]
table {%
1 0.141245918367347
1 0.131955102040816
};
\addplot [black]
table {%
1 0.268627551020408
1 0.386928571428571
};
\addplot [black]
table {%
0.85 0.131955102040816
1.15 0.131955102040816
};
\addplot [black]
table {%
0.85 0.386928571428571
1.15 0.386928571428571
};
\path [draw=black, fill=white, opacity=0.7]
(axis cs:1.7,0.243177551020408)
--(axis cs:2.3,0.243177551020408)
--(axis cs:2.3,0.245108673469388)
--(axis cs:1.7,0.245108673469388)
--(axis cs:1.7,0.243177551020408)
--cycle;
\addplot [black]
table {%
2 0.243177551020408
2 0.243177551020408
};
\addplot [black]
table {%
2 0.245108673469388
2 0.245373469387755
};
\addplot [black]
table {%
1.85 0.243177551020408
2.15 0.243177551020408
};
\addplot [black]
table {%
1.85 0.245373469387755
2.15 0.245373469387755
};
\path [draw=black, fill=white, opacity=0.7]
(axis cs:2.7,0.408132653061224)
--(axis cs:3.3,0.408132653061224)
--(axis cs:3.3,0.473797704081632)
--(axis cs:2.7,0.473797704081632)
--(axis cs:2.7,0.408132653061224)
--cycle;
\addplot [black]
table {%
3 0.408132653061224
3 0.350676530612245
};
\addplot [black]
table {%
3 0.473797704081632
3 0.55269081632653
};
\addplot [black]
table {%
2.85 0.350676530612245
3.15 0.350676530612245
};
\addplot [black]
table {%
2.85 0.55269081632653
3.15 0.55269081632653
};
\addplot [thick, black]
table {%
0.7 0.222119897959184
1.3 0.222119897959184
};
\addplot [thick, black]
table {%
1.7 0.244897959183673
2.3 0.244897959183673
};
\addplot [thick, black]
table {%
2.7 0.458953571428571
3.3 0.458953571428571
};
\end{axis}

\end{tikzpicture}}
        \subcaption*{(b) Standard Bertrand}
        \label{fig:standard_delta_boxplot}
    \end{subfigure}
    \hspace{0.01\linewidth}
    \begin{subfigure}[t]{0.31\linewidth}
        \centering
        \resizebox{\linewidth}{!}{
\begin{tikzpicture}

\definecolor{darkgray176}{RGB}{176,176,176}
\definecolor{darkorange25512714}{RGB}{255,127,14}
\definecolor{forestgreen4416044}{RGB}{44,160,44}
\definecolor{steelblue31119180}{RGB}{31,119,180}
\scriptsize

\begin{axis}[
width=0.9\textwidth, 
height=0.64\textwidth, 
tick align=outside,
tick pos=left,
x grid style={darkgray176},
xmin=0.5, xmax=3.5,
xtick style={color=black},
xtick={1,2,3},
xticklabels={DQN,PPO,TQL},
y grid style={darkgray176},
ylabel={\(\displaystyle \Delta\)},
ymajorgrids,
ymin=0.130032418367347, ymax=0.514121663265306,
ytick style={color=black}
]
\path [draw=black, fill=white, opacity=0.7]
(axis cs:0.7,0.252794795918367)
--(axis cs:1.3,0.252794795918367)
--(axis cs:1.3,0.329151020408163)
--(axis cs:0.7,0.329151020408163)
--(axis cs:0.7,0.252794795918367)
--cycle;
\addplot [black]
table {%
1 0.252794795918367
1 0.147491020408163
};
\addplot [black]
table {%
1 0.329151020408163
1 0.360331836734694
};
\addplot [black]
table {%
0.85 0.147491020408163
1.15 0.147491020408163
};
\addplot [black]
table {%
0.85 0.360331836734694
1.15 0.360331836734694
};
\path [draw=black, fill=white, opacity=0.7]
(axis cs:1.7,0.334625612244898)
--(axis cs:2.3,0.334625612244898)
--(axis cs:2.3,0.336772959183673)
--(axis cs:1.7,0.336772959183673)
--(axis cs:1.7,0.334625612244898)
--cycle;
\addplot [black]
table {%
2 0.334625612244898
2 0.331531632653061
};
\addplot [black]
table {%
2 0.336772959183673
2 0.339381428571428
};
\addplot [black]
table {%
1.85 0.331531632653061
2.15 0.331531632653061
};
\addplot [black]
table {%
1.85 0.339381428571428
2.15 0.339381428571428
};
\path [draw=black, fill=white, opacity=0.7]
(axis cs:2.7,0.458919132653061)
--(axis cs:3.3,0.458919132653061)
--(axis cs:3.3,0.489352959183673)
--(axis cs:2.7,0.489352959183673)
--(axis cs:2.7,0.458919132653061)
--cycle;
\addplot [black]
table {%
3 0.458919132653061
3 0.413269795918367
};
\addplot [black]
table {%
3 0.489352959183673
3 0.49666306122449
};
\addplot [black]
table {%
2.85 0.413269795918367
3.15 0.413269795918367
};
\addplot [black]
table {%
2.85 0.49666306122449
3.15 0.49666306122449
};
\addplot [thick, black]
table {%
0.7 0.293095306122449
1.3 0.293095306122449
};
\addplot [thick, black]
table {%
1.7 0.33665387755102
2.3 0.33665387755102
};
\addplot [thick, black]
table {%
2.7 0.459071224489796
3.3 0.459071224489796
};
\end{axis}

\end{tikzpicture}}
        \subcaption*{(c) Edgeworth Bertrand}
        \label{fig:edgeworth_delta_boxplot}
    \end{subfigure}
    \caption{Normalized profit comparisons across three Bertrand models. \(\Delta\) values closer to 0 indicate profits near the Nash equilibrium, reflecting more competitive market behavior, while values approaching 1 correspond to profits near the monopoly price, indicating stronger supra-competitive tendencies.}

    \label{fig:bertrand_models_delta_boxplot}
\end{figure}
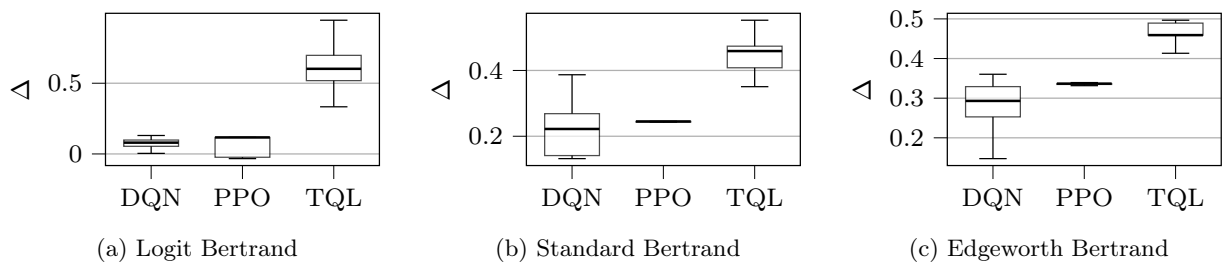

\textit{\textbf{Insight 3.} Homogeneous \gls{acr:drl} agents (\gls{acr:dqn}, \gls{acr:ppo}) adapt their strategies across demand models, showing strong competitiveness near Nash price in Logit Bertrand but reduced stability in Standard and Edgeworth Bertrand. \gls{acr:tql} agents consistently favor pricing strategies close to the monopoly price and hence benefit collusion.}


\subsection{Comparing \gls{acr:drl} and \gls{acr:tql} in Competitive Pricing}
While traditional research on algorithmic pricing has primarily focused on \gls{acr:tql}, recent studies increasingly consider advanced \gls{acr:drl} algorithms such as \gls{acr:dqn} and \gls{acr:ppo}.
Against this background, this section evaluates \gls{acr:drl} algorithms in a competitive pricing environment to determine whether they outperform \gls{acr:tql} in achieving superior pricing strategies and profitability.

To ensure a fair comparison, the \gls{acr:tql} agent is pretrained before interacting with a \gls{acr:drl} competitor. This design choice accounts for fundamental differences in how these algorithms learn: TQL updates its strategy at every timestep from the start, whereas DQN first collects data for its replay buffer before batch learning. As a result, their learning rhythms are inherently misaligned, and once DQN begins updating, it benefits from using more data per step. Furthermore, our setup reflects a realistic market dynamic where an incumbent firm relies on a well-established TQL-based strategy, while a newcomer leverages \gls{acr:drl} to adapt dynamically and compete effectively.
Specifically, we simulate a duopoly where one firm employs a pretrained \gls{acr:tql} strategy, and the other adopts a dynamically learning \gls{acr:drl} algorithm (\gls{acr:dqn} or \gls{acr:ppo}). The \gls{acr:tql} strategy, pretrained over 1,000,000 timesteps with two homogeneous agents, converges to a stable pricing strategy. During interactions, \gls{acr:tql} operates without further updates, ensuring a consistent baseline. 
The interaction phase spans 100,000 timesteps, divided into 100 epochs of 1,000 timesteps each. \gls{acr:drl} agents start without prior knowledge and adapt their strategies based on \gls{acr:tql}'s behavior. Results include price trajectories for \gls{acr:tql} vs. \gls{acr:ppo} and \gls{acr:tql} vs. \gls{acr:dqn}, along with normalized profit \(\Delta\) distributions over the final 10,000 timesteps.\medskip

\noindent\textbf{Main Findings}\quad In the Logit Bertrand setting, \gls{acr:tql}'s prices remain stable, clustering around 1.8 for \gls{acr:tql} vs. \gls{acr:dqn} (Figure~\ref{fig:Combined_price_learning_curve_tql_drl_logit}(a)) and 1.75 for \gls{acr:tql} vs. \gls{acr:ppo} (Figure~ \ref{fig:Combined_price_learning_curve_tql_drl_logit}(b)). These prices align closely with the monopoly price, reflecting \gls{acr:tql}'s preference for high-price strategies. In contrast, \gls{acr:dqn} and \gls{acr:ppo} gradually reduce their prices, stabilizing around a price of 1.6. \gls{acr:dqn} exhibits minimal fluctuations by the end, showcasing a learned competitive low-price strategy. \gls{acr:ppo}, while initially volatile, converges to a similarly aggressive low-price strategy. Both \gls{acr:drl} algorithms demonstrate superior adaptability by consistently underpricing \gls{acr:tql}. The profit differences further highlight \gls{acr:drl}'s advantages. Figure~\ref{fig:delta_tql_drl_logit} shows that \gls{acr:dqn} and \gls{acr:ppo} achieve \(\Delta\) values predominantly above 1.0, reflecting stable and higher profitability compared to \gls{acr:tql}, whose \(\Delta\) values cluster around 0 with high variability.

The results in the Standard Bertrand and Edgeworth Bertrand models, shown in Appendix~\ref{appendix:supplementary_results}, confirm these observations: in the Standard Bertrand model, \gls{acr:tql} again consistently maintains a high-price strategy near the monopoly price. Contrarily, \gls{acr:dqn} and \gls{acr:ppo} converge to competitive low-price strategies, achieving superior profitability. 
\begin{figure}[!ht]
	\centering
	\begin{minipage}[t]{0.48\linewidth}
		\centering
		\includegraphics[width=1.0\linewidth]{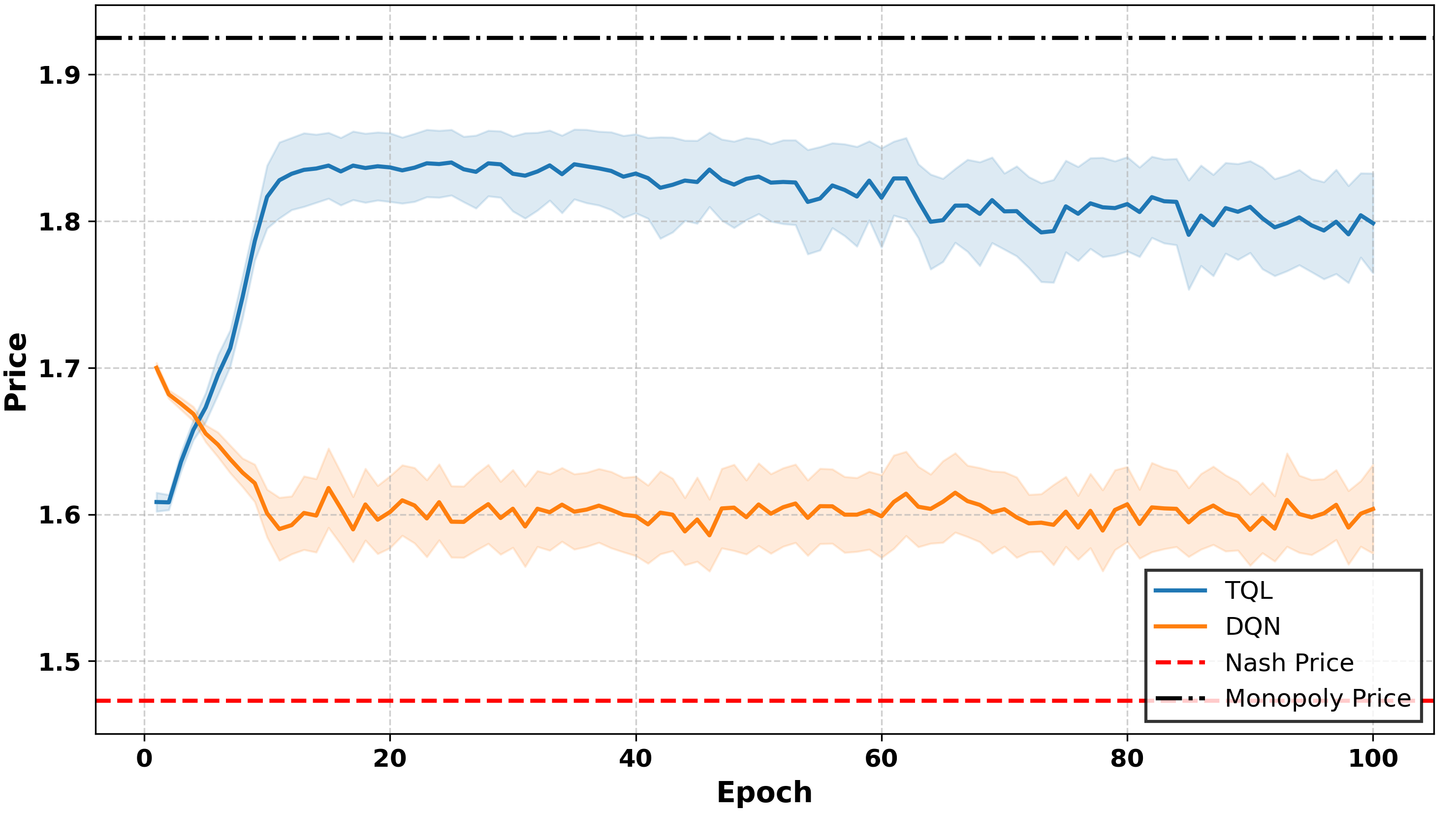}
        \subcaption*{(a) TQL vs DQN: Price Trend (with 95\% CI)}
        \label{TQL_DQN_price_per_iteration_logit}
	\end{minipage}
	\hfill
	\begin{minipage}[t]{0.48\linewidth}
		\centering
		\includegraphics[width=1.0\linewidth]{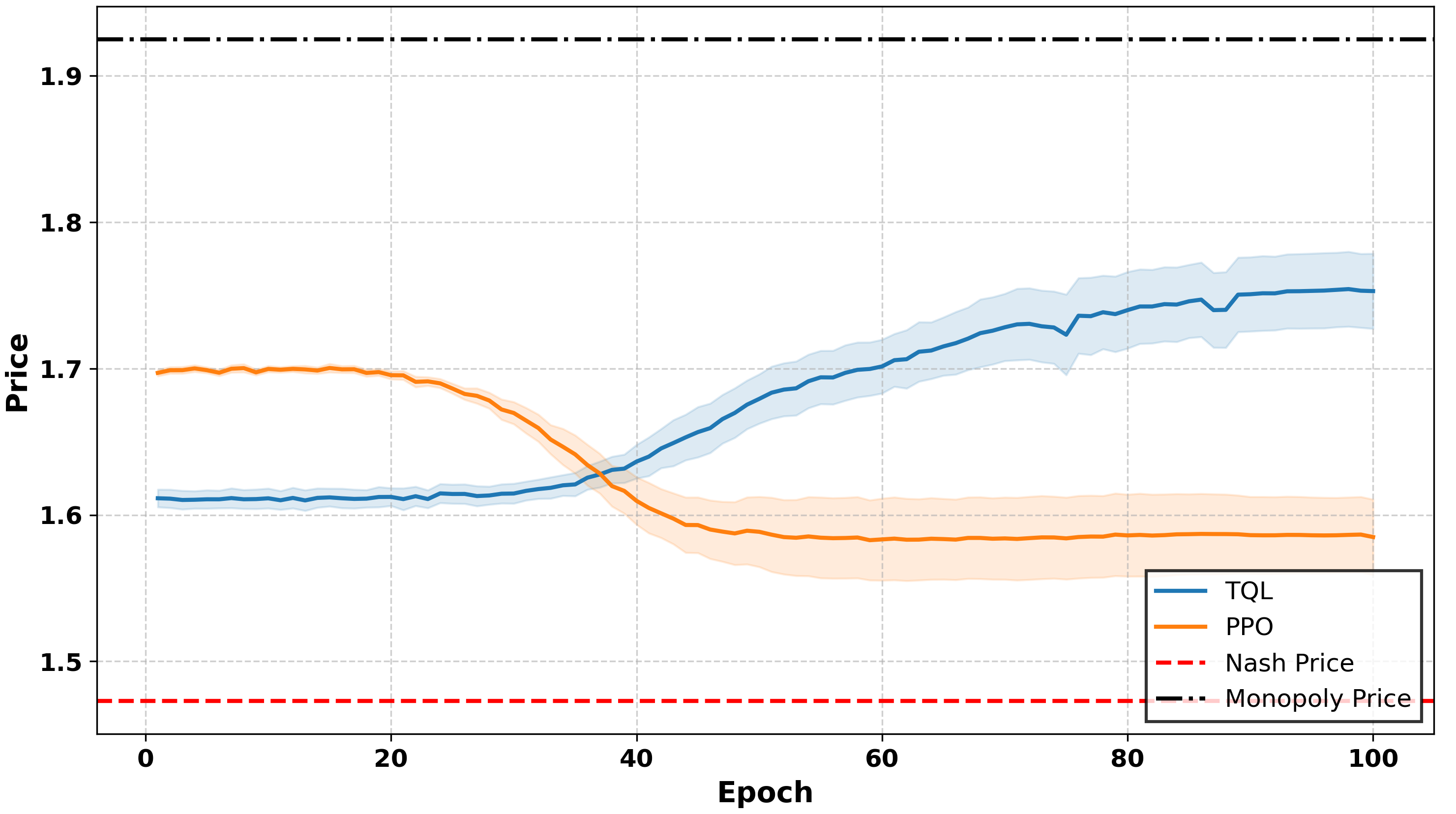}
        \subcaption*{(b) TQL vs PPO: Price Trend (with 95\% CI)}

        \label{TQL_PPO_price_per_iteration_logit}
	\end{minipage}
        \caption{Logit Bertrand: Price learning curves for the Logit Bertrand model, showing competition between \gls{acr:tql} and \gls{acr:dqn} (left) and between \gls{acr:tql} and \gls{acr:ppo} (right). Results are based on 20 independent runs, with mean values and 95\% confidence intervals displayed.}
    \label{fig:Combined_price_learning_curve_tql_drl_logit}
\end{figure} 
The Edgeworth Bertrand model 
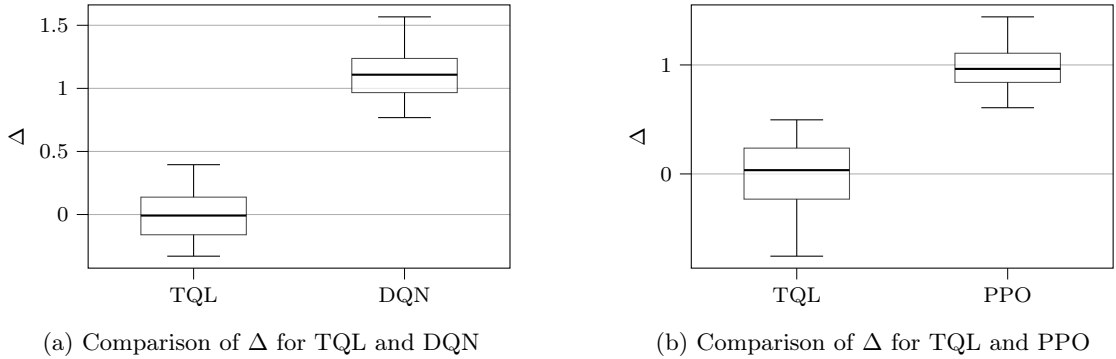
\begin{figure}[!ht]
	\centering
	\begin{minipage}{0.48\linewidth}
		\centering
\begin{tikzpicture}

\definecolor{darkgray176}{RGB}{176,176,176}
\definecolor{darkorange25512714}{RGB}{255,127,14}
\definecolor{steelblue31119180}{RGB}{31,119,180}
\scriptsize

\begin{axis}[
width=0.9\textwidth, 
height=0.64\textwidth, 
tick align=outside,
tick pos=left,
x grid style={darkgray176},
xmin=0.5, xmax=2.5,
xtick style={color=black},
xtick={1,2},
xticklabels={TQL,DQN},
y grid style={darkgray176},
ylabel={\(\displaystyle \Delta\)},
ymajorgrids,
ymin=-0.424962187383692, ymax=1.66153591900822,
ytick style={color=black}
]
\path [draw=black, fill=white, opacity=0.7]
(axis cs:0.75,-0.160343440262597)
--(axis cs:1.25,-0.160343440262597)
--(axis cs:1.25,0.137800482213879)
--(axis cs:0.75,0.137800482213879)
--(axis cs:0.75,-0.160343440262597)
--cycle;
\addplot [black]
table {%
1 -0.160343440262597
1 -0.330121364365878
};
\addplot [black]
table {%
1 0.137800482213879
1 0.395090868977581
};
\addplot [black]
table {%
0.875 -0.330121364365878
1.125 -0.330121364365878
};
\addplot [black]
table {%
0.875 0.395090868977581
1.125 0.395090868977581
};
\path [draw=black, fill=white, opacity=0.7]
(axis cs:1.75,0.966422602955375)
--(axis cs:2.25,0.966422602955375)
--(axis cs:2.25,1.23711815658151)
--(axis cs:1.75,1.23711815658151)
--(axis cs:1.75,0.966422602955375)
--cycle;
\addplot [black]
table {%
2 0.966422602955375
2 0.768174298121732
};
\addplot [black]
table {%
2 1.23711815658151
2 1.56669509599041
};
\addplot [black]
table {%
1.875 0.768174298121732
2.125 0.768174298121732
};
\addplot [black]
table {%
1.875 1.56669509599041
2.125 1.56669509599041
};
\addplot [thick, black]
table {%
0.75 -0.00796919777116109
1.25 -0.00796919777116109
};
\addplot [thick, black]
table {%
1.75 1.10831807511545
2.25 1.10831807511545
};
\end{axis}

\end{tikzpicture}
        \subcaption*{(a) Comparison of \(\displaystyle \Delta\) for TQL and DQN}
        \label{Delta_tql_dqn_logit}
	\end{minipage}
	\begin{minipage}{0.48\linewidth}
		\centering
\begin{tikzpicture}

\definecolor{darkgray176}{RGB}{176,176,176}
\definecolor{darkorange25512714}{RGB}{255,127,14}
\definecolor{steelblue31119180}{RGB}{31,119,180}
\scriptsize

\begin{axis}[
width=0.9\textwidth, 
height=0.64\textwidth, 
tick align=outside,
tick pos=left,
x grid style={darkgray176},
xmin=0.5, xmax=2.5,
xtick style={color=black},
xtick={1,2},
xticklabels={TQL,PPO},
y grid style={darkgray176},
ylabel={\(\displaystyle \Delta\)},
ymajorgrids,
ymin=-0.864476283614648, ymax=1.55086020812368,
ytick style={color=black}
]
\path [draw=black, fill=white, opacity=0.7]
(axis cs:0.75,-0.23138307600085)
--(axis cs:1.25,-0.23138307600085)
--(axis cs:1.25,0.236780245874345)
--(axis cs:0.75,0.236780245874345)
--(axis cs:0.75,-0.23138307600085)
--cycle;
\addplot [black]
table {%
1 -0.23138307600085
1 -0.754688261262906
};
\addplot [black]
table {%
1 0.236780245874345
1 0.496320826811037
};
\addplot [black]
table {%
0.875 -0.754688261262906
1.125 -0.754688261262906
};
\addplot [black]
table {%
0.875 0.496320826811037
1.125 0.496320826811037
};
\path [draw=black, fill=white, opacity=0.7]
(axis cs:1.75,0.839818162338601)
--(axis cs:2.25,0.839818162338601)
--(axis cs:2.25,1.1073313453385)
--(axis cs:1.75,1.1073313453385)
--(axis cs:1.75,0.839818162338601)
--cycle;
\addplot [black]
table {%
2 0.839818162338601
2 0.607439177276334
};
\addplot [black]
table {%
2 1.1073313453385
2 1.44107218577193
};
\addplot [black]
table {%
1.875 0.607439177276334
2.125 0.607439177276334
};
\addplot [black]
table {%
1.875 1.44107218577193
2.125 1.44107218577193
};
\addplot [thick, black]
table {%
0.75 0.0345503224603321
1.25 0.0345503224603321
};
\addplot [thick, black]
table {%
1.75 0.9637169327331
2.25 0.9637169327331
};
\end{axis}

\end{tikzpicture}
        \subcaption*{(b) Comparison of \(\displaystyle \Delta\) for TQL and PPO}
        \label{Delta_tql_ppo_logit}
	\end{minipage}
        \caption{Boxplots of normalized profit \(\Delta\) for the Logit Bertrand model, comparing competition between \gls{acr:tql} and \gls{acr:dqn} (left) and between \gls{acr:tql} and \gls{acr:ppo} (right). \gls{acr:dqn} and \gls{acr:ppo} both achieve higher profits with lower   variance
         compared to \gls{acr:tql}.}
        \label{fig:delta_tql_drl_logit}
\end{figure}
exhibits slightly higher volatility in pricing and profit trends, but \gls{acr:drl} algorithms still outperform \gls{acr:tql}, further demonstrating their robustness across different market conditions.\medskip

\noindent\textbf{Implications}\quad These results highlight the limitations of \gls{acr:tql}, stemming from its static nature and inability to dynamically adapt to competitors' strategies. By contrast, \gls{acr:drl} algorithms excel in dynamic environments, learning competitive strategies that result in lower prices and higher profitability. The consistent performance of \gls{acr:dqn} and \gls{acr:ppo} across different Bertrand models underscores their adaptability and effectiveness in automated pricing.

For practitioners, our findings emphasize the importance of transitioning from traditional \gls{acr:tql} to advanced \gls{acr:drl} techniques to achieve competitive pricing. The \gls{acr:drl} algorithm's ability to outperform \gls{acr:tql} in both pricing and profit stability demonstrates its practicality for modern markets.

\textit{\textbf{Insight 2.} The studied \gls{acr:drl} algorithms (\gls{acr:dqn}, \gls{acr:ppo}) consistently outperform \gls{acr:tql} by converging to lower prices and achieving higher, more stable profits. This highlights the superior adaptability of \gls{acr:drl} methods over \gls{acr:tql}, making them better suited to achieve competitive pricing.}

\subsection{Pricing behavior of heterogeneous \gls{acr:drl} Agents}
Understanding how different \gls{acr:drl} algorithms interact in competitive markets provides valuable insights into their strategic adaptability and market influence. Against this background, we analyze interactions between two heterogeneous agents: one using \gls{acr:dqn} and the other employing \gls{acr:ppo}, within the Bertrand competition model. By studying algorithmic diversity, we assess its impact on pricing strategies, profitability, and competitive dynamics.

Specifically, our experiment features two parallel market environments, each designed to focus on a specific algorithm's learning process. In the \gls{acr:dqn} environment, Agent 0 employs \gls{acr:ppo}, while Agent 1 uses \gls{acr:dqn}. In the \gls{acr:ppo} environment, Agent 0 uses \gls{acr:ppo}, and Agent 1 adopts \gls{acr:dqn}. Policy weights are exchanged every 1,000 timesteps between the two environments to enhance adaptability. For example, updated \gls{acr:ppo} policy weights from the \gls{acr:ppo} environment transfer to Agent 0 in the \gls{acr:dqn} environment, while \gls{acr:dqn} weights from the \gls{acr:dqn} environment transfer to Agent 1 in the \gls{acr:ppo} environment. This exchange allows agents to integrate complementary strategies learned in the alternate environment.

One comment on this experimental design is in order. We intentionally separate the learning of \gls{acr:dqn} and \gls{acr:ppo} into independent environments to avoid conflicts arising from their distinct learning mechanisms. \gls{acr:dqn}, as an off-policy algorithm, performs frequent updates via experience replay, enabling continuous strategy refinement. \gls{acr:ppo}, an on-policy algorithm, requires complete trajectory collection for updates, leading to less frequent but larger adjustments. Mixing the algorithms in a single environment could disrupt \gls{acr:ppo}'s stability due to \gls{acr:dqn}'s frequent updates. This setup replicates real-world market conditions, where companies adapt their algorithms gradually based on long-term strategies rather than immediate competitor reactions.\medskip

\noindent\textbf{Main Findings}\quad In the \gls{acr:ppo} environment, as shown in Figure~\ref{PPO_env_price_profit_learning_curve_logit}(a), the \gls{acr:ppo} agent stabilizes near the Nash price, adopting a competitive low-price strategy. Meanwhile, the \gls{acr:dqn} agent gradually increases its price toward the monopoly price. Figure~\ref{PPO_env_price_profit_learning_curve_logit}(b) reveals that \gls{acr:ppo} achieves profits near the monopoly level, while \gls{acr:dqn}'s profits decrease and stabilize near the Nash level. 

In the \gls{acr:dqn} environment (Figure~\ref{DQN_env_price_profit_learning_curve_logit}), both agents converge to similar mean prices, but \gls{acr:ppo} exhibits more stable pricing. Initially, \gls{acr:dqn} achieves slightly higher profits, but \gls{acr:ppo} matches this performance in later iterations while maintaining lower variance.

Results in the Standard and Edgeworth Bertrand models, presented in Appendix~\ref{appendix:supplementary_results}, align with Logit Bertrand findings, with some variations. In the \gls{acr:ppo} environment, \gls{acr:ppo} consistently adopts a low-price strategy, stabilizing around 0.4 and achieving profits near the monopoly level. \gls{acr:dqn}, by contrast, converges to higher prices and lower profits near the Nash level. In the \gls{acr:dqn} environment, \gls{acr:dqn} stabilizes at slightly lower prices than \gls{acr:ppo} in the Standard Bertrand model, achieving higher profits. This role reversal highlights the influence of the learning environment on competitive dynamics. In the Edgeworth Bertrand model, the profit gap between \gls{acr:ppo} and \gls{acr:dqn} narrows significantly, showing overlapping profit curves. Increased market complexity in the Edgeworth Bertrand model diminishes algorithmic performance differences, fostering more balanced competition.\medskip
\begin{figure}[!ht]
	\centering
	\begin{minipage}[t]{0.48\linewidth}
		\centering
        \includegraphics[width=1.0\linewidth]{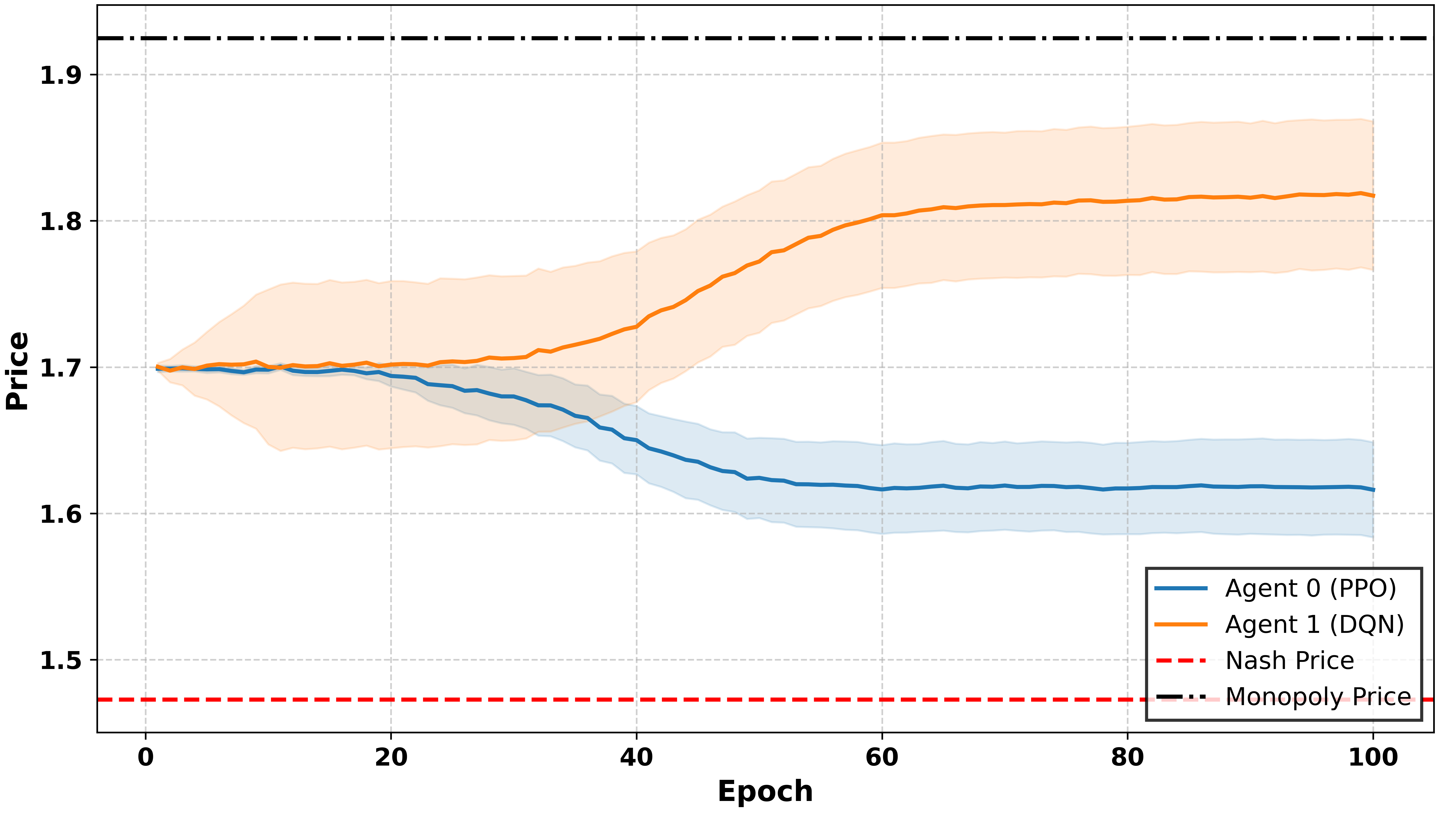}  
        \subcaption*{(a) Price per Epoch in PPO Environment}\label{PPO_env_price_learning_curve_logit}
	\end{minipage}
	\hfill
	\begin{minipage}[t]{0.48\linewidth}
		\centering
        \includegraphics[width=1.0\linewidth]{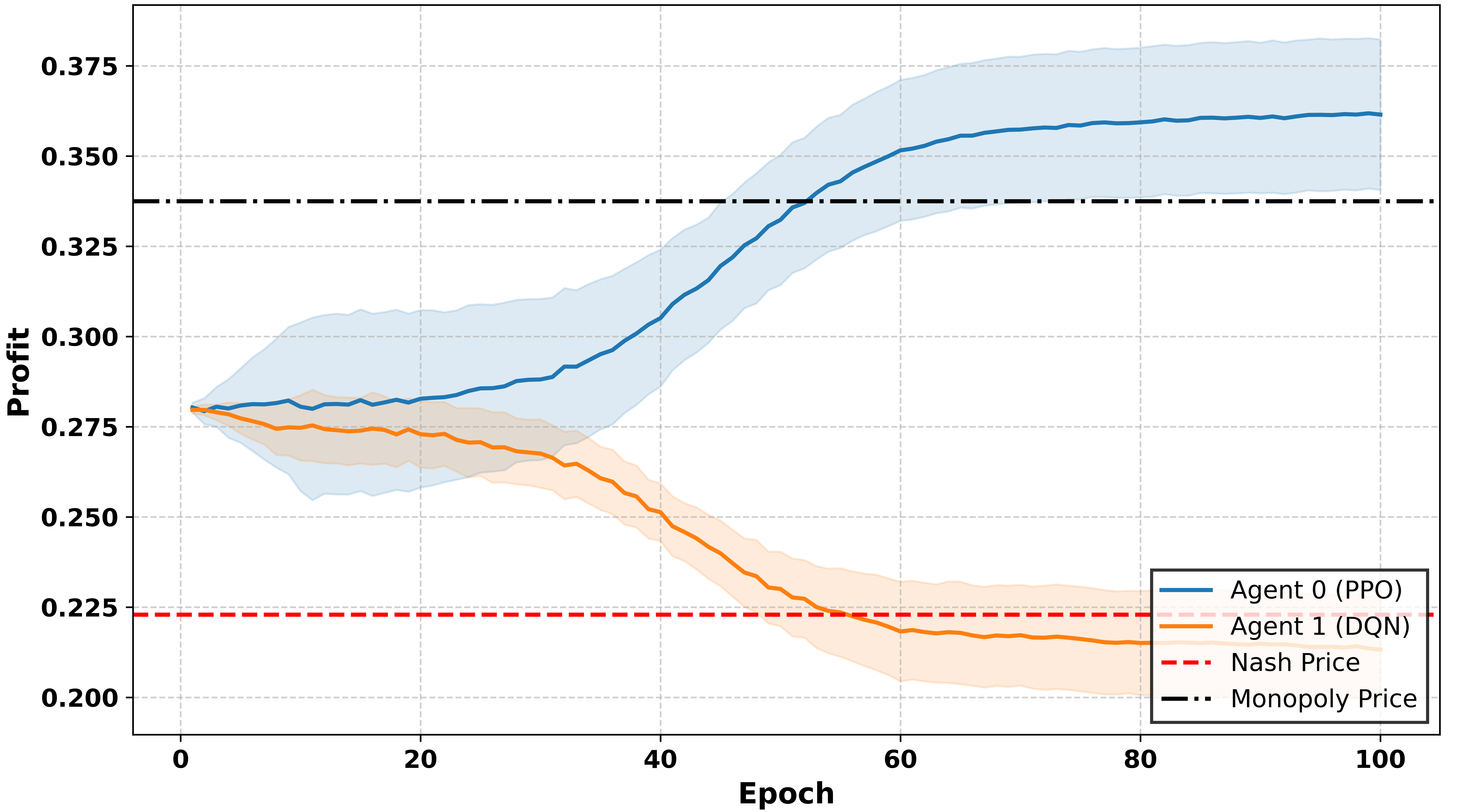}  
        \subcaption*{(b) Profit per Epoch in PPO Environment}\label{PPO_env_profit_learning_curve_logit}
	\end{minipage}
        \caption{Results for the Logit Bertrand model with a PPO environment. The left panel illustrates price dynamics, while the right panel depicts profit outcomes. Results are based on the mean values with 95\% confidence intervals from 20 independent runs.}
        \label{PPO_env_price_profit_learning_curve_logit}
\end{figure} 
\begin{figure}[!ht]
	\centering
	\begin{minipage}[t]{0.48\linewidth}
		\centering
        \includegraphics[width=1.0\linewidth]{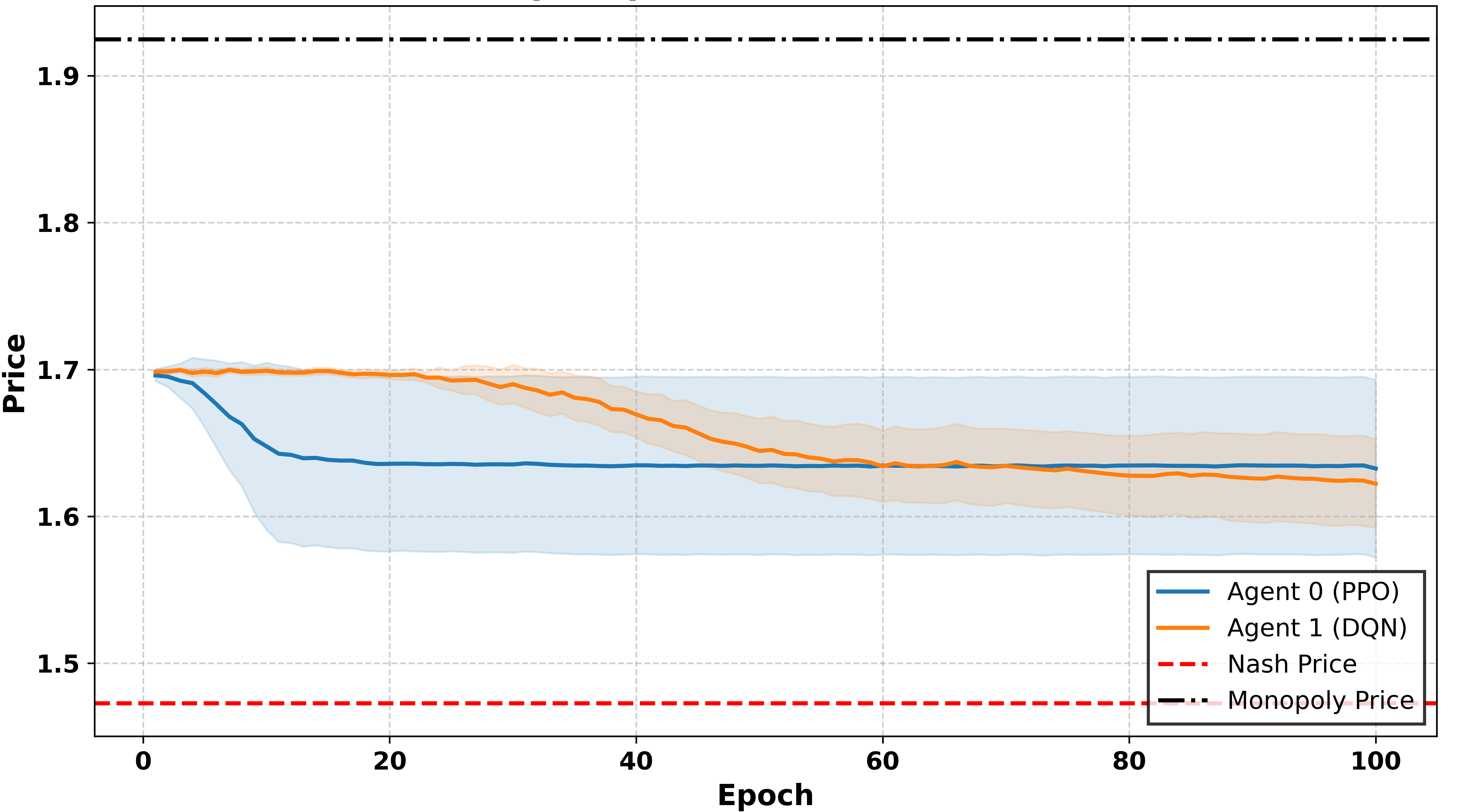}
        
        \subcaption*{(a) Price per Epoch in DQN Environment}
        \label{DQN_env_price_learning_curve_logit}
	\end{minipage}
	\hfill
	\begin{minipage}[t]{0.48\linewidth}
		\centering
        \includegraphics[width=1.0\linewidth]{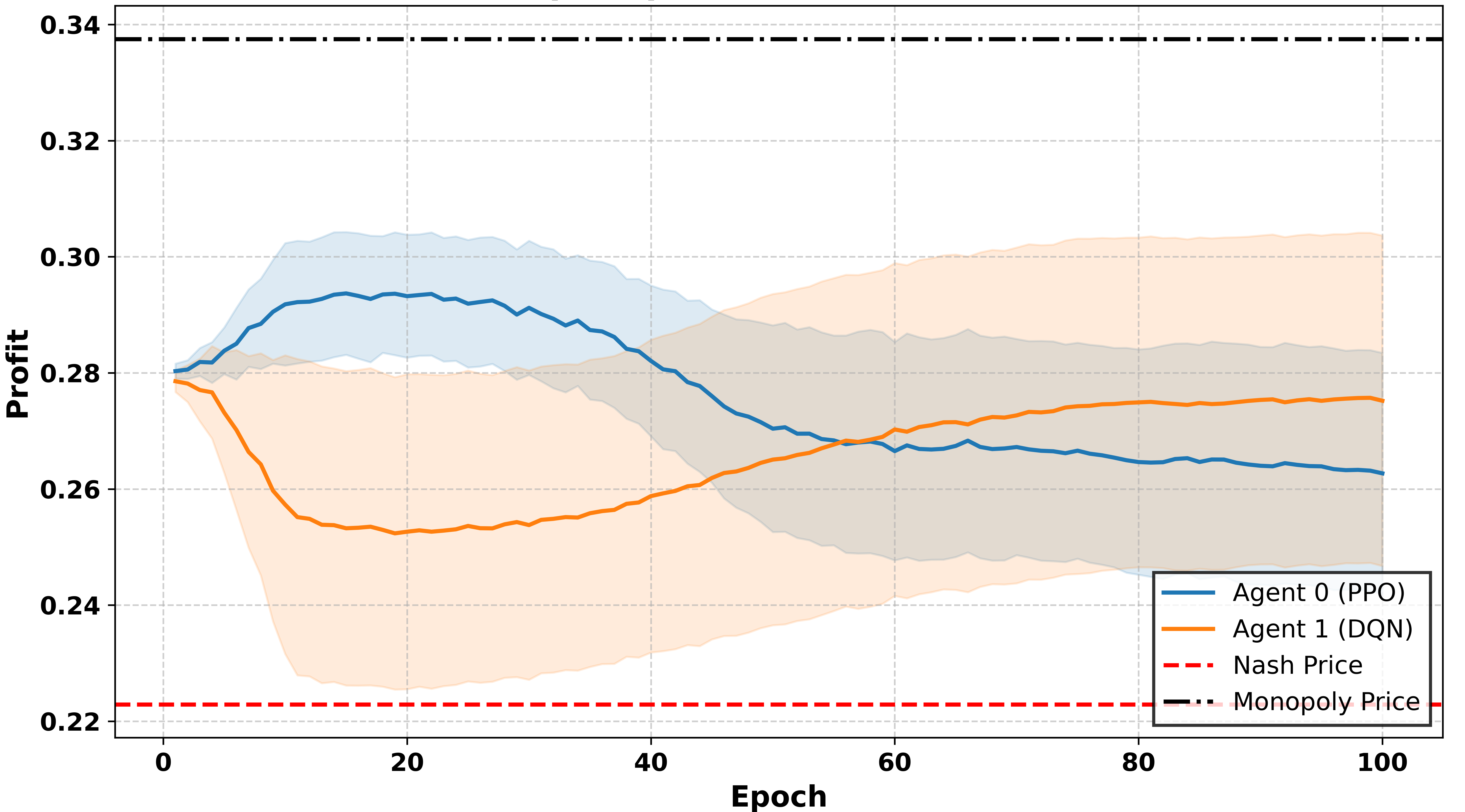}
        \subcaption*{(b) Profit per Epoch in DQN Environment}
        \label{DQN_env_profit_learning_curve_logit}
	\end{minipage}
        \caption{Results for the Logit Bertrand model with a DQN environment. The left panel illustrates pricing dynamics, while the right panel depicts profit outcomes. Results are based on the mean values with 95\% confidence intervals from 20 independent runs.}
        \label{DQN_env_price_profit_learning_curve_logit}
\end{figure} 

\noindent\textbf{Implications}\quad These results demonstrate \gls{acr:ppo}'s consistent advantage in adopting competitive low-price strategies and achieving higher profits across most scenarios. \gls{acr:dqn}, while less consistent, shows environment-specific strengths, particularly in the \gls{acr:dqn} environment within the Standard Bertrand model. The narrowing profit gap in the Edgeworth Bertrand model suggests that greater market complexity reduces algorithmic differences, creating more balanced competitive outcomes. These findings highlight the need to consider market dynamics when deploying \gls{acr:drl} algorithms in real-world settings.

\textit{\textbf{Insight 4.} Across the three Bertrand demand models, \gls{acr:ppo} consistently adopts lower pricing strategies and achieves higher profits, while \gls{acr:dqn}'s performance varies by environment. In more dynamic settings like the Edgeworth Bertrand model, the profit gap between the two algorithms narrows, reflecting the impact of market complexity on their competitiveness.}

\subsection{Impact of State-Space Design on \gls{acr:rl} Pricing Strategies}
In the following, we analyze how state-space definitions affect homogeneous \gls{acr:rl} agents, specifically \gls{acr:tql}, \gls{acr:dqn}, and \gls{acr:ppo}. While \gls{acr:dqn} and \gls{acr:ppo} exhibit minimal differences compared to the complete information setting, \gls{acr:tql} is more sensitive to state definitions. Therefore, this subsection focuses on how state-space design influences \gls{acr:tql}'s convergence behavior, particularly regarding the availability of opponent information.

State-space design plays a crucial role in pricing strategies. In a \textit{complete information setting}, each agent’s state $ s_t $ at time $ t $ includes a finite memory of past pricing decisions over $ l < t $ steps, capturing both its own and its opponent’s historical prices. Formally,
$$
    \mathcal{S} = \{s_t \mid s_t = (p_{i,t-k})_{i=0,1, k=1,\dots,l} \}
$$
where $ l $ denotes the memory length, yielding a state-space dimensionality of $ 2l $. This setting reflects a scenario where agents have full awareness of market dynamics.

However, in real-world competitive environments, access to competitor pricing data is often restricted. To model such conditions, we define a \textit{limited information setting}, where each agent observes only its own past prices:
$$
    \mathcal{S} = \{s_t \mid s_t = (p_{i,t-k})_{k=1,\dots,l} \}
$$
Here, the state space reduces to a dimensionality of $ l $, limiting the agent’s ability to infer competitor strategies. Comparing these two settings allows us to assess how information richness affects learning dynamics and competitive behavior.

We evaluate three memory lengths ($ l = 1, 2, 3 $) under both complete and limited information settings, yielding six unique configurations. Each configuration involves two homogeneous \gls{acr:tql} agents interacting over 1,000,000 timesteps per run, with results averaged over 20 random seeds. To assess long-term behavior, we focus on the final 10,000 timesteps, analyzing \gls{acr:rpdi} and $ \Delta $, specifically the normalized price and normalized profit. This setup enables a systematic evaluation of how state-space definitions influence learning and market dynamics.\medskip

\noindent\textbf{Main Findings}\quad For the Logit Bertrand model, Figures~\ref{fig:state_price_profit_diff_boxplot_logit} reveal distinct pricing and profit behaviors across state-space definitions. When agents observe both their own and opponent pricing histories ($ k1 $, $ k2 $, $ k3 $), normalized prices decrease as memory length increases. However, despite this decline, prices remain above Nash equilibrium, indicating supra-competitive behavior.

Conversely, when agents rely solely on self-observed pricing histories ($ self\_k1 $, $ self\_k2 $, $ self\_k3 $), prices \textbf{increase} with memory length. Notably, under shorter memory settings ($ self\_k1 $ and $ self\_k2 $), prices converge near Nash equilibrium, suggesting that limited information leads to more competitive pricing strategies (see Figure~\ref{fig:pricing_curve_comparison}).

These trends hold across alternative Bertrand models (Standard and Edgeworth), with minor variations (See Appendix~\ref{appendix:supplementary_results}). The consistency across different pricing models reinforces the generalizability of these findings.\medskip

\noindent\textbf{Implications}\quad The results indicate that the availability of opponent information plays a crucial role in shaping pricing behavior among homogeneous \gls{acr:tql} agents. When agents can observe both their own and their opponent’s past prices, supra-competitive pricing persists, though it gradually declines with increasing memory length. 
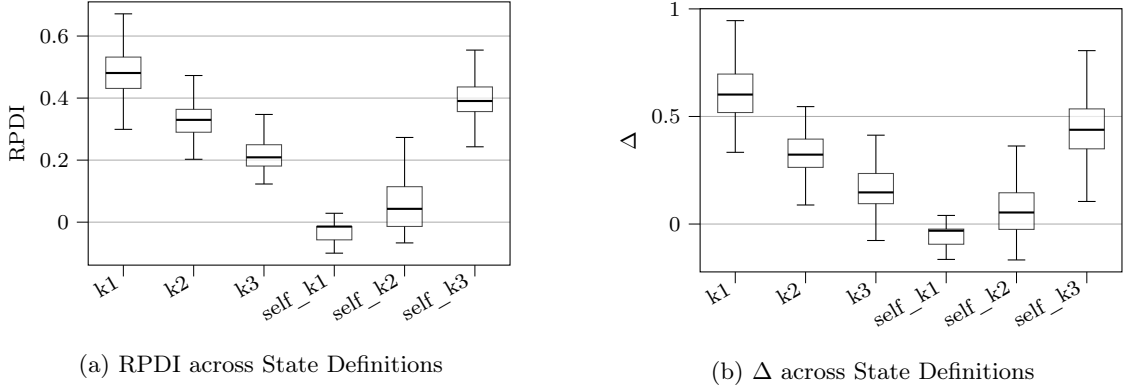
\begin{figure}[!ht]
	\centering
	\begin{minipage}{0.48\linewidth}
		\centering
\begin{tikzpicture}

\definecolor{darkgray176}{RGB}{176,176,176}
\definecolor{skyblue}{RGB}{135,206,235}
\scriptsize

\begin{axis}[
width=0.9\textwidth, 
height=0.64\textwidth, 
tick align=outside,
tick pos=left,
x grid style={darkgray176},
xmin=0.5, xmax=6.5,
xtick style={color=black},
xtick={1,2,3,4,5,6},
xticklabel style={rotate=30.0,anchor=east},
xticklabels={k1,k2,k3,self\_k1,self\_k2,self\_k3},
y grid style={darkgray176},
ylabel={RPDI},
ymajorgrids,
ymin=-0.138529571458096, ymax=0.710146727233428,
ytick style={color=black}
]
\path [draw=black, fill=white, opacity=0.7]
(axis cs:0.75,0.431118676574994)
--(axis cs:1.25,0.431118676574994)
--(axis cs:1.25,0.532156161797989)
--(axis cs:0.75,0.532156161797989)
--(axis cs:0.75,0.431118676574994)
--cycle;
\addplot [black]
table {%
1 0.431118676574994
1 0.299173086470092
};
\addplot [black]
table {%
1 0.532156161797989
1 0.671570531838359
};
\addplot [black]
table {%
0.875 0.299173086470092
1.125 0.299173086470092
};
\addplot [black]
table {%
0.875 0.671570531838359
1.125 0.671570531838359
};
\path [draw=black, fill=white, opacity=0.7]
(axis cs:1.75,0.289878366523003)
--(axis cs:2.25,0.289878366523003)
--(axis cs:2.25,0.363644624585085)
--(axis cs:1.75,0.363644624585085)
--(axis cs:1.75,0.289878366523003)
--cycle;
\addplot [black]
table {%
2 0.289878366523003
2 0.202681163055225
};
\addplot [black]
table {%
2 0.363644624585085
2 0.472740248284307
};
\addplot [black]
table {%
1.875 0.202681163055225
2.125 0.202681163055225
};
\addplot [black]
table {%
1.875 0.472740248284307
2.125 0.472740248284307
};
\path [draw=black, fill=white, opacity=0.7]
(axis cs:2.75,0.180909186998642)
--(axis cs:3.25,0.180909186998642)
--(axis cs:3.25,0.249634822216924)
--(axis cs:2.75,0.249634822216924)
--(axis cs:2.75,0.180909186998642)
--cycle;
\addplot [black]
table {%
3 0.180909186998642
3 0.122991315856878
};
\addplot [black]
table {%
3 0.249634822216924
3 0.347213308571562
};
\addplot [black]
table {%
2.875 0.122991315856878
3.125 0.122991315856878
};
\addplot [black]
table {%
2.875 0.347213308571562
3.125 0.347213308571562
};
\path [draw=black, fill=white, opacity=0.7]
(axis cs:3.75,-0.0570480743000723)
--(axis cs:4.25,-0.0570480743000723)
--(axis cs:4.25,-0.014187778089956)
--(axis cs:3.75,-0.014187778089956)
--(axis cs:3.75,-0.0570480743000723)
--cycle;
\addplot [black]
table {%
4 -0.0570480743000723
4 -0.0999533760630266
};
\addplot [black]
table {%
4 -0.014187778089956
4 0.0286339419205368
};
\addplot [black]
table {%
3.875 -0.0999533760630266
4.125 -0.0999533760630266
};
\addplot [black]
table {%
3.875 0.0286339419205368
4.125 0.0286339419205368
};
\path [draw=black, fill=white, opacity=0.7]
(axis cs:4.75,-0.0138491648203009)
--(axis cs:5.25,-0.0138491648203009)
--(axis cs:5.25,0.114358820576247)
--(axis cs:4.75,0.114358820576247)
--(axis cs:4.75,-0.0138491648203009)
--cycle;
\addplot [black]
table {%
5 -0.0138491648203009
5 -0.0668978628533855
};
\addplot [black]
table {%
5 0.114358820576247
5 0.273026998480101
};
\addplot [black]
table {%
4.875 -0.0668978628533855
5.125 -0.0668978628533855
};
\addplot [black]
table {%
4.875 0.273026998480101
5.125 0.273026998480101
};
\path [draw=black, fill=white, opacity=0.7]
(axis cs:5.75,0.356527316560769)
--(axis cs:6.25,0.356527316560769)
--(axis cs:6.25,0.435893552412685)
--(axis cs:5.75,0.435893552412685)
--(axis cs:5.75,0.356527316560769)
--cycle;
\addplot [black]
table {%
6 0.356527316560769
6 0.242946138559812
};
\addplot [black]
table {%
6 0.435893552412685
6 0.554701804767031
};
\addplot [black]
table {%
5.875 0.242946138559812
6.125 0.242946138559812
};
\addplot [black]
table {%
5.875 0.554701804767031
6.125 0.554701804767031
};
\addplot [thick, black]
table {%
0.75 0.48083267690028
1.25 0.48083267690028
};
\addplot [thick, black]
table {%
1.75 0.329789726881978
2.25 0.329789726881978
};
\addplot [thick, black]
table {%
2.75 0.208973369179454
3.25 0.208973369179454
};
\addplot [thick, black]
table {%
3.75 -0.0142284974094331
4.25 -0.0142284974094331
};
\addplot [thick, black]
table {%
4.75 0.0429199929485112
5.25 0.0429199929485112
};
\addplot [thick, black]
table {%
5.75 0.390581524580675
6.25 0.390581524580675
};
\end{axis}

\end{tikzpicture}
        \subcaption*{(a) RPDI across State Definitions}
        \label{fig:combined_normalized_price_boxplot_logit}
	\end{minipage}
	\begin{minipage}{0.48\linewidth}
		\centering
\begin{tikzpicture}

\definecolor{darkgray176}{RGB}{176,176,176}
\definecolor{skyblue}{RGB}{135,206,235}
\scriptsize

\begin{axis}[
width=0.9\textwidth, 
height=0.64\textwidth, 
tick align=outside,
tick pos=left,
x grid style={darkgray176},
xmin=0.5, xmax=6.5,
xtick style={color=black},
xtick={1,2,3,4,5,6},
xticklabel style={rotate=30.0,anchor=east},
xticklabels={k1,k2,k3,self\_k1,self\_k2,self\_k3},
y grid style={darkgray176},
ylabel={\(\displaystyle \Delta\)},
ymajorgrids,
ymin=-0.22249291223058, ymax=1.00110100879576,
ytick style={color=black}
]
\path [draw=black, fill=white, opacity=0.7]
(axis cs:0.75,0.518343467263942)
--(axis cs:1.25,0.518343467263942)
--(axis cs:1.25,0.697089797960653)
--(axis cs:0.75,0.697089797960653)
--(axis cs:0.75,0.518343467263942)
--cycle;
\addplot [black]
table {%
1 0.518343467263942
1 0.333363453670027
};
\addplot [black]
table {%
1 0.697089797960653
1 0.945483103294562
};
\addplot [black]
table {%
0.875 0.333363453670027
1.125 0.333363453670027
};
\addplot [black]
table {%
0.875 0.945483103294562
1.125 0.945483103294562
};
\path [draw=black, fill=white, opacity=0.7]
(axis cs:1.75,0.263365699611802)
--(axis cs:2.25,0.263365699611802)
--(axis cs:2.25,0.394790685716386)
--(axis cs:1.75,0.394790685716386)
--(axis cs:1.75,0.263365699611802)
--cycle;
\addplot [black]
table {%
2 0.263365699611802
2 0.0885256331892064
};
\addplot [black]
table {%
2 0.394790685716386
2 0.545835154972044
};
\addplot [black]
table {%
1.875 0.0885256331892064
2.125 0.0885256331892064
};
\addplot [black]
table {%
1.875 0.545835154972044
2.125 0.545835154972044
};
\path [draw=black, fill=white, opacity=0.7]
(axis cs:2.75,0.0947371290940359)
--(axis cs:3.25,0.0947371290940359)
--(axis cs:3.25,0.234982753882407)
--(axis cs:2.75,0.234982753882407)
--(axis cs:2.75,0.0947371290940359)
--cycle;
\addplot [black]
table {%
3 0.0947371290940359
3 -0.0765293023674325
};
\addplot [black]
table {%
3 0.234982753882407
3 0.413029871911697
};
\addplot [black]
table {%
2.875 -0.0765293023674325
3.125 -0.0765293023674325
};
\addplot [black]
table {%
2.875 0.413029871911697
3.125 0.413029871911697
};
\path [draw=black, fill=white, opacity=0.7]
(axis cs:3.75,-0.0939278865418398)
--(axis cs:4.25,-0.0939278865418398)
--(axis cs:4.25,-0.023526036247046)
--(axis cs:3.75,-0.023526036247046)
--(axis cs:3.75,-0.0939278865418398)
--cycle;
\addplot [black]
table {%
4 -0.0939278865418398
4 -0.164441536799298
};
\addplot [black]
table {%
4 -0.023526036247046
4 0.0399827794675713
};
\addplot [black]
table {%
3.875 -0.164441536799298
4.125 -0.164441536799298
};
\addplot [black]
table {%
3.875 0.0399827794675713
4.125 0.0399827794675713
};
\path [draw=black, fill=white, opacity=0.7]
(axis cs:4.75,-0.0248615311996299)
--(axis cs:5.25,-0.0248615311996299)
--(axis cs:5.25,0.145355813711492)
--(axis cs:4.75,0.145355813711492)
--(axis cs:4.75,-0.0248615311996299)
--cycle;
\addplot [black]
table {%
5 -0.0248615311996299
5 -0.166875006729382
};
\addplot [black]
table {%
5 0.145355813711492
5 0.362773150514542
};
\addplot [black]
table {%
4.875 -0.166875006729382
5.125 -0.166875006729382
};
\addplot [black]
table {%
4.875 0.362773150514542
5.125 0.362773150514542
};
\path [draw=black, fill=white, opacity=0.7]
(axis cs:5.75,0.349628334663927)
--(axis cs:6.25,0.349628334663927)
--(axis cs:6.25,0.535237001757824)
--(axis cs:5.75,0.535237001757824)
--(axis cs:5.75,0.349628334663927)
--cycle;
\addplot [black]
table {%
6 0.349628334663927
6 0.104923313034082
};
\addplot [black]
table {%
6 0.535237001757824
6 0.80602060775286
};
\addplot [black]
table {%
5.875 0.104923313034082
6.125 0.104923313034082
};
\addplot [black]
table {%
5.875 0.80602060775286
6.125 0.80602060775286
};
\addplot [thick, black]
table {%
0.75 0.601911361124162
1.25 0.601911361124162
};
\addplot [thick, black]
table {%
1.75 0.32260978796785
2.25 0.32260978796785
};
\addplot [thick, black]
table {%
2.75 0.147054225703875
3.25 0.147054225703875
};
\addplot [thick, black]
table {%
3.75 -0.0310795431062481
4.25 -0.0310795431062481
};
\addplot [thick, black]
table {%
4.75 0.0536958351484615
5.25 0.0536958351484615
};
\addplot [thick, black]
table {%
5.75 0.438162857546208
6.25 0.438162857546208
};
\end{axis}

\end{tikzpicture}
        \subcaption*{(b) \(\displaystyle \Delta\) across State Definitions}
        \label{fig:combined_delta_boxplot_logit}
	\end{minipage}
        \caption{Logit Bertrand - Boxplots comparing normalized price and profit across different state definitions. Results are averaged over 20 runs. For states with opponent information (\(k1\), \(k2\), \(k3\)), prices decrease with longer memory but remain supra-competitive, while profits also decline. In contrast, states without opponent information (\(self\_k1\), \(self\_k2\), \(self\_k3\)) lead to competitive pricing and higher profits with longer memory.}
        \label{fig:state_price_profit_diff_boxplot_logit}
\end{figure}
In contrast, 
\begin{figure}[!ht]
	\centering
	\begin{minipage}[t]{0.48\linewidth}
		\centering
        \includegraphics[width=1.0\linewidth]{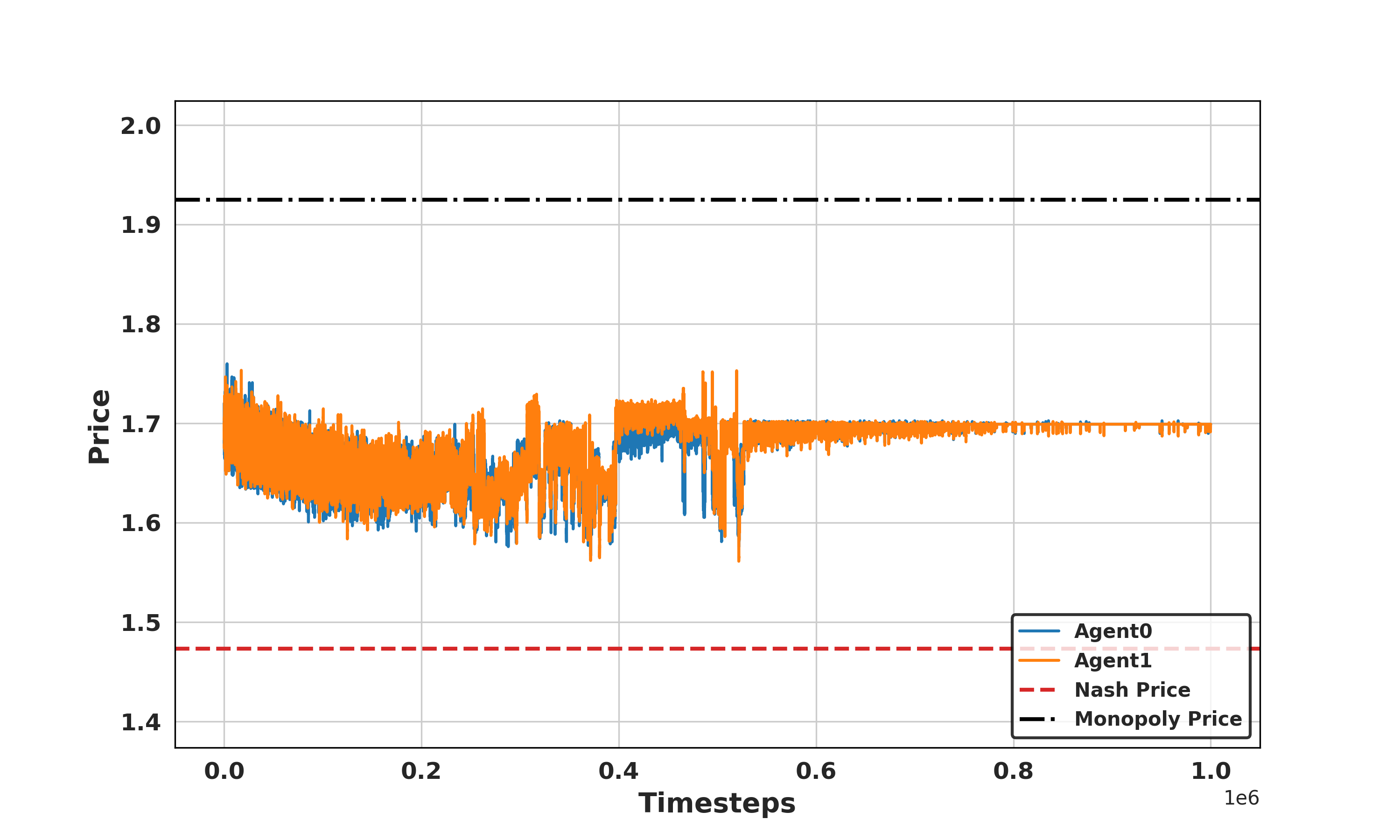}
        \subcaption*{(a) Pricing curves for homogeneous TQL agents under complete information ( \(k1\))}
        \label{fig:complete_information_pricing_curve}
	\end{minipage}
	\hfill
	\begin{minipage}[t]{0.48\linewidth}
		\centering
        \includegraphics[width=1.0\linewidth]{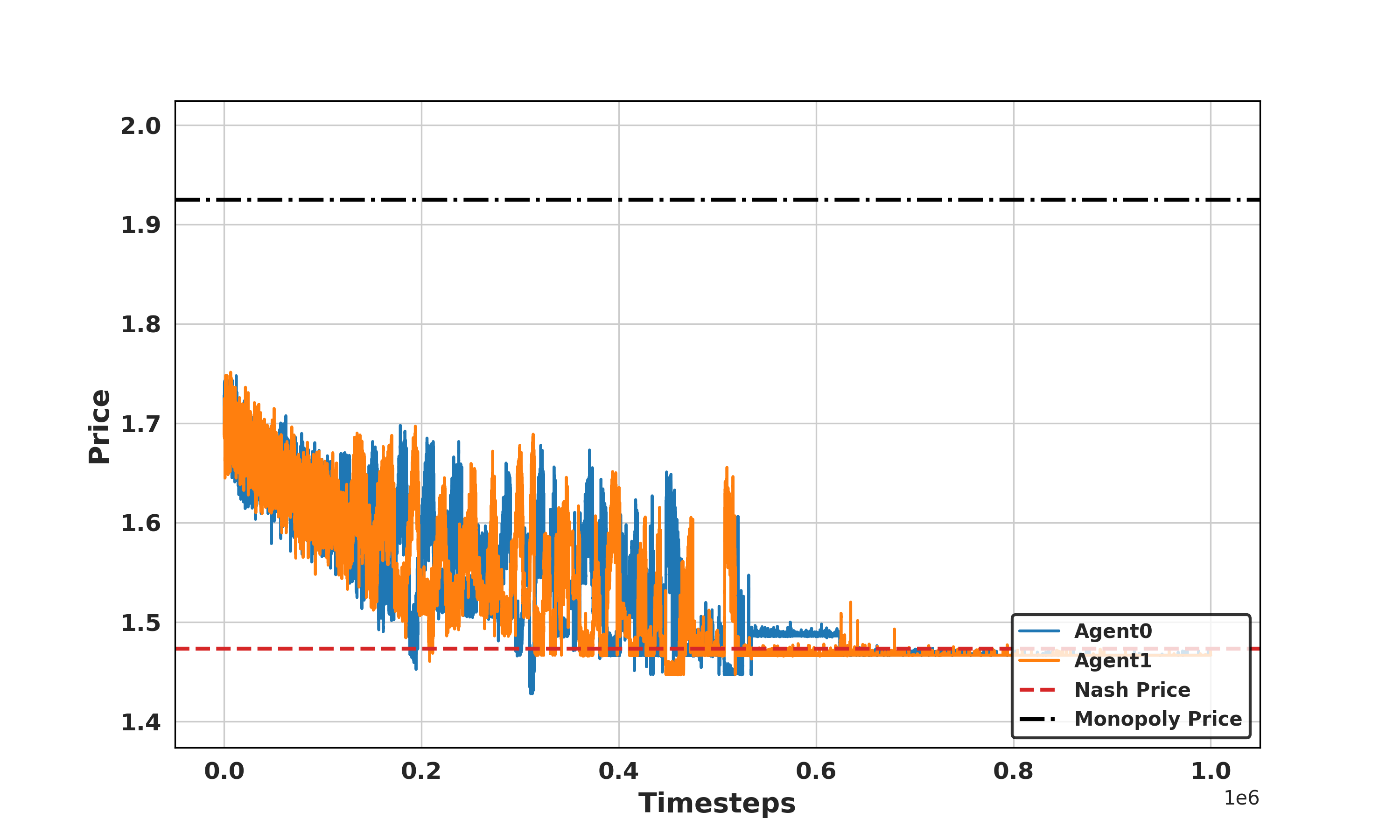}
        \subcaption*{(b) Pricing curves for homogeneous TQL agents under limited information (\(self\_k1\))}
        \label{fig:limited_information_pricing_curve}
	\end{minipage}
        \caption{Logit Bertrand - Comparison of pricing curves for homogeneous TQL agents under complete information (\(k1\)) and limited information (\(self\_k1\)) settings (\(l=1\)). The curves show that access to opponent information in the state definition leads to supra-competitive pricing, while limited information results in prices converging closer to Nash price levels.}

        \label{fig:pricing_curve_comparison}
\end{figure} 
when agents rely solely on their own pricing history, they tend to adopt more competitive strategies, with prices converging near the Nash equilibrium. 

\textit{\textbf{Insight 5.} When \gls{acr:tql} agents have access to opponent pricing information, prices decrease with memory length but remain above Nash equilibrium, sustaining supra-competitive behavior. Without opponent information, prices increase with memory but tend to converge near Nash equilibrium for shorter memory lengths. This suggests that richer state definitions can sustain higher, less competitive prices, while limited-information settings encourage more competitive pricing strategies.}

\subsection{Discussion}

In the following, we analyze the factors driving differences in pricing behavior among the \gls{acr:rl} algorithms studied. Specifically, we address two key questions: (1) why \gls{acr:drl} algorithms, particularly \gls{acr:dqn} and \gls{acr:ppo}, resist collusion compared to the more collusion-prone \gls{acr:tql}; and (2) why \gls{acr:ppo} consistently achieves lower prices than \gls{acr:dqn} under similar competitive conditions.\medskip

\noindent\textbf{Collusion Propensity: \gls{acr:tql} vs. \gls{acr:drl}}\quad \gls{acr:tql} fosters collusion due to strong temporal correlations in its learning process, which enable it to identify and stabilize high-profit focal points. Since \gls{acr:tql} updates its Q-values immediately after each experience, it quickly recognizes strategies leading to supra-competitive pricing. When states and actions remain relatively unchanged, the non-stationarity in learning decreases, allowing the algorithm to maintain these high-price strategies over time. This characteristic makes \gls{acr:tql} prone to tacit collusion in price competition, as firms can repeatedly adjust strategies without explicit coordination.

In contrast, \gls{acr:drl} algorithms, such as \gls{acr:dqn} and \gls{acr:ppo}, are less susceptible to collusion due to their more dynamic training mechanisms. \gls{acr:dqn} utilizes experience replay, where updates are applied to randomly sampled experiences rather than immediate observations. This process disrupts temporal correlations, making it difficult to stabilize collusive pricing. Similarly, \gls{acr:ppo} updates its strategy incrementally and avoids rapid convergence to high-price strategies. While \gls{acr:ppo} does not introduce as much randomness as \gls{acr:dqn}, its update mechanism still prevents the rapid locking of high-price strategies, promoting continuous adaptation.

Additionally, non-stationarity in multi-agent learning environments further limits collusion in \gls{acr:drl}. Unlike \gls{acr:tql}, which relies on consistent pricing patterns, \gls{acr:drl} algorithms prioritize adapting to evolving competitor strategies rather than settling on fixed price agreements. Since deep neural networks update policies using randomly sampled past experiences, their ability to establish supra-competitive pricing strategies is restricted. The continuous disruption of temporal correlations between states prevents \gls{acr:drl} from stabilizing on collusive pricing strategies, making competitive behavior more prevalent in multi-agent pricing environments.

In summary, \gls{acr:tql} encourages collusion through strong temporal correlation and immediate updates, while \gls{acr:drl} algorithms, with their dynamic learning mechanisms and exploratory behaviors, maintain a competitive equilibrium and reduce collusion tendencies.\medskip

\noindent\textbf{Pricing Strategies: Why \gls{acr:ppo} Undercuts \gls{acr:dqn}}\quad In Bertrand competition, \gls{acr:ppo} tends to set lower prices than \gls{acr:dqn}, ultimately achieving higher profits. This outcome stems from differences in the algorithms' learning mechanisms and convergence properties.

Policy gradient methods, such as \gls{acr:ppo}, naturally favor competitive pricing. Although multi-agent policy gradient convergence remains an open research question, prior studies \citep{hambly2023policy, shi2020multi} suggest that such methods can reach Nash equilibria under certain conditions. This theoretical foundation supports \gls{acr:ppo}'s ability to converge toward aggressive pricing strategies in competitive markets.

In contrast, \gls{acr:dqn}, an off-policy algorithm relies on \(\epsilon\)-greedy exploration and experience replay. This approach limits \gls{acr:dqn}'s ability to fully explore the strategy space, increasing the risk of overfitting to local optima such as supra-competitive pricing. Unlike \gls{acr:ppo}, which continuously refines its policy via gradient updates, \gls{acr:dqn} may fail to adequately explore lower-pricing strategies due to its reliance on past experiences, leading to higher price outcomes in some scenarios.

Moreover, \gls{acr:ppo} incorporates policy clipping, which prevents overly large updates and ensures stable learning. This stability allows \gls{acr:ppo} to steadily refine its pricing strategy without abrupt fluctuations. In contrast, \gls{acr:dqn}, which relies on experience replay, may suffer from outdated experiences, causing convergence instability and occasional strategy oscillation. These factors make it harder for \gls{acr:dqn} to consistently achieve the lower-price strategies found by \gls{acr:ppo}.

Thus, \gls{acr:ppo}'s robust policy optimization and efficient exploration enable it to identify more competitive pricing strategies, while \gls{acr:dqn}'s reliance on past experiences and limited exploration can lead to higher pricing levels. These differences highlight \gls{acr:ppo}'s advantage in competitive pricing environments, reinforcing its ability to achieve lower prices compared to \gls{acr:dqn}.

\section{Conclusion}
\label{sec:conclusion}

This study contributes to the growing discourse on algorithmic pricing by examining the competitive and collusive behavior of different \gls{acr:rl} algorithms in oligopolistic markets. Through extensive simulations based on Bertrand competition models, we compare the pricing behaviors of \gls{acr:tql}, \gls{acr:dqn}, and \gls{acr:ppo}, offering a nuanced perspective on the risks of algorithmic collusion.

Our findings reveal critical distinctions between traditional and \gls{acr:drl} methods. \gls{acr:tql} consistently exhibits a strong tendency for supra-competitive pricing, with agents learning to sustain collusion when provided with opponent information. Moreover, variations in learning rates within \gls{acr:tql} agents lead to an inherent asymmetry, 
as faster-learning agents systematically achieve higher profits. In contrast, \gls{acr:drl} agents--particularly \gls{acr:ppo} and \gls{acr:dqn}--demonstrate more competitive behavior, consistently stabilizing at lower price levels closer to the Nash equilibrium. Notably, when pre-trained \gls{acr:tql} agents compete against \gls{acr:drl} agents, the latter quickly dominate, reinforcing the advantages of deep learning-based pricing strategies. Furthermore, when heterogeneous \gls{acr:drl} agents interact, we observe a further reduction in collusive outcomes, highlighting the role of algorithmic diversity in fostering competition.

These results have significant implications for businesses, regulators, and researchers. From a business perspective, the findings underscore the importance of selecting the right algorithmic pricing strategies, as DRL-based methods yield more competitive and stable outcomes. For regulators, our work suggests that fostering diversity in pricing algorithms may serve as a natural countermeasure against tacit collusion, potentially informing future antitrust policies.

For future research, several promising research directions emerge. First, future studies could explore more complex market structures, including multi-product competition and multi-agent reinforcement learning (MARL) frameworks with dynamic entry and exit of firms. Second, investigating how algorithmic pricing strategies evolve under different regulatory interventions--such as pricing caps or transparency mandates--would provide further insights into policy implications. Finally, studying the sensitivity and dynamics of learning-based pricing strategies with respect to varying demand elasiticity models and preference heterogeneity remains an open question, warranting empirical validation through field experiments or real-world data.

\subsection*{Acknowledgements}
This work was funded by the Deutsche Forschungsgemeinschaft (DFG, German Research Foundation) -
Projektnummer 277991500.

\subsection*{Declaration of generative AI and AI-assisted technologies in the writing process}

During the preparation of this work the author(s) used ChatGPT4o in order to improve the paper's writing style and grammar. After using this tool, the authors reviewed and edited the content as needed and take full responsibility for the content of the publication.


\bibliographystyle{model5-names}  

\newpage
\onehalfspacing
\begin{appendices}
	\normalsize
	\section{Supplementary results}
\label{appendix:supplementary_results}


\subsection{Learning Rate-Induced Market Imbalances in Tabular Q-Learning}

In both the Standard Bertrand and Edgeworth Bertrand models, the observed results align with these established trends. As illustrated in Figures~\ref{fig:normalized_price_profit_diff_boxplot_standard} and \ref{fig:normalized_price_profit_diff_boxplot_edgeworth}, agents with higher learning rates tend to adopt lower pricing strategies. However, this effect is less pronounced compared to the Logit Bertrand model. The Edgeworth Bertrand model exhibits greater volatility in pricing and profitability differences due to its inherently dynamic nature. Despite this, agents with higher learning rates consistently attain greater profitability, particularly in scenarios with substantial learning rate disparities (e.g., 0.5\_0.01 and 0.1\_0.01).

\begin{figure}[htbp]
	\centering
	\begin{minipage}{0.48\linewidth}
		\centering
\begin{tikzpicture}

\definecolor{darkgray176}{RGB}{176,176,176}
\definecolor{skyblue}{RGB}{135,206,235}
\scriptsize

\begin{axis}[
width=0.9\textwidth, 
height=0.64\textwidth, 
tick align=outside,
tick pos=left,
x grid style={darkgray176},
xlabel={Alpha Combinations},
xmin=0.5, xmax=6.5,
xtick style={color=black},
xtick={1,2,3,4,5,6},
xticklabel style={rotate=30.0},
xticklabels={05\_001,05\_005,05\_01,01\_001,01\_005,005\_001},
y grid style={darkgray176},
ylabel={RPDI (Agent 0 - Agent 1)},
ymajorgrids,
ymin=-0.415347857142857, ymax=0.356019285714286,
ytick style={color=black}
]
\path [draw=black, fill=white, thick]
(axis cs:0.75,-0.20245)
--(axis cs:1.25,-0.20245)
--(axis cs:1.25,-0.0356535714285714)
--(axis cs:0.75,-0.0356535714285714)
--(axis cs:0.75,-0.20245)
--cycle;
\addplot [black]
table {%
1 -0.20245
1 -0.380285714285714
};
\addplot [black]
table {%
1 -0.0356535714285714
1 0.2144
};
\addplot [black]
table {%
0.875 -0.380285714285714
1.125 -0.380285714285714
};
\addplot [black]
table {%
0.875 0.2144
1.125 0.2144
};
\path [draw=black, fill=white, thick]
(axis cs:1.75,0.0395357142857142)
--(axis cs:2.25,0.0395357142857142)
--(axis cs:2.25,0.21435)
--(axis cs:1.75,0.21435)
--(axis cs:1.75,0.0395357142857142)
--cycle;
\addplot [black]
table {%
2 0.0395357142857142
2 -0.166571428571429
};
\addplot [black]
table {%
2 0.21435
2 0.320957142857143
};
\addplot [black]
table {%
1.875 -0.166571428571429
2.125 -0.166571428571429
};
\addplot [black]
table {%
1.875 0.320957142857143
2.125 0.320957142857143
};
\path [draw=black, fill=white, thick]
(axis cs:2.75,-0.000225000000000003)
--(axis cs:3.25,-0.000225000000000003)
--(axis cs:3.25,0.090275)
--(axis cs:2.75,0.090275)
--(axis cs:2.75,-0.000225000000000003)
--cycle;
\addplot [black]
table {%
3 -0.000225000000000003
3 -0.0708571428571428
};
\addplot [black]
table {%
3 0.090275
3 0.178542857142857
};
\addplot [black]
table {%
2.875 -0.0708571428571428
3.125 -0.0708571428571428
};
\addplot [black]
table {%
2.875 0.178542857142857
3.125 0.178542857142857
};
\path [draw=black, fill=white, thick]
(axis cs:3.75,-0.178496428571429)
--(axis cs:4.25,-0.178496428571429)
--(axis cs:4.25,-0.0532928571428572)
--(axis cs:3.75,-0.0532928571428572)
--(axis cs:3.75,-0.178496428571429)
--cycle;
\addplot [black]
table {%
4 -0.178496428571429
4 -0.2497
};
\addplot [black]
table {%
4 -0.0532928571428572
4 0.0476000000000001
};
\addplot [black]
table {%
3.875 -0.2497
4.125 -0.2497
};
\addplot [black]
table {%
3.875 0.0476000000000001
4.125 0.0476000000000001
};
\path [draw=black, fill=white, thick]
(axis cs:4.75,-0.0743928571428571)
--(axis cs:5.25,-0.0743928571428571)
--(axis cs:5.25,0.0736642857142856)
--(axis cs:4.75,0.0736642857142856)
--(axis cs:4.75,-0.0743928571428571)
--cycle;
\addplot [black]
table {%
5 -0.0743928571428571
5 -0.19
};
\addplot [black]
table {%
5 0.0736642857142856
5 0.2857
};
\addplot [black]
table {%
4.875 -0.19
5.125 -0.19
};
\addplot [black]
table {%
4.875 0.2857
5.125 0.2857
};
\path [draw=black, fill=white, thick]
(axis cs:5.75,-0.142821428571429)
--(axis cs:6.25,-0.142821428571429)
--(axis cs:6.25,-8.92857142856807e-05)
--(axis cs:5.75,-8.92857142856807e-05)
--(axis cs:5.75,-0.142821428571429)
--cycle;
\addplot [black]
table {%
6 -0.142821428571429
6 -0.284328571428571
};
\addplot [black]
table {%
6 -8.92857142856807e-05
6 0.107171428571429
};
\addplot [black]
table {%
5.875 -0.284328571428571
6.125 -0.284328571428571
};
\addplot [black]
table {%
5.875 0.107171428571429
6.125 0.107171428571429
};
\addplot [thick, black]
table {%
0.75 -0.0832928571428572
1.25 -0.0832928571428572
};
\addplot [thick, black]
table {%
1.75 0.104471428571429
2.25 0.104471428571429
};
\addplot [thick, black]
table {%
2.75 0.000392857142857139
3.25 0.000392857142857139
};
\addplot [thick, black]
table {%
3.75 -0.1049
4.25 -0.1049
};
\addplot [thick, black]
table {%
4.75 9.28571428572278e-05
5.25 9.28571428572278e-05
};
\addplot [thick, black]
table {%
5.75 -0.11895
6.25 -0.11895
};
\end{axis}

\end{tikzpicture}
        \subcaption*{(a) RPDI Differences}
        \label{fig:normalized_price_diff_boxplot_logit}
	\end{minipage}
	\begin{minipage}{0.48\linewidth}
		\centering
\begin{tikzpicture}

\definecolor{darkgray176}{RGB}{176,176,176}
\definecolor{skyblue}{RGB}{135,206,235}
\scriptsize

\begin{axis}[
width=0.9\textwidth, 
height=0.64\textwidth, 
tick align=outside,
tick pos=left,
x grid style={darkgray176},
xlabel={Alpha Combinations},
xmin=0.5, xmax=6.5,
xtick style={color=black},
xtick={1,2,3,4,5,6},
xticklabel style={rotate=30.0},
xticklabels={05\_001,05\_005,05\_01,01\_001,01\_005,005\_001},
y grid style={darkgray176},
ylabel={\(\displaystyle \Delta\) (Agent 0 - Agent 1)},
ymajorgrids,
ymin=-0.127273367346939, ymax=0.657810102040816,
ytick style={color=black}
]
\path [draw=black, fill=white, thick]
(axis cs:0.75,0.459222959183673)
--(axis cs:1.25,0.459222959183673)
--(axis cs:1.25,0.570258673469388)
--(axis cs:0.75,0.570258673469388)
--(axis cs:0.75,0.459222959183673)
--cycle;
\addplot [black]
table {%
1 0.459222959183673
1 0.387508163265306
};
\addplot [black]
table {%
1 0.570258673469388
1 0.622124489795918
};
\addplot [black]
table {%
0.875 0.387508163265306
1.125 0.387508163265306
};
\addplot [black]
table {%
0.875 0.622124489795918
1.125 0.622124489795918
};
\path [draw=black, fill=white, thick]
(axis cs:1.75,0.241102040816326)
--(axis cs:2.25,0.241102040816326)
--(axis cs:2.25,0.3059)
--(axis cs:1.75,0.3059)
--(axis cs:1.75,0.241102040816326)
--cycle;
\addplot [black]
table {%
2 0.241102040816326
2 0.147083673469388
};
\addplot [black]
table {%
2 0.3059
2 0.362020408163265
};
\addplot [black]
table {%
1.875 0.147083673469388
2.125 0.147083673469388
};
\addplot [black]
table {%
1.875 0.362020408163265
2.125 0.362020408163265
};
\path [draw=black, fill=white, thick]
(axis cs:2.75,2.80612244897699e-05)
--(axis cs:3.25,2.80612244897699e-05)
--(axis cs:3.25,0.0770841836734695)
--(axis cs:2.75,0.0770841836734695)
--(axis cs:2.75,2.80612244897699e-05)
--cycle;
\addplot [black]
table {%
3 2.80612244897699e-05
3 -0.0915877551020409
};
\addplot [black]
table {%
3 0.0770841836734695
3 0.183126530612245
};
\addplot [black]
table {%
2.875 -0.0915877551020409
3.125 -0.0915877551020409
};
\addplot [black]
table {%
2.875 0.183126530612245
3.125 0.183126530612245
};
\path [draw=black, fill=white, thick]
(axis cs:3.75,0.422372959183673)
--(axis cs:4.25,0.422372959183673)
--(axis cs:4.25,0.490790306122449)
--(axis cs:3.75,0.490790306122449)
--(axis cs:3.75,0.422372959183673)
--cycle;
\addplot [black]
table {%
4 0.422372959183673
4 0.366865306122449
};
\addplot [black]
table {%
4 0.490790306122449
4 0.577789795918367
};
\addplot [black]
table {%
3.875 0.366865306122449
4.125 0.366865306122449
};
\addplot [black]
table {%
3.875 0.577789795918367
4.125 0.577789795918367
};
\path [draw=black, fill=white, thick]
(axis cs:4.75,0.128869897959184)
--(axis cs:5.25,0.128869897959184)
--(axis cs:5.25,0.206542346938775)
--(axis cs:4.75,0.206542346938775)
--(axis cs:4.75,0.128869897959184)
--cycle;
\addplot [black]
table {%
5 0.128869897959184
5 0.0559530612244897
};
\addplot [black]
table {%
5 0.206542346938775
5 0.251571428571429
};
\addplot [black]
table {%
4.875 0.0559530612244897
5.125 0.0559530612244897
};
\addplot [black]
table {%
4.875 0.251571428571429
5.125 0.251571428571429
};
\path [draw=black, fill=white, thick]
(axis cs:5.75,0.319534693877551)
--(axis cs:6.25,0.319534693877551)
--(axis cs:6.25,0.428443367346939)
--(axis cs:5.75,0.428443367346939)
--(axis cs:5.75,0.319534693877551)
--cycle;
\addplot [black]
table {%
6 0.319534693877551
6 0.249375510204082
};
\addplot [black]
table {%
6 0.428443367346939
6 0.459132653061224
};
\addplot [black]
table {%
5.875 0.249375510204082
6.125 0.249375510204082
};
\addplot [black]
table {%
5.875 0.459132653061224
6.125 0.459132653061224
};
\addplot [thick, black]
table {%
0.75 0.530418367346939
1.25 0.530418367346939
};
\addplot [thick, black]
table {%
1.75 0.285706122448979
2.25 0.285706122448979
};
\addplot [thick, black]
table {%
2.75 0.000324489795918398
3.25 0.000324489795918398
};
\addplot [thick, black]
table {%
3.75 0.458898979591836
4.25 0.458898979591836
};
\addplot [thick, black]
table {%
4.75 0.171819387755102
5.25 0.171819387755102
};
\addplot [thick, black]
table {%
5.75 0.388994897959184
6.25 0.388994897959184
};
\end{axis}

\end{tikzpicture}
        \subcaption*{(b) \(\displaystyle \Delta\) Differences}
        \label{fig:normalized_profit_diff_boxplot_standard}
	\end{minipage}
        \caption{Standard Bertrand: RPDI and $\Delta$ differences between Agent 0 and Agent 1. A negative RPDI difference indicates that Agent 0 sets lower prices , while a positive $\Delta$ difference indicates that Agent 0 gets a higher profit compared to Agent 1.}
        \label{fig:normalized_price_profit_diff_boxplot_standard}
\end{figure}
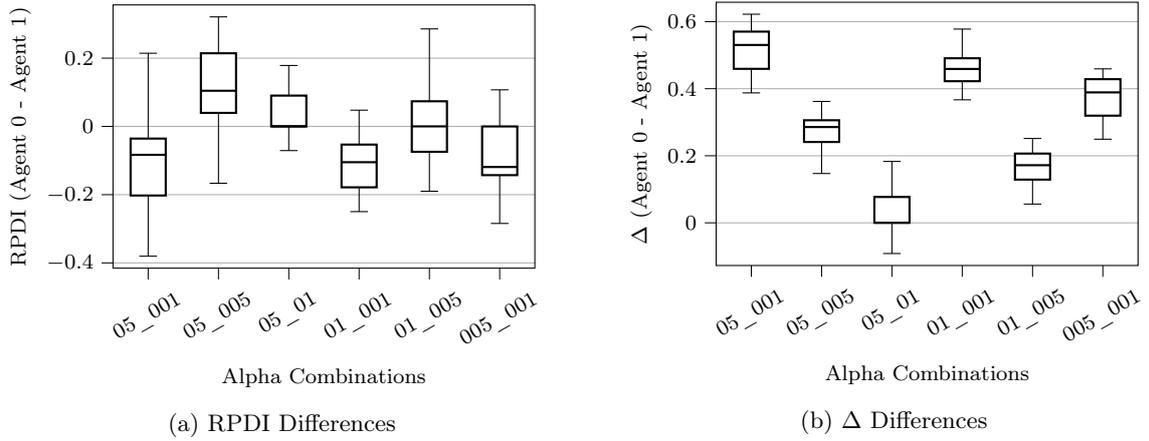

\begin{figure}[htbp]
	\centering
	\begin{minipage}{0.48\linewidth}
		\centering
\begin{tikzpicture}

\definecolor{darkgray176}{RGB}{176,176,176}
\definecolor{skyblue}{RGB}{135,206,235}
\scriptsize

\begin{axis}[
width=0.9\textwidth, 
height=0.64\textwidth, 
tick align=outside,
tick pos=left,
x grid style={darkgray176},
xlabel={Alpha Combinations},
xmin=0.5, xmax=6.5,
xtick style={color=black},
xtick={1,2,3,4,5,6},
xticklabel style={rotate=30.0},
xticklabels={05\_001,05\_005,05\_01,01\_001,01\_005,005\_001},
y grid style={darkgray176},
ylabel={RPDI (Agent 0 - Agent 1)},
ymajorgrids,
ymin=-0.403019285714286, ymax=0.616633571428572,
ytick style={color=black}
]
\path [draw=black, fill=white, thick]
(axis cs:0.75,-0.0716214285714287)
--(axis cs:1.25,-0.0716214285714287)
--(axis cs:1.25,-0.0713107142857143)
--(axis cs:0.75,-0.0713107142857143)
--(axis cs:0.75,-0.0716214285714287)
--cycle;
\addplot [black]
table {%
1 -0.0716214285714287
1 -0.071642857142857
};
\addplot [black]
table {%
1 -0.0713107142857143
1 -0.0713000000000001
};
\addplot [black]
table {%
0.875 -0.071642857142857
1.125 -0.071642857142857
};
\addplot [black]
table {%
0.875 -0.0713000000000001
1.125 -0.0713000000000001
};
\path [draw=black, fill=white, thick]
(axis cs:1.75,0.0213285714285713)
--(axis cs:2.25,0.0213285714285713)
--(axis cs:2.25,0.344517857142857)
--(axis cs:1.75,0.344517857142857)
--(axis cs:1.75,0.0213285714285713)
--cycle;
\addplot [black]
table {%
2 0.0213285714285713
2 -0.143028571428571
};
\addplot [black]
table {%
2 0.344517857142857
2 0.570285714285714
};
\addplot [black]
table {%
1.875 -0.143028571428571
2.125 -0.143028571428571
};
\addplot [black]
table {%
1.875 0.570285714285714
2.125 0.570285714285714
};
\path [draw=black, fill=white, thick]
(axis cs:2.75,7.14285714295548e-06)
--(axis cs:3.25,7.14285714295548e-06)
--(axis cs:3.25,0.151753571428572)
--(axis cs:2.75,0.151753571428572)
--(axis cs:2.75,7.14285714295548e-06)
--cycle;
\addplot [black]
table {%
3 7.14285714295548e-06
3 -0.0953000000000003
};
\addplot [black]
table {%
3 0.151753571428572
3 0.333114285714286
};
\addplot [black]
table {%
2.875 -0.0953000000000003
3.125 -0.0953000000000003
};
\addplot [black]
table {%
2.875 0.333114285714286
3.125 0.333114285714286
};
\path [draw=black, fill=white, thick]
(axis cs:3.75,-0.142853571428571)
--(axis cs:4.25,-0.142853571428571)
--(axis cs:4.25,-0.0714178571428569)
--(axis cs:3.75,-0.0714178571428569)
--(axis cs:3.75,-0.142853571428571)
--cycle;
\addplot [black]
table {%
4 -0.142853571428571
4 -0.238085714285714
};
\addplot [black]
table {%
4 -0.0714178571428569
4 -4.28571428569002e-05
};
\addplot [black]
table {%
3.875 -0.238085714285714
4.125 -0.238085714285714
};
\addplot [black]
table {%
3.875 -4.28571428569002e-05
4.125 -4.28571428569002e-05
};
\path [draw=black, fill=white, thick]
(axis cs:4.75,-0.142775)
--(axis cs:5.25,-0.142775)
--(axis cs:5.25,0.0534642857142857)
--(axis cs:4.75,0.0534642857142857)
--(axis cs:4.75,-0.142775)
--cycle;
\addplot [black]
table {%
5 -0.142775
5 -0.356671428571429
};
\addplot [black]
table {%
5 0.0534642857142857
5 0.163457142857143
};
\addplot [black]
table {%
4.875 -0.356671428571429
5.125 -0.356671428571429
};
\addplot [black]
table {%
4.875 0.163457142857143
5.125 0.163457142857143
};
\path [draw=black, fill=white, thick]
(axis cs:5.75,-0.115192857142857)
--(axis cs:6.25,-0.115192857142857)
--(axis cs:6.25,-0.042757142857143)
--(axis cs:5.75,-0.042757142857143)
--(axis cs:5.75,-0.115192857142857)
--cycle;
\addplot [black]
table {%
6 -0.115192857142857
6 -0.213571428571429
};
\addplot [black]
table {%
6 -0.042757142857143
6 0.0475285714285714
};
\addplot [black]
table {%
5.875 -0.213571428571429
6.125 -0.213571428571429
};
\addplot [black]
table {%
5.875 0.0475285714285714
6.125 0.0475285714285714
};
\addplot [thick, black]
table {%
0.75 -0.0715499999999999
1.25 -0.0715499999999999
};
\addplot [thick, black]
table {%
1.75 0.196171428571429
2.25 0.196171428571429
};
\addplot [thick, black]
table {%
2.75 0.000257142857142845
3.25 0.000257142857142845
};
\addplot [thick, black]
table {%
3.75 -0.08335
4.25 -0.08335
};
\addplot [thick, black]
table {%
4.75 -0.0179571428571428
5.25 -0.0179571428571428
};
\addplot [thick, black]
table {%
5.75 -0.0714857142857143
6.25 -0.0714857142857143
};
\end{axis}

\end{tikzpicture}

        \subcaption*{(a) RPDI Differences}
        \label{fig:normalized_price_diff_boxplot_edgeworth}
	\end{minipage}
	\begin{minipage}{0.48\linewidth}
		\centering
\begin{tikzpicture}

\definecolor{darkgray176}{RGB}{176,176,176}
\definecolor{skyblue}{RGB}{135,206,235}
\scriptsize

\begin{axis}[
width=0.9\textwidth, 
height=0.64\textwidth, 
tick align=outside,
tick pos=left,
x grid style={darkgray176},
xlabel={Alpha Combinations},
xmin=0.5, xmax=6.5,
xtick style={color=black},
xtick={1,2,3,4,5,6},
xticklabel style={rotate=30.0},
xticklabels={05\_001,05\_005,05\_01,01\_001,01\_005,005\_001},
y grid style={darkgray176},
ylabel={\(\displaystyle \Delta\) (Agent 0 - Agent 1)},
ymajorgrids,
ymin=-0.0307804285714287, ymax=0.642680428571428,
ytick style={color=black}
]
\path [draw=black, fill=white, thick]
(axis cs:0.75,0.428478469387755)
--(axis cs:1.25,0.428478469387755)
--(axis cs:1.25,0.540048673469388)
--(axis cs:0.75,0.540048673469388)
--(axis cs:0.75,0.428478469387755)
--cycle;
\addplot [black]
table {%
1 0.428478469387755
1 0.428326530612245
};
\addplot [black]
table {%
1 0.540048673469388
1 0.612068571428571
};
\addplot [black]
table {%
0.875 0.428326530612245
1.125 0.428326530612245
};
\addplot [black]
table {%
0.875 0.612068571428571
1.125 0.612068571428571
};
\path [draw=black, fill=white, thick]
(axis cs:1.75,0.144910204081632)
--(axis cs:2.25,0.144910204081632)
--(axis cs:2.25,0.306616632653061)
--(axis cs:1.75,0.306616632653061)
--(axis cs:1.75,0.144910204081632)
--cycle;
\addplot [black]
table {%
2 0.144910204081632
2 0.000533469387755114
};
\addplot [black]
table {%
2 0.306616632653061
2 0.489609795918367
};
\addplot [black]
table {%
1.875 0.000533469387755114
2.125 0.000533469387755114
};
\addplot [black]
table {%
1.875 0.489609795918367
2.125 0.489609795918367
};
\path [draw=black, fill=white, thick]
(axis cs:2.75,0.000145714285714305)
--(axis cs:3.25,0.000145714285714305)
--(axis cs:3.25,0.161887653061224)
--(axis cs:2.75,0.161887653061224)
--(axis cs:2.75,0.000145714285714305)
--cycle;
\addplot [black]
table {%
3 0.000145714285714305
3 -0.000168571428571518
};
\addplot [black]
table {%
3 0.161887653061224
3 0.223804081632653
};
\addplot [black]
table {%
2.875 -0.000168571428571518
3.125 -0.000168571428571518
};
\addplot [black]
table {%
2.875 0.223804081632653
3.125 0.223804081632653
};
\path [draw=black, fill=white, thick]
(axis cs:3.75,0.41964)
--(axis cs:4.25,0.41964)
--(axis cs:4.25,0.498233469387755)
--(axis cs:3.75,0.498233469387755)
--(axis cs:3.75,0.41964)
--cycle;
\addplot [black]
table {%
4 0.41964
4 0.342805714285714
};
\addplot [black]
table {%
4 0.498233469387755
4 0.580384897959183
};
\addplot [black]
table {%
3.875 0.342805714285714
4.125 0.342805714285714
};
\addplot [black]
table {%
3.875 0.580384897959183
4.125 0.580384897959183
};
\path [draw=black, fill=white, thick]
(axis cs:4.75,0.108213367346939)
--(axis cs:5.25,0.108213367346939)
--(axis cs:5.25,0.190547755102041)
--(axis cs:4.75,0.190547755102041)
--(axis cs:4.75,0.108213367346939)
--cycle;
\addplot [black]
table {%
5 0.108213367346939
5 0.000115102040816328
};
\addplot [black]
table {%
5 0.190547755102041
5 0.309563265306122
};
\addplot [black]
table {%
4.875 0.000115102040816328
5.125 0.000115102040816328
};
\addplot [black]
table {%
4.875 0.309563265306122
5.125 0.309563265306122
};
\path [draw=black, fill=white, thick]
(axis cs:5.75,0.331224489795918)
--(axis cs:6.25,0.331224489795918)
--(axis cs:6.25,0.428462448979592)
--(axis cs:5.75,0.428462448979592)
--(axis cs:5.75,0.331224489795918)
--cycle;
\addplot [black]
table {%
6 0.331224489795918
6 0.284888571428571
};
\addplot [black]
table {%
6 0.428462448979592
6 0.497725714285714
};
\addplot [black]
table {%
5.875 0.284888571428571
6.125 0.284888571428571
};
\addplot [black]
table {%
5.875 0.497725714285714
6.125 0.497725714285714
};
\addplot [thick, black]
table {%
0.75 0.489829183673469
1.25 0.489829183673469
};
\addplot [thick, black]
table {%
1.75 0.263337551020408
2.25 0.263337551020408
};
\addplot [thick, black]
table {%
2.75 0.10008
3.25 0.10008
};
\addplot [thick, black]
table {%
3.75 0.489281632653061
4.25 0.489281632653061
};
\addplot [thick, black]
table {%
4.75 0.165459183673469
5.25 0.165459183673469
};
\addplot [thick, black]
table {%
5.75 0.403311632653061
6.25 0.403311632653061
};
\end{axis}

\end{tikzpicture}
        
        \subcaption*{(b) \(\displaystyle \Delta\) Differences}
        \label{fig:normalized_profit_diff_boxplot_edgeworth}
	\end{minipage}
        \caption{Edgeworth Bertrand: RPDI and $\Delta$ differences between Agent 0 and Agent 1. A negative RPDI difference indicates that Agent 0 sets lower prices , while a positive $\Delta$ difference indicates that Agent 0 gets a higher profit compared to Agent 1.}
        \label{fig:normalized_price_profit_diff_boxplot_edgeworth}
\end{figure}
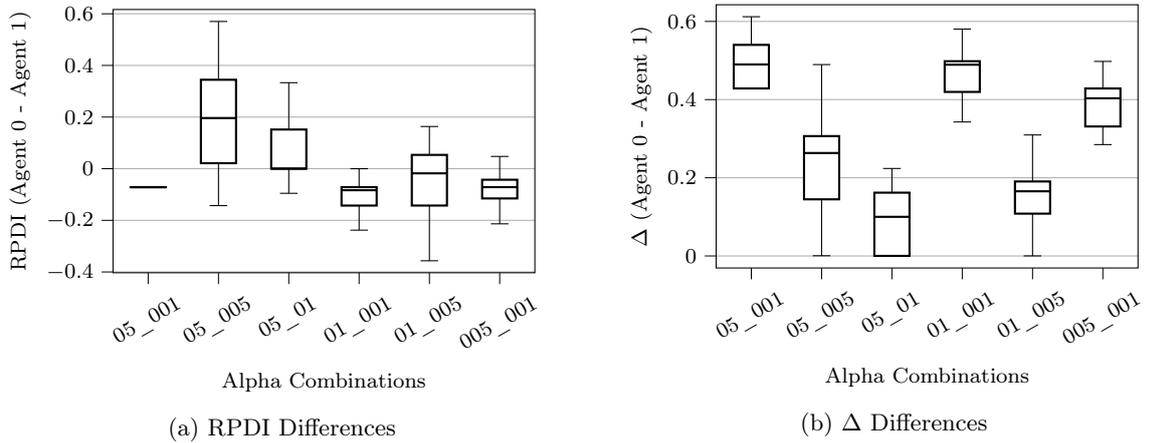

\subsection{\gls{acr:drl} vs \gls{acr:tql}: Superior Pricing Strategies}

The experimental results from the Standard Bertrand and Edgeworth Bertrand models (Figures~\ref{Combined_price_learning_curve_tql_drl_standard} -\ref{fig:delta_tql_drl_edgeworth}) further support the main findings from the Logit Bertrand model. In the Standard Bertrand model, \gls{acr:tql} consistently adopts high-price strategies near the monopoly price, while \gls{acr:dqn} and \gls{acr:ppo} converge to more competitive low-price strategies, achieving higher and more stable profits. In the Edgeworth Bertrand model, pricing dynamics are less smooth, but final price and profit distributions are more stable.  \gls{acr:drl} algorithms still outperform \gls{acr:tql}, demonstrating robustness across different market conditions.

\begin{figure}[htbp]
	\centering

	\begin{minipage}[t]{0.48\linewidth}
		\centering
		\includegraphics[width=1.0\linewidth]{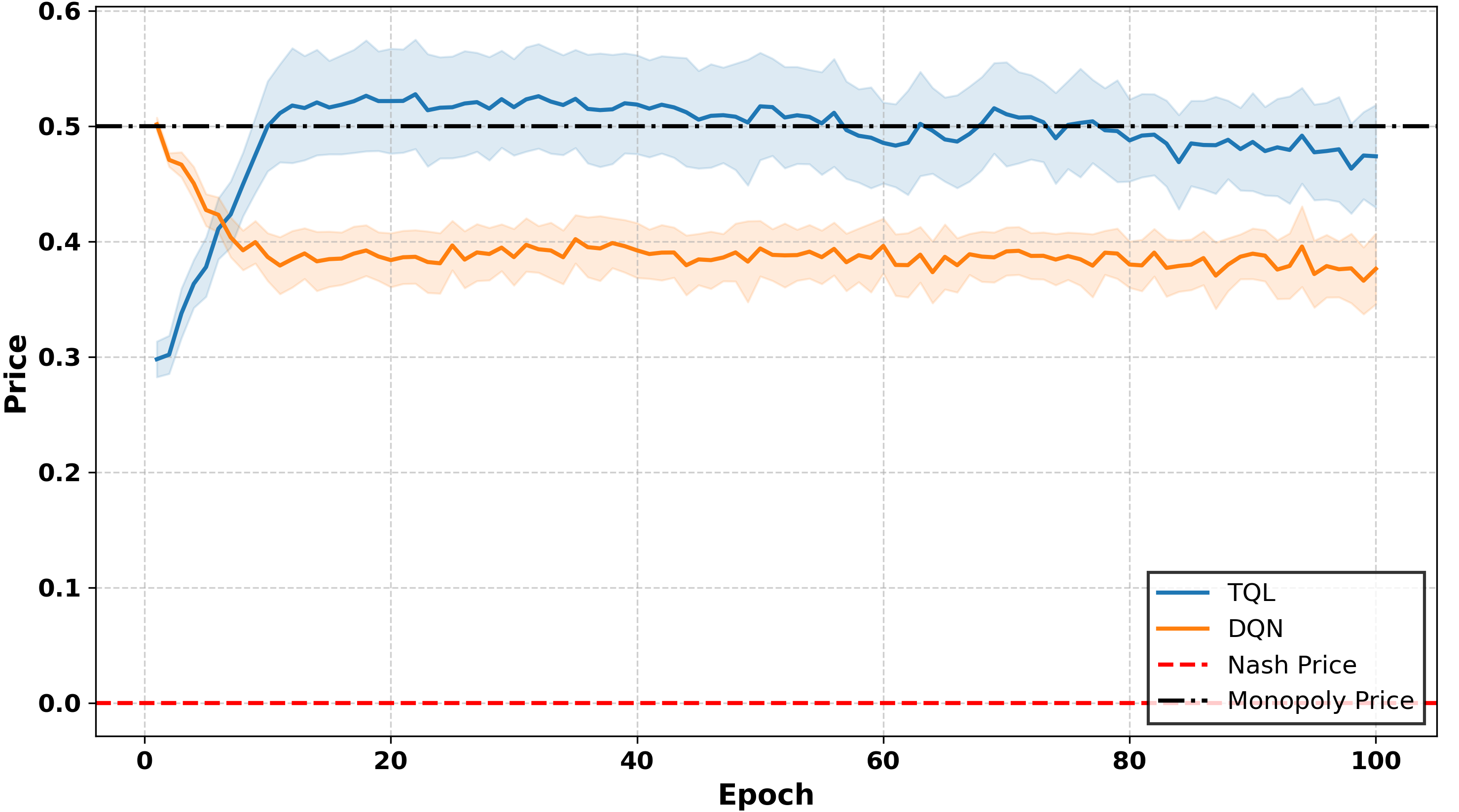}
        \subcaption*{(a) TQL vs DQN: Price Trend (with 95\% CI)}
        \label{TQL_PPO_price_per_iteration_standard}
	\end{minipage}
	\hfill
    \begin{minipage}[t]{0.48\linewidth}
		\centering
		\includegraphics[width=1.0\linewidth]{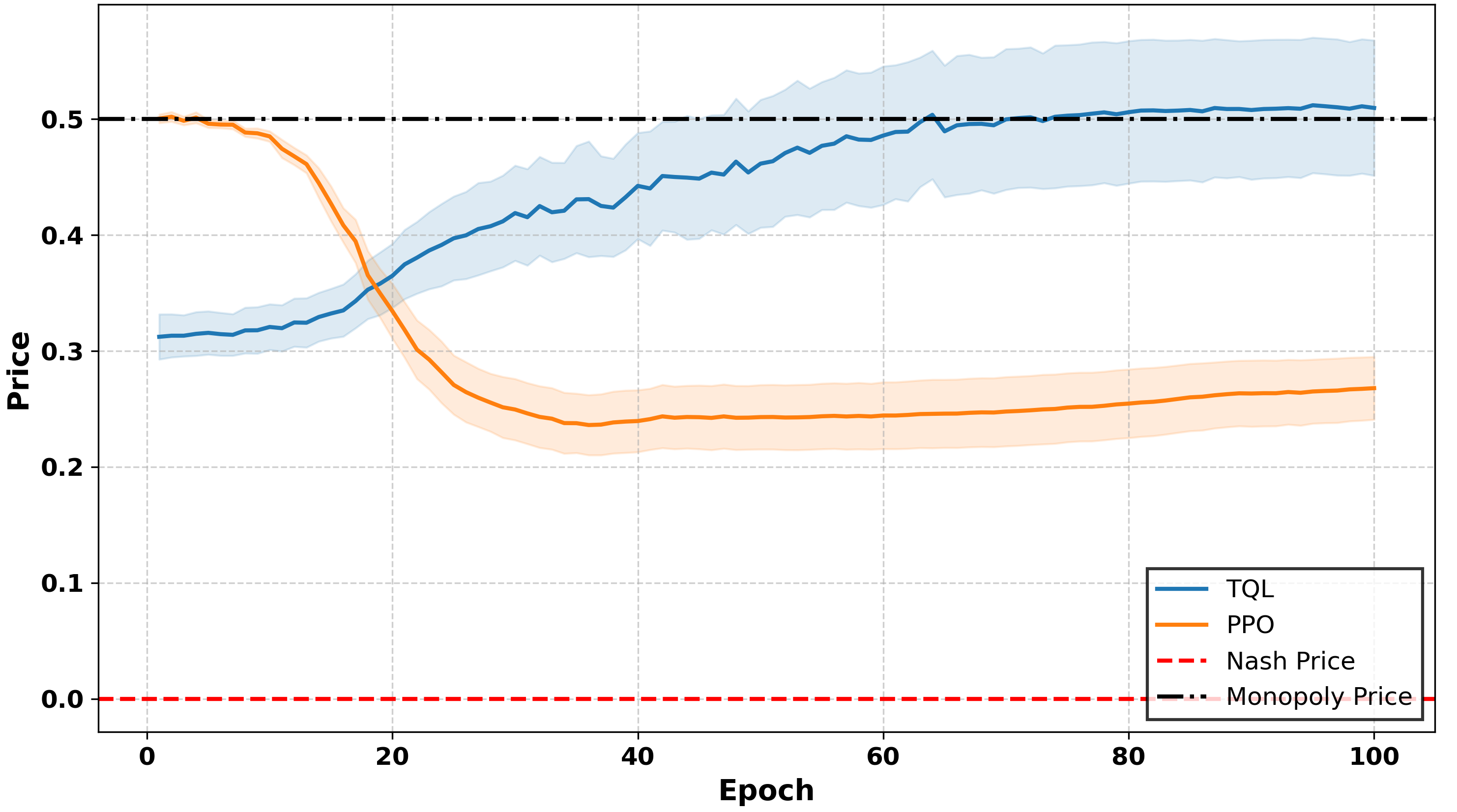}
        \subcaption*{(b) TQL vs PPO: Price Trend (with 95\% CI)}
        \label{TQL_DQN_price_per_iteration_standard}
	\end{minipage}
        \caption{Price learning curves for the Standard Bertrand model, showing competition between \gls{acr:tql} and \gls{acr:dqn} (left) and between \gls{acr:tql} and \gls{acr:ppo} (right). Results are based on 20 independent runs, with mean values and 95\% confidence intervals displayed.}
    \label{Combined_price_learning_curve_tql_drl_standard}
\end{figure}

\begin{figure}[htbp]
	\centering
	\begin{minipage}{0.48\linewidth}
		\centering
\begin{tikzpicture}

\definecolor{darkgray176}{RGB}{176,176,176}
\definecolor{darkorange25512714}{RGB}{255,127,14}
\definecolor{steelblue31119180}{RGB}{31,119,180}
\scriptsize

\begin{axis}[
width=0.9\textwidth, 
height=0.64\textwidth, 
tick align=outside,
tick pos=left,
x grid style={darkgray176},
xmin=0.5, xmax=2.5,
xtick style={color=black},
xtick={1,2},
xticklabels={TQL,DQN},
y grid style={darkgray176},
ylabel={\(\displaystyle \Delta\)},
ymajorgrids,
ymin=-0.0319270408163265, ymax=0.989041326530612,
ytick style={color=black}
]
\path [draw=black, fill=white, opacity=0.7]
(axis cs:0.75,0.0191839285714286)
--(axis cs:1.25,0.0191839285714286)
--(axis cs:1.25,0.0971660714285714)
--(axis cs:0.75,0.0971660714285714)
--(axis cs:0.75,0.0191839285714286)
--cycle;
\addplot [black]
table {%
1 0.0191839285714286
1 0.014480612244898
};
\addplot [black]
table {%
1 0.0971660714285714
1 0.16080612244898
};
\addplot [black]
table {%
0.875 0.014480612244898
1.125 0.014480612244898
};
\addplot [black]
table {%
0.875 0.16080612244898
1.125 0.16080612244898
};
\path [draw=black, fill=white, opacity=0.7]
(axis cs:1.75,0.688502040816326)
--(axis cs:2.25,0.688502040816326)
--(axis cs:2.25,0.891936224489795)
--(axis cs:1.75,0.891936224489795)
--(axis cs:1.75,0.688502040816326)
--cycle;
\addplot [black]
table {%
2 0.688502040816326
2 0.439479591836735
};
\addplot [black]
table {%
2 0.891936224489795
2 0.942633673469387
};
\addplot [black]
table {%
1.875 0.439479591836735
2.125 0.439479591836735
};
\addplot [black]
table {%
1.875 0.942633673469387
2.125 0.942633673469387
};
\addplot [thick, black]
table {%
0.75 0.0534551020408163
1.25 0.0534551020408163
};
\addplot [thick, black]
table {%
1.75 0.80481581632653
2.25 0.80481581632653
};
\end{axis}

\end{tikzpicture}
        \subcaption*{(a) Comparison of \(\displaystyle \Delta\) for TQL and DQN}
        \label{Delta_tql_dqn_standard}
	\end{minipage}
	\begin{minipage}{0.48\linewidth}
		\centering
\begin{tikzpicture}

\definecolor{darkgray176}{RGB}{176,176,176}
\definecolor{darkorange25512714}{RGB}{255,127,14}
\definecolor{steelblue31119180}{RGB}{31,119,180}
\scriptsize

\begin{axis}[
width=0.9\textwidth, 
height=0.64\textwidth, 
tick align=outside,
tick pos=left,
x grid style={darkgray176},
xmin=0.5, xmax=2.5,
xtick style={color=black},
xtick={1,2},
xticklabels={TQL,PPO},
y grid style={darkgray176},
ylabel={\(\displaystyle \Delta\)},
ymajorgrids,
ymin=-0.0487602040816326, ymax=1.02396428571428,
ytick style={color=black}
]
\path [draw=black, fill=white, opacity=0.7]
(axis cs:0.75,0.000279336734693877)
--(axis cs:1.25,0.000279336734693877)
--(axis cs:1.25,0.0832099489795917)
--(axis cs:0.75,0.0832099489795917)
--(axis cs:0.75,0.000279336734693877)
--cycle;
\addplot [black]
table {%
1 0.000279336734693877
1 0
};
\addplot [black]
table {%
1 0.0832099489795917
1 0.16845
};
\addplot [black]
table {%
0.875 0
1.125 0
};
\addplot [black]
table {%
0.875 0.16845
1.125 0.16845
};
\path [draw=black, fill=white, opacity=0.7]
(axis cs:1.75,0.62296887755102)
--(axis cs:2.25,0.62296887755102)
--(axis cs:2.25,0.799353826530612)
--(axis cs:1.75,0.799353826530612)
--(axis cs:1.75,0.62296887755102)
--cycle;
\addplot [black]
table {%
2 0.62296887755102
2 0.46929387755102
};
\addplot [black]
table {%
2 0.799353826530612
2 0.975204081632652
};
\addplot [black]
table {%
1.875 0.46929387755102
2.125 0.46929387755102
};
\addplot [black]
table {%
1.875 0.975204081632652
2.125 0.975204081632652
};
\addplot [thick, black]
table {%
0.75 0.00171785714285714
1.25 0.00171785714285714
};
\addplot [thick, black]
table {%
1.75 0.684975510204081
2.25 0.684975510204081
};
\end{axis}

\end{tikzpicture}
        \subcaption*{(b) Comparison of \(\displaystyle \Delta\) for TQL and PPO}
        \label{Delta_tql_ppo_standard}
	\end{minipage}
        \caption{Boxplots of normalized profit \(\Delta\) for the Standard Bertrand model, comparing competition between \gls{acr:tql} and \gls{acr:dqn} (left) and between \gls{acr:tql} and \gls{acr:ppo} (right). \gls{acr:dqn} and \gls{acr:ppo} both achieve higher and more stable profits compared to \gls{acr:tql}.}
        \label{fig:delta_tql_drl_standard}
\end{figure}
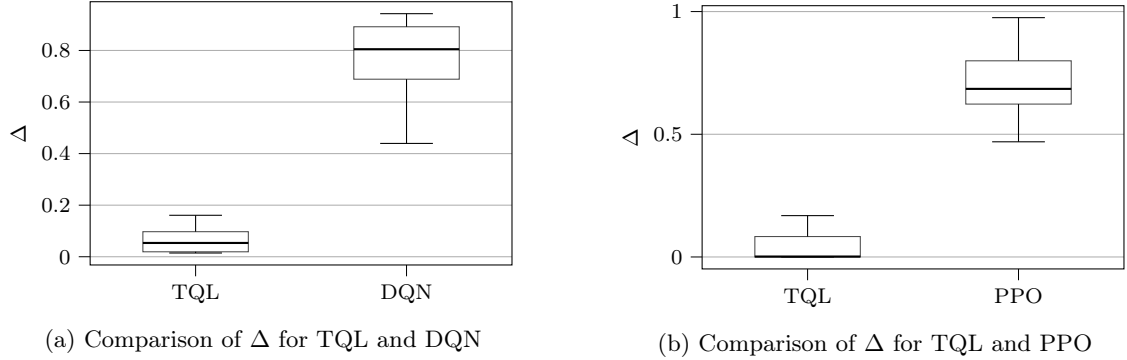

\begin{figure}[htbp]
	\centering

	\begin{minipage}[t]{0.48\linewidth}
		\centering
		\includegraphics[width=1.0\linewidth]{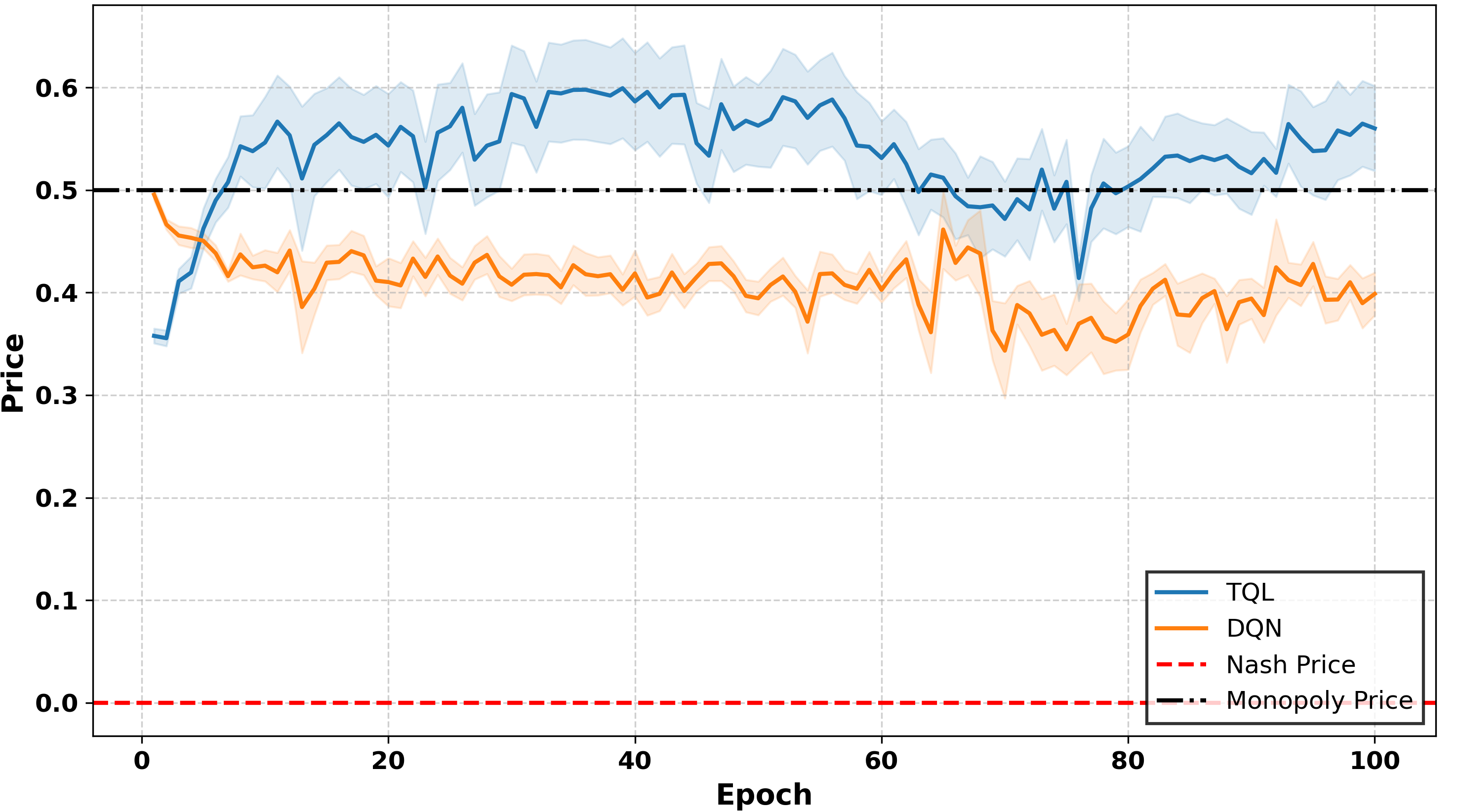}
        \subcaption*{(a) TQL vs DQN: Price Trend (with 95\% CI)}\label{TQL_PPO_price_per_iteration_edgeworth}
	\end{minipage}
	\hfill
    \begin{minipage}[t]{0.48\linewidth}
		\centering
		\includegraphics[width=1.0\linewidth]{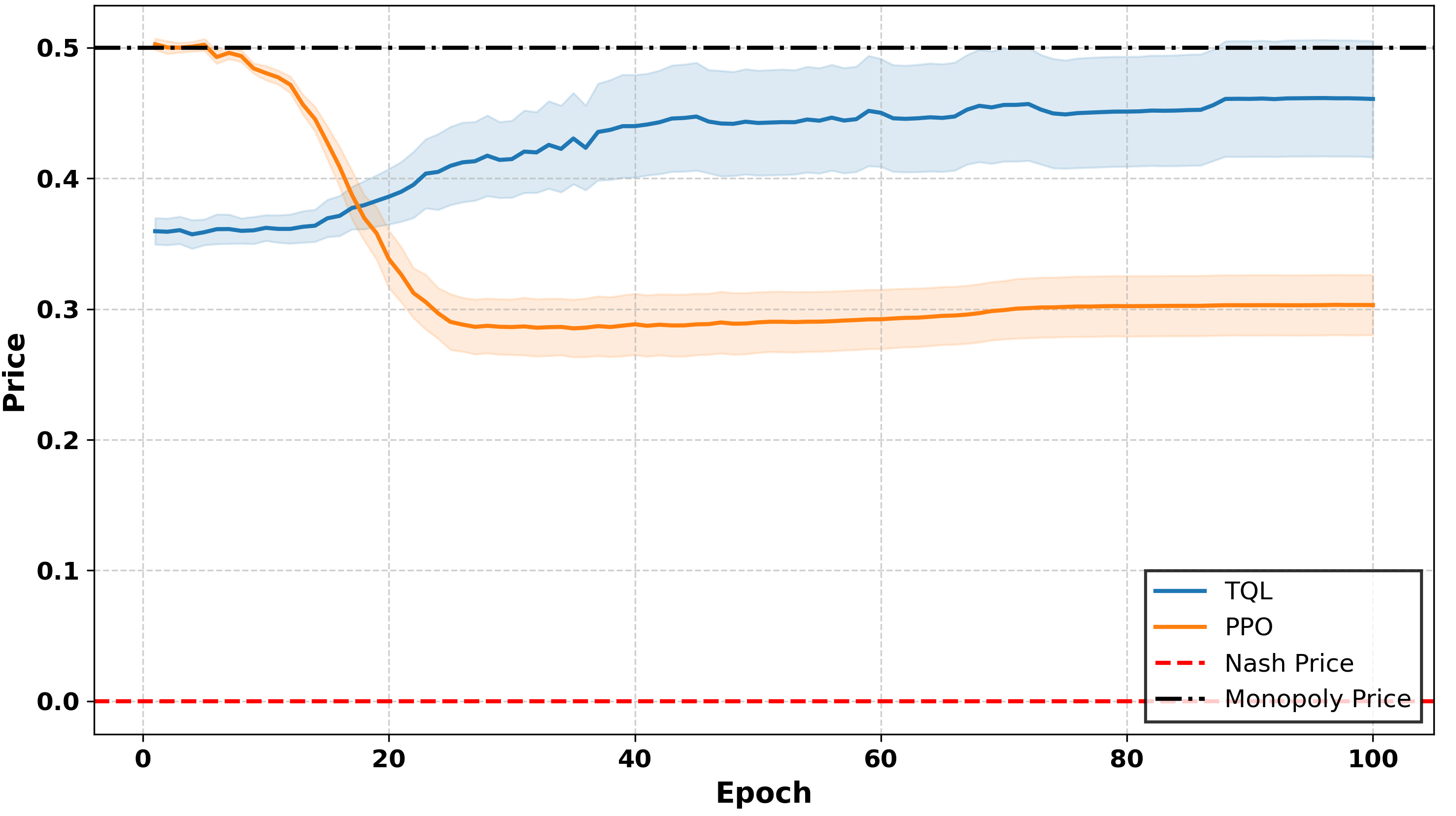}
        \subcaption*{(b) TQL vs PPO: Price Trend (with 95\% CI)}
        \label{TQL_DQN_price_per_iteration_edgeworth}
	\end{minipage}
        \caption{Price learning curves for the Edgeworth Bertrand model, showing competition between \gls{acr:tql} and \gls{acr:dqn} (left) and between \gls{acr:tql} and \gls{acr:ppo} (right). Results are based on 20 independent runs, with mean values and 95\% confidence intervals displayed.}
    \label{Combined_price_learning_curve_tql_drl_edgeworth}
\end{figure}

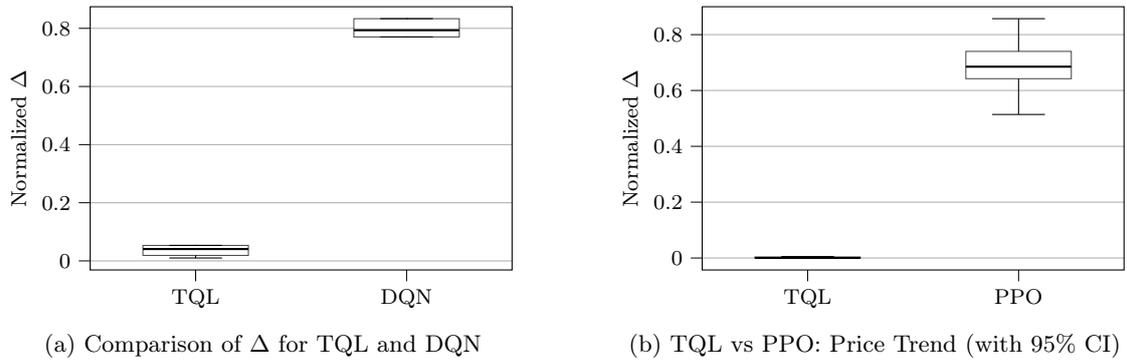
\begin{figure}[htbp]
	\centering
	\begin{minipage}{0.48\linewidth}
		\centering
\begin{tikzpicture}

\definecolor{darkgray176}{RGB}{176,176,176}
\definecolor{darkorange25512714}{RGB}{255,127,14}
\definecolor{steelblue31119180}{RGB}{31,119,180}
\scriptsize

\begin{axis}[
width=0.9\textwidth, 
height=0.64\textwidth, 
tick align=outside,
tick pos=left,
x grid style={darkgray176},
xmin=0.5, xmax=2.5,
xtick style={color=black},
xtick={1,2},
xticklabels={TQL,DQN},
y grid style={darkgray176},
ylabel={Normalized \(\displaystyle \Delta\)},
ymajorgrids,
ymin=-0.0311755204081633, ymax=0.874358173469388,
ytick style={color=black}
]
\path [draw=black, fill=white, opacity=0.7]
(axis cs:0.75,0.0188444897959184)
--(axis cs:1.25,0.0188444897959184)
--(axis cs:1.25,0.0532855102040816)
--(axis cs:0.75,0.0532855102040816)
--(axis cs:0.75,0.0188444897959184)
--cycle;
\addplot [black]
table {%
1 0.0188444897959184
1 0.00998510204081633
};
\addplot [black]
table {%
1 0.0532855102040816
1 0.0532855102040816
};
\addplot [black]
table {%
0.875 0.00998510204081633
1.125 0.00998510204081633
};
\addplot [black]
table {%
0.875 0.0532855102040816
1.125 0.0532855102040816
};
\path [draw=black, fill=white, opacity=0.7]
(axis cs:1.75,0.770372653061224)
--(axis cs:2.25,0.770372653061224)
--(axis cs:2.25,0.833197551020408)
--(axis cs:1.75,0.833197551020408)
--(axis cs:1.75,0.770372653061224)
--cycle;
\addplot [black]
table {%
2 0.770372653061224
2 0.770372653061224
};
\addplot [black]
table {%
2 0.833197551020408
2 0.833197551020408
};
\addplot [black]
table {%
1.875 0.770372653061224
2.125 0.770372653061224
};
\addplot [black]
table {%
1.875 0.833197551020408
2.125 0.833197551020408
};
\addplot [thick, black]
table {%
0.75 0.0408391836734694
1.25 0.0408391836734694
};
\addplot [thick, black]
table {%
1.75 0.793683469387755
2.25 0.793683469387755
};
\end{axis}

\end{tikzpicture}
        \subcaption*{(a) Comparison of \(\displaystyle \Delta\) for TQL and DQN}
        \label{Delta_tql_dqn_edgeworth}
	\end{minipage}
	\begin{minipage}{0.48\linewidth}
		\centering
\begin{tikzpicture}

\definecolor{darkgray176}{RGB}{176,176,176}
\definecolor{darkorange25512714}{RGB}{255,127,14}
\definecolor{steelblue31119180}{RGB}{31,119,180}
\scriptsize

\begin{axis}[
width=0.9\textwidth, 
height=0.64\textwidth, 
tick align=outside,
tick pos=left,
x grid style={darkgray176},
xmin=0.5, xmax=2.5,
xtick style={color=black},
xtick={1,2},
xticklabels={TQL,PPO},
y grid style={darkgray176},
ylabel={Normalized \(\displaystyle \Delta\)},
ymajorgrids,
ymin=-0.0428571428571428, ymax=0.9,
ytick style={color=black}
]
\path [draw=black, fill=white, opacity=0.7]
(axis cs:0.75,2.52551020408163e-05)
--(axis cs:1.25,2.52551020408163e-05)
--(axis cs:1.25,0.00344709183673469)
--(axis cs:0.75,0.00344709183673469)
--(axis cs:0.75,2.52551020408163e-05)
--cycle;
\addplot [black]
table {%
1 2.52551020408163e-05
1 0
};
\addplot [black]
table {%
1 0.00344709183673469
1 0.00520204081632653
};
\addplot [black]
table {%
0.875 0
1.125 0
};
\addplot [black]
table {%
0.875 0.00520204081632653
1.125 0.00520204081632653
};
\path [draw=black, fill=white, opacity=0.7]
(axis cs:1.75,0.642190204081633)
--(axis cs:2.25,0.642190204081633)
--(axis cs:2.25,0.740023112244898)
--(axis cs:1.75,0.740023112244898)
--(axis cs:1.75,0.642190204081633)
--cycle;
\addplot [black]
table {%
2 0.642190204081633
2 0.514081632653061
};
\addplot [black]
table {%
2 0.740023112244898
2 0.857142857142857
};
\addplot [black]
table {%
1.875 0.514081632653061
2.125 0.514081632653061
};
\addplot [black]
table {%
1.875 0.857142857142857
2.125 0.857142857142857
};
\addplot [thick, black]
table {%
0.75 0.000417448979591837
1.25 0.000417448979591837
};
\addplot [thick, black]
table {%
1.75 0.685262040816326
2.25 0.685262040816326
};
\end{axis}

\end{tikzpicture}
        \subcaption*{(b) TQL vs PPO: Price Trend (with 95\% CI)}
        \label{Delta_tql_ppo_edgeworth}
	\end{minipage}
        \caption{Boxplots of normalized profit \(\Delta\) for the Edgeworth Bertrand model, comparing competition between \gls{acr:tql} and \gls{acr:dqn} (left) and between \gls{acr:tql} and \gls{acr:ppo} (right). \gls{acr:dqn} and \gls{acr:ppo} both achieve higher and more stable profits compared to \gls{acr:tql}.}
        \label{fig:delta_tql_drl_edgeworth}
\end{figure}

\newpage
\subsection{Heterogeneous \gls{acr:drl} Interaction}

Figures~\ref{PPO_env_price_profit_learning_curve_standard}-\ref{DQN_env_price_profit_learning_curve_standard} in the Standard Bertrand model show that, \gls{acr:ppo} consistently adopts a low-price strategy in the \gls{acr:ppo} environment, stabilizing near 0.4 and achieving profits close to the monopoly level. Meanwhile, \gls{acr:dqn} converges to higher prices with profits near the Nash level. In the \gls{acr:dqn} environment, \gls{acr:ppo} still sets slightly lower prices and earns slightly higher profits than \gls{acr:dqn}, suggesting that the learned strategies remain consistent across environments.

Figures~\ref{PPO_env_price_profit_learning_curve_edgeworth}-\ref{DQN_env_price_profit_learning_curve_edgeworth} in the Edgeworth Bertrand model show that, increased market complexity narrows the profit gap between the two algorithms. In the \gls{acr:ppo} environment, \gls{acr:ppo} maintains its low-price strategy with higher profits. However, in the \gls{acr:dqn} environment, the profit curves of \gls{acr:dqn} and \gls{acr:ppo} overlap, reflecting more balanced competition in this model.


\begin{figure}[htbp]
	\centering
	\begin{minipage}[t]{0.48\linewidth}
		\centering
		\includegraphics[width=1.0\linewidth]{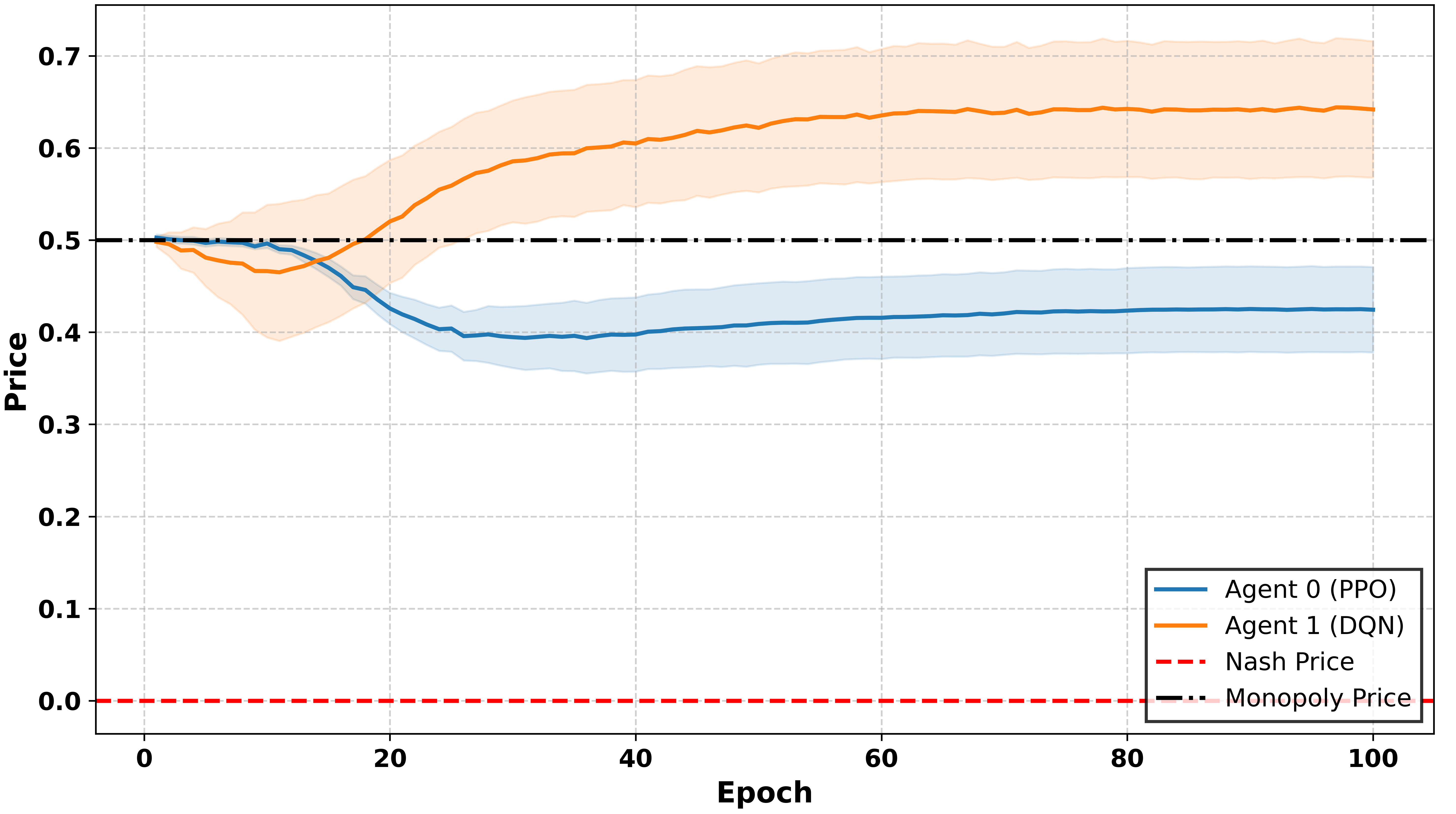} 
        \subcaption*{(a) Price per Epoch in PPO Environment}
        \label{PPO_env_price_learning_curve_standard}
	\end{minipage}
	\hfill
	\begin{minipage}[t]{0.48\linewidth}
		\centering
		\includegraphics[width=1.0\linewidth]{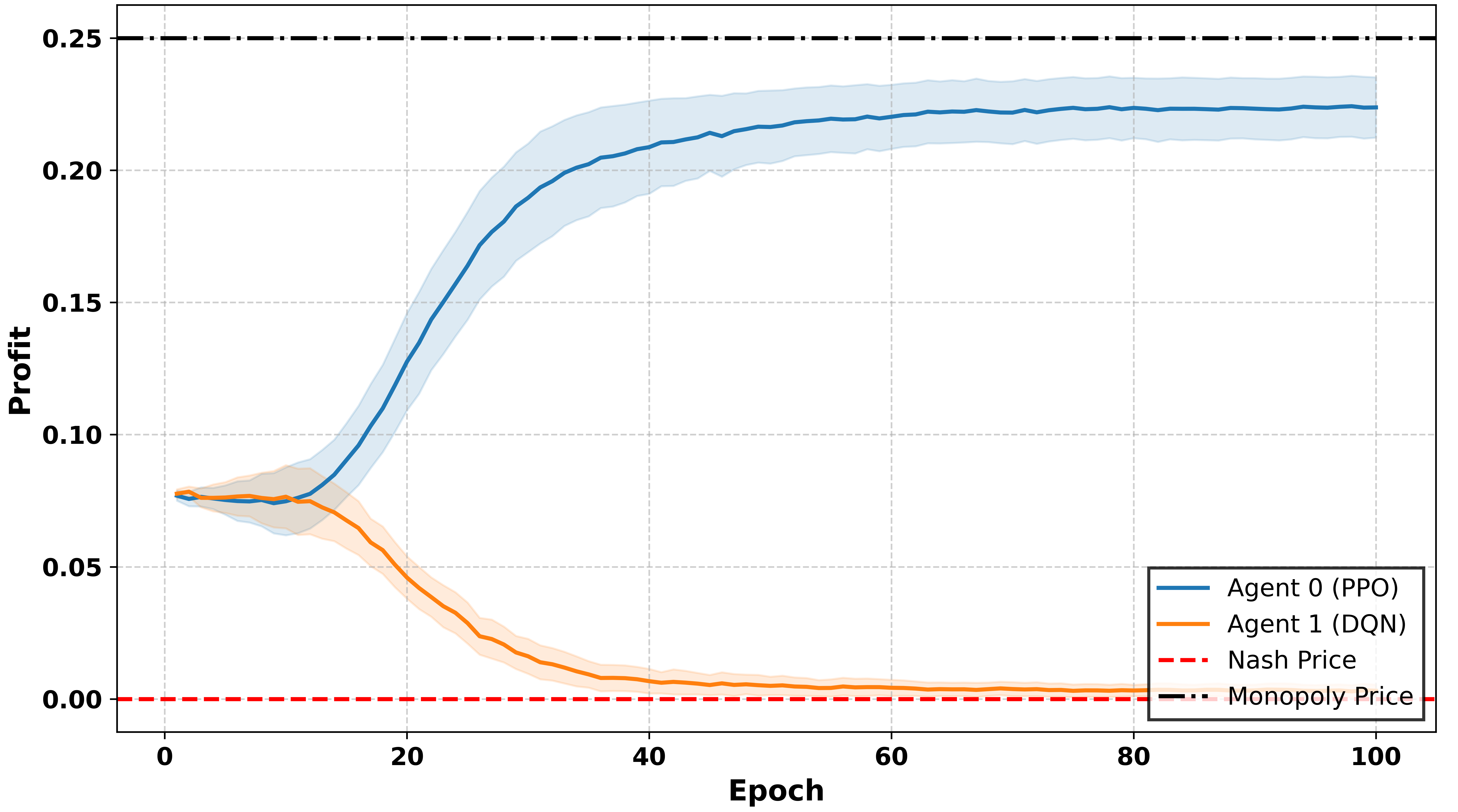} 
        \subcaption*{(b) Profit per Epoch in PPO Environment}\label{PPO_env_profit_learning_curve_standard}
	\end{minipage}
        \caption{Results for the Standard Bertrand model with a PPO environment. The left panel illustrates price dynamics, while the right panel depicts profit outcomes. Results are based on the mean values with 95\% confidence intervals from 20 independent runs. 
        }        \label{PPO_env_price_profit_learning_curve_standard}
\end{figure}

\begin{figure}[htbp]
	\centering
	\begin{minipage}[t]{0.48\linewidth}
		\centering
		\includegraphics[width=1.0\linewidth]{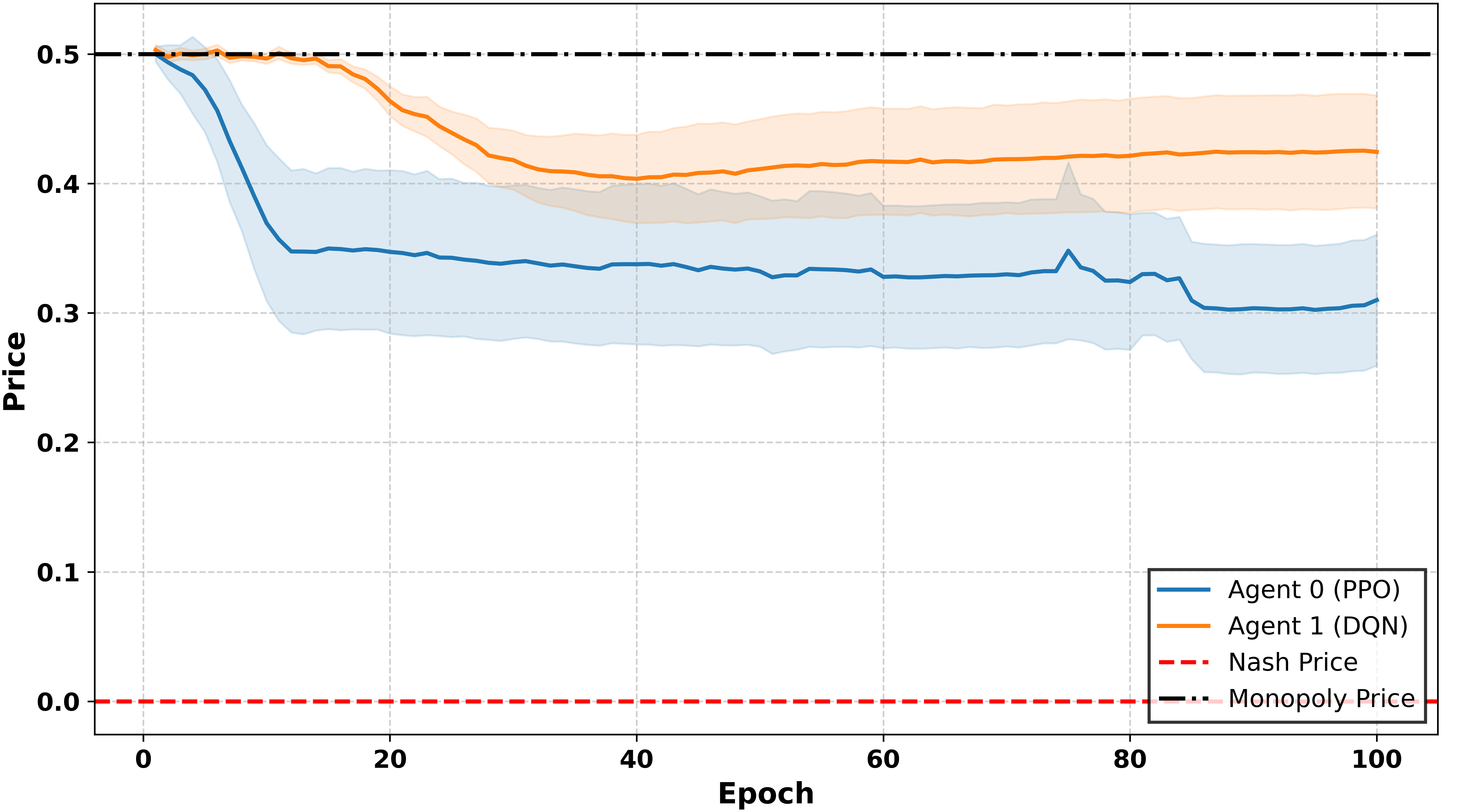}
        \subcaption*{(a) Price per Epoch in DQN Environment}
        \label{DQN_env_price_learning_curve_standard}
	\end{minipage}
	\hfill
	\begin{minipage}[t]{0.48\linewidth}
		\centering
		\includegraphics[width=1.0\linewidth]{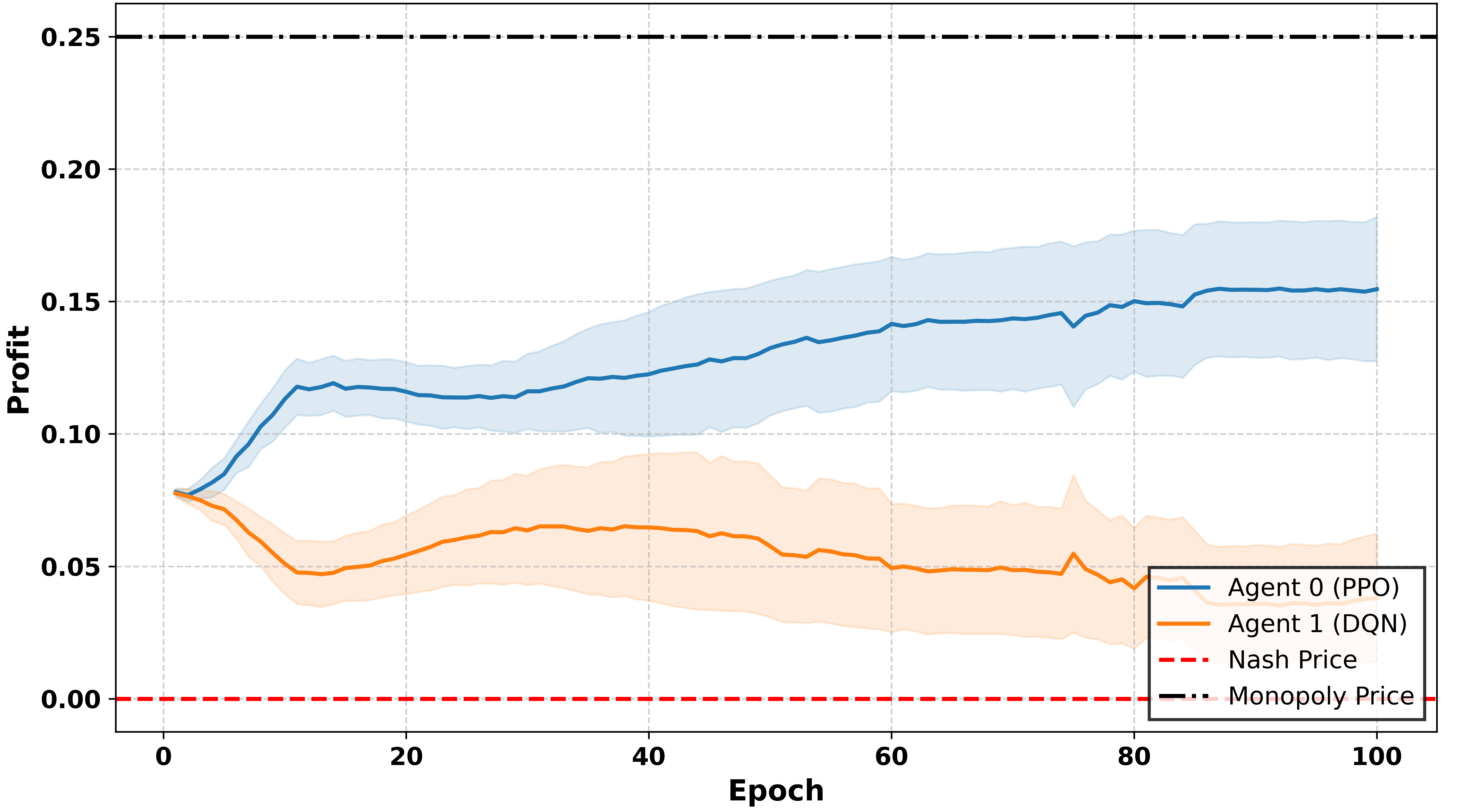}
        \subcaption*{(b) Profit per Epoch in DQN Environment}
        \label{DQN_env_profit_learning_curve_standard}
	\end{minipage}
    \caption{Results for the Standard Bertrand model with a DQN environment. The left panel illustrates price dynamics, while the right panel depicts profit outcomes. Results are based on the mean values with 95\% confidence intervals from 20 independent runs. 
    }
    \label{DQN_env_price_profit_learning_curve_standard}
\end{figure} 


\begin{figure}[htbp]
	\centering
	\begin{minipage}[t]{0.48\linewidth}
		\centering
		\includegraphics[width=1.0\linewidth]{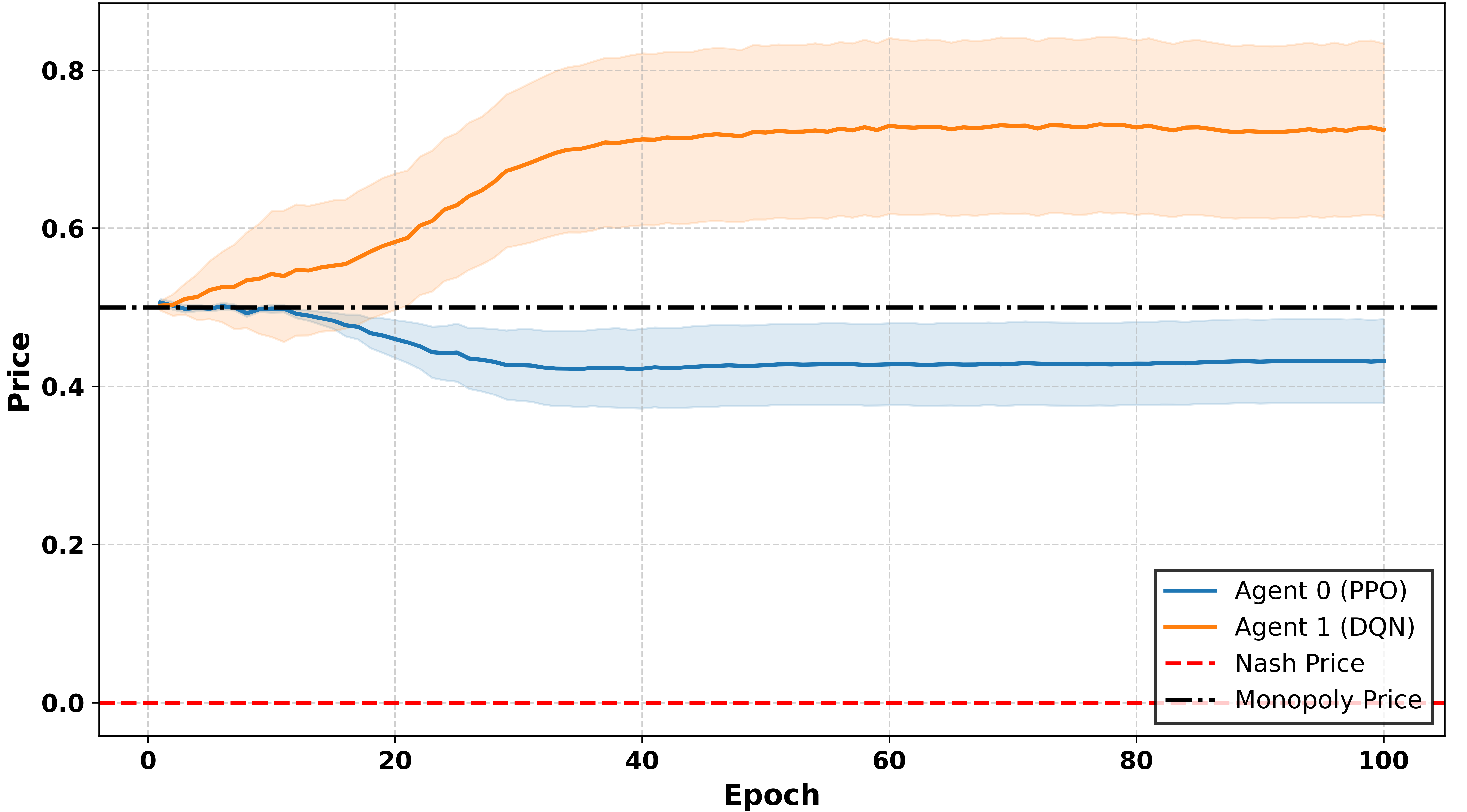}
        \subcaption*{(a) Price per Epoch in PPO Environment}
        \label{PPO_env_price_learning_curve_edgeworth}
	\end{minipage}
	\hfill
	\begin{minipage}[t]{0.48\linewidth}
		\centering
		\includegraphics[width=1.0\linewidth]{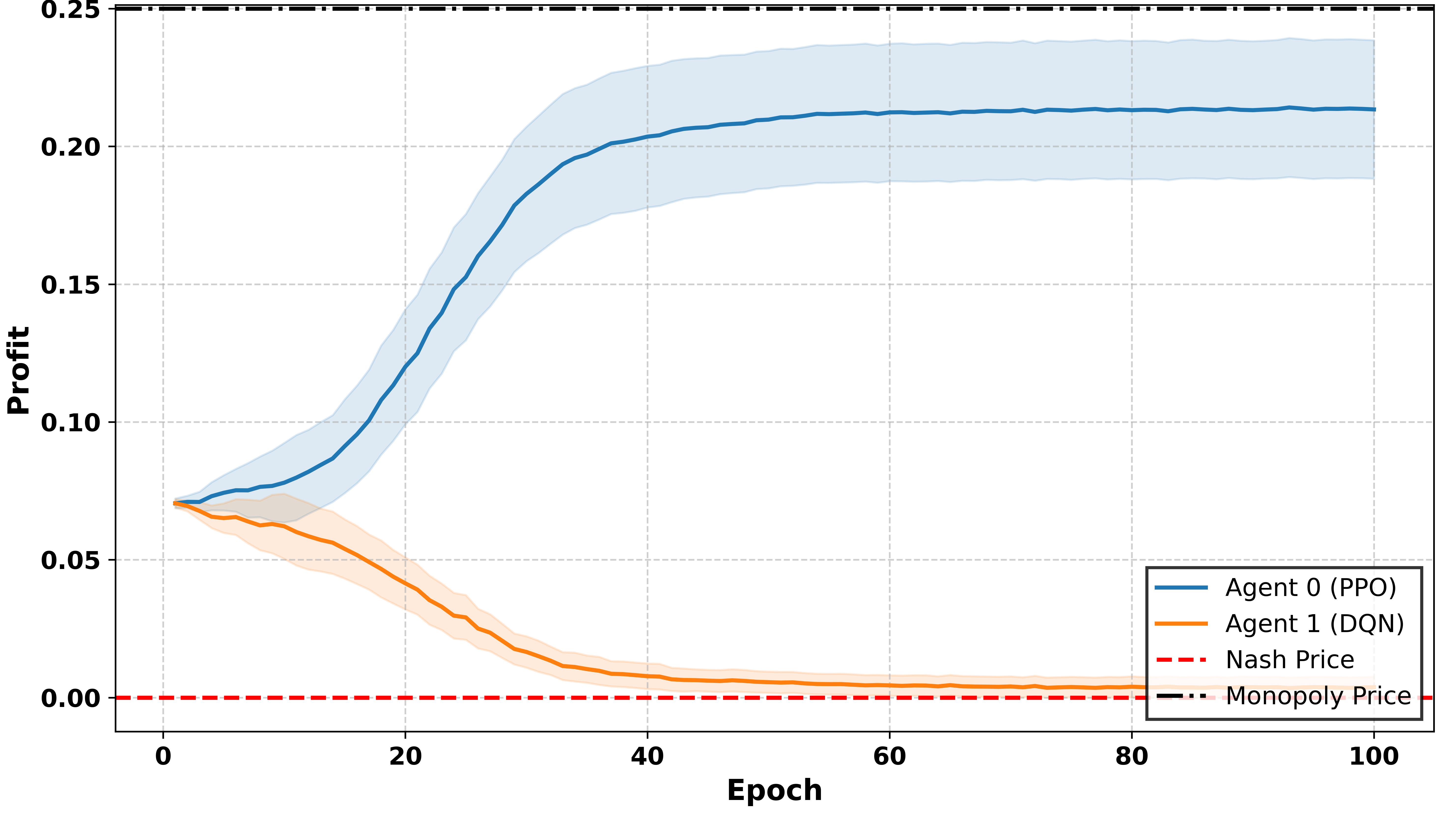}  
        \subcaption*{(b) Profit per Epoch in PPO Environment}\label{PPO_env_profit_learning_curve_edgeworth}
	\end{minipage}
        \caption{Results for the Edgeworth Bertrand model with a PPO environment. The left panel illustrates price dynamics, while the right panel depicts profit outcomes. Results are based on the mean values with 95\% confidence intervals from 20 independent runs. 
        }
      \label{PPO_env_price_profit_learning_curve_edgeworth}
\end{figure}

\begin{figure}[htbp]
	\centering
	\begin{minipage}[t]{0.48\linewidth}
		\centering
		\includegraphics[width=1.0\linewidth]{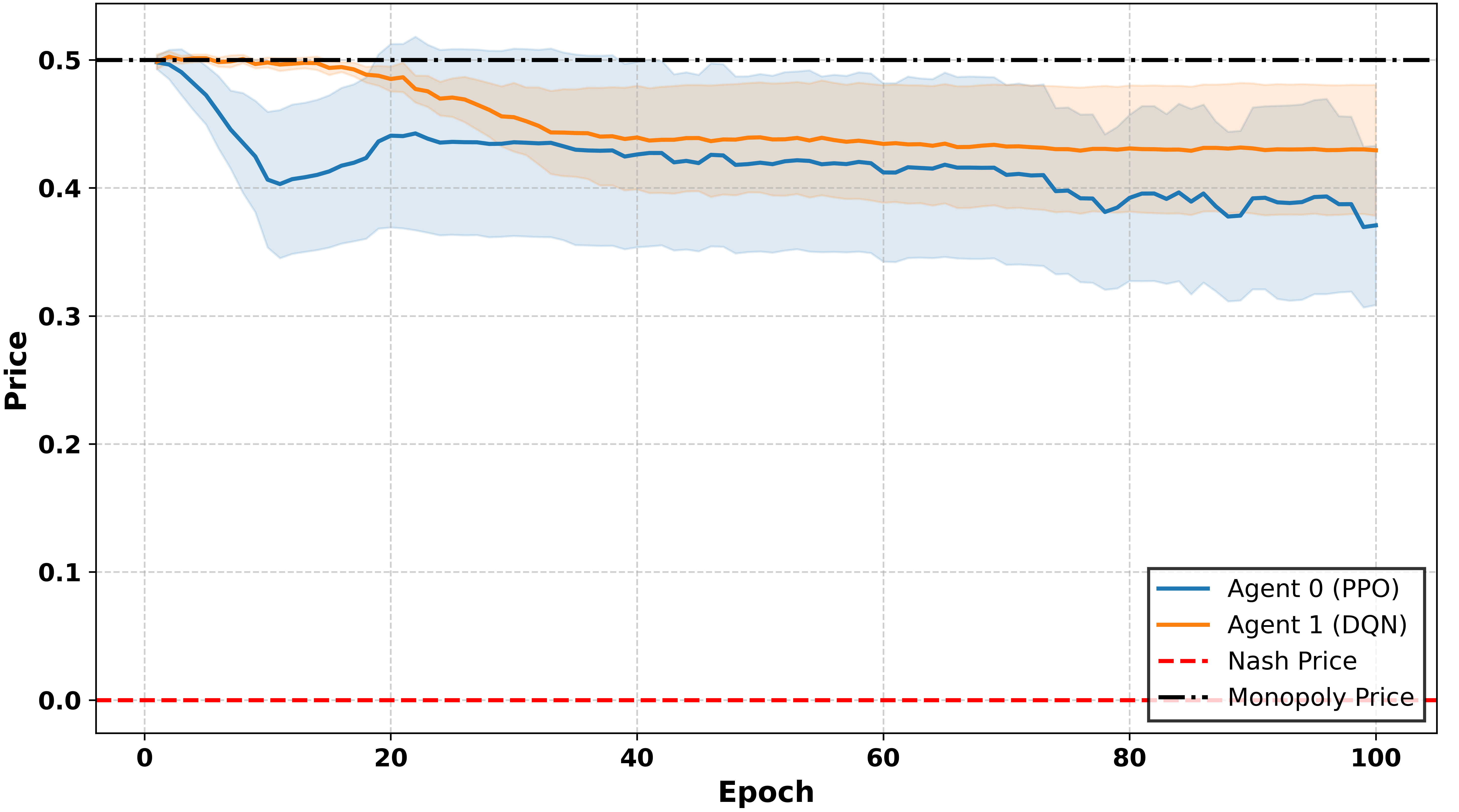}
        \subcaption*{(a) Price per Epoch in DQN Environment}
        \label{DQN_env_price_learning_curve_edgeworth}
	\end{minipage}
	\hfill
	\begin{minipage}[t]{0.48\linewidth}
		\centering
		\includegraphics[width=1.0\linewidth]{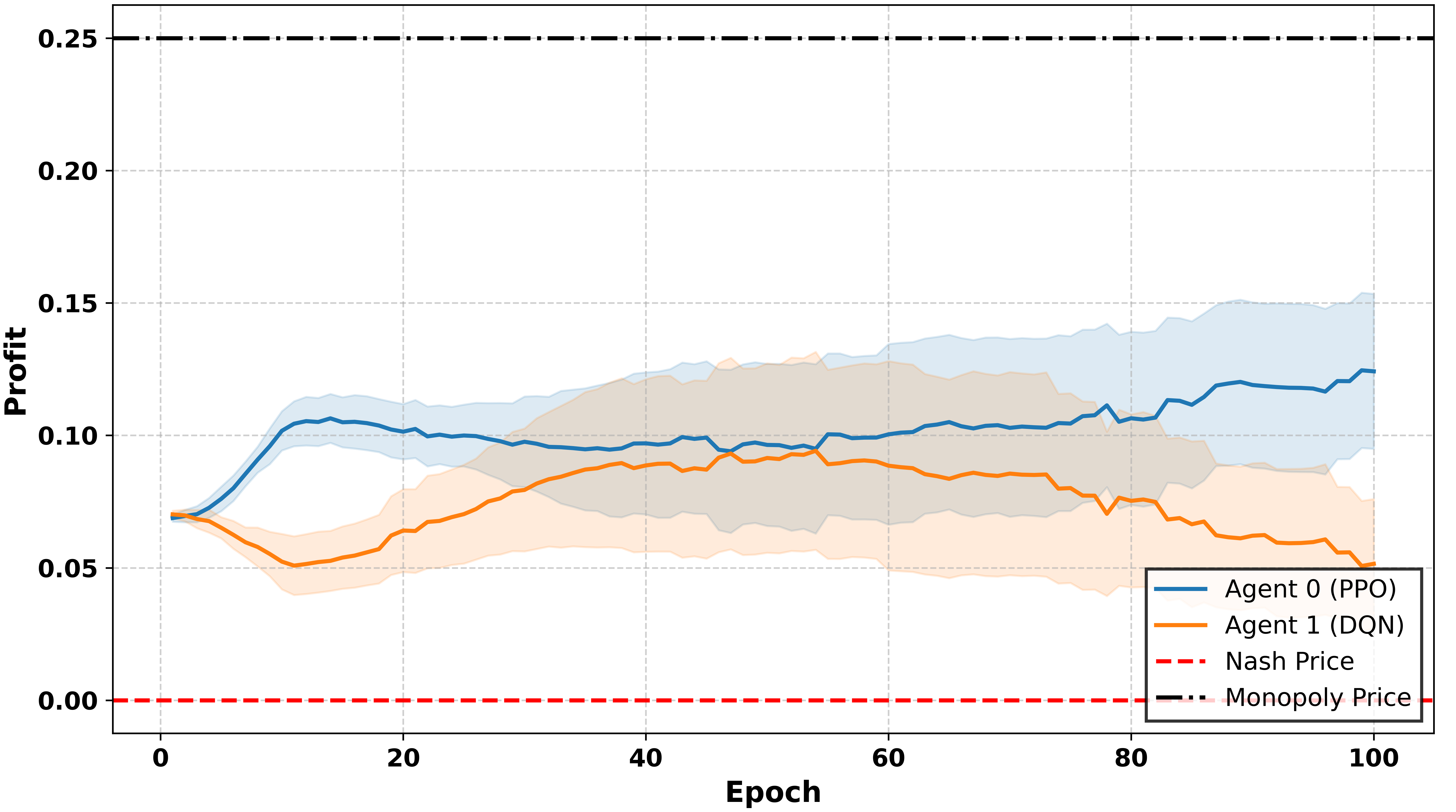}
        \subcaption*{(b) Profit per Epoch in DQN Environment}
        \label{DQN_env_profit_learning_curve_edgworth}
	\end{minipage}
    \caption{Results for the Edgeworth Bertrand model with a DQN environment. The left panel illustrates price dynamics, while the right panel depicts profit outcomes. Results are based on the mean values with 95\% confidence intervals from 20 independent runs.
    }
    \label{DQN_env_price_profit_learning_curve_edgeworth}
\end{figure}

\FloatBarrier

\subsection{Impact of State Definitions on TQL Agent Behavior}

In the Standard Bertrand model (Figure~\ref{fig:state_price_profit_diff_boxplot_standard}), when state definitions exclude opponent information (e.g., \(self\_k1\)), pricing remains competitive and stabilizes near the Nash price. As memory length increases (e.g., \(self\_k2\) and \(self\_k3\)), prices gradually rise, reflecting higher pricing strategies. Conversely, when state definitions include complete opponent information (e.g., \(k1\), \(k2\), \(k3\)), greater state information leads to lower prices, highlighting the influence of memory length on pricing behavior.

In the Edgeworth Bertrand model (Figure~\ref{fig:state_price_profit_diff_boxplot_edgeworth}), under complete opponent information (e.g., \(k1\), \(k2\), \(k3\)), increasing memory length leads to lower prices and lower profits. In contrast, under limited information (self\_k1, self\_k2, self\_k3), this trend is less pronounced.

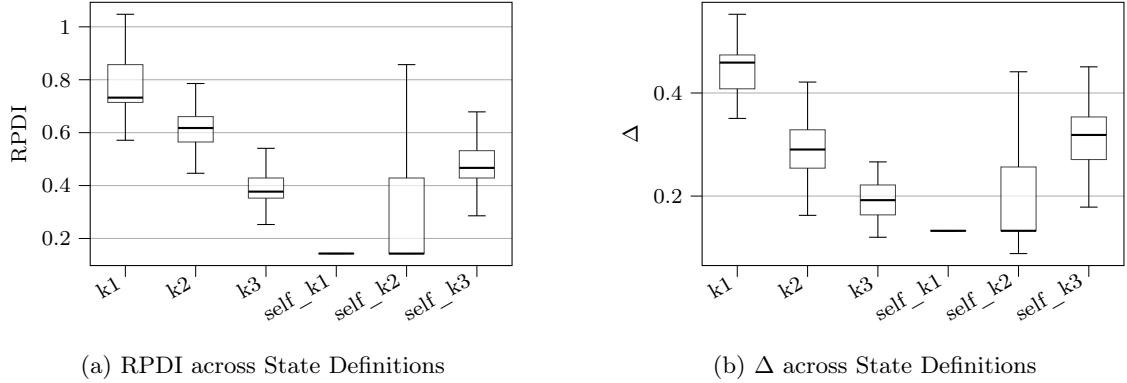
\begin{figure}[htbp]
	\centering
	\begin{minipage}{0.48\linewidth}
		\centering
\begin{tikzpicture}

\definecolor{darkgray176}{RGB}{176,176,176}
\definecolor{skyblue}{RGB}{135,206,235}
\scriptsize

\begin{axis}[
width=0.9\textwidth, 
height=0.64\textwidth, 
tick align=outside,
tick pos=left,
x grid style={darkgray176},
xmin=0.5, xmax=6.5,
xtick style={color=black},
xtick={1,2,3,4,5,6},
xticklabel style={rotate=30.0,anchor=east},
xticklabels={k1,k2,k3,self\_k1,self\_k2,self\_k3},
y grid style={darkgray176},
ylabel={RPDI},
ymajorgrids,
ymin=0.0976292857142857, ymax=1.09264214285714,
ytick style={color=black}
]
\path [draw=black, fill=white, opacity=0.7]
(axis cs:0.75,0.714164285714285)
--(axis cs:1.25,0.714164285714285)
--(axis cs:1.25,0.857196428571429)
--(axis cs:0.75,0.857196428571429)
--(axis cs:0.75,0.714164285714285)
--cycle;
\addplot [black]
table {%
1 0.714164285714285
1 0.571285714285714
};
\addplot [black]
table {%
1 0.857196428571429
1 1.04741428571429
};
\addplot [black]
table {%
0.875 0.571285714285714
1.125 0.571285714285714
};
\addplot [black]
table {%
0.875 1.04741428571429
1.125 1.04741428571429
};
\path [draw=black, fill=white, opacity=0.7]
(axis cs:1.75,0.564735714285714)
--(axis cs:2.25,0.564735714285714)
--(axis cs:2.25,0.660732142857143)
--(axis cs:1.75,0.660732142857143)
--(axis cs:1.75,0.564735714285714)
--cycle;
\addplot [black]
table {%
2 0.564735714285714
2 0.446428571428571
};
\addplot [black]
table {%
2 0.660732142857143
2 0.785728571428571
};
\addplot [black]
table {%
1.875 0.446428571428571
2.125 0.446428571428571
};
\addplot [black]
table {%
1.875 0.785728571428571
2.125 0.785728571428571
};
\path [draw=black, fill=white, opacity=0.7]
(axis cs:2.75,0.352517857142857)
--(axis cs:3.25,0.352517857142857)
--(axis cs:3.25,0.428571428571428)
--(axis cs:2.75,0.428571428571428)
--(axis cs:2.75,0.352517857142857)
--cycle;
\addplot [black]
table {%
3 0.352517857142857
3 0.252785714285714
};
\addplot [black]
table {%
3 0.428571428571428
3 0.540771428571428
};
\addplot [black]
table {%
2.875 0.252785714285714
3.125 0.252785714285714
};
\addplot [black]
table {%
2.875 0.540771428571428
3.125 0.540771428571428
};
\path [draw=black, fill=white, opacity=0.7]
(axis cs:3.75,0.142971428571428)
--(axis cs:4.25,0.142971428571428)
--(axis cs:4.25,0.143157142857143)
--(axis cs:3.75,0.143157142857143)
--(axis cs:3.75,0.142971428571428)
--cycle;
\addplot [black]
table {%
4 0.142971428571428
4 0.142857142857143
};
\addplot [black]
table {%
4 0.143157142857143
4 0.1433
};
\addplot [black]
table {%
3.875 0.142857142857143
4.125 0.142857142857143
};
\addplot [black]
table {%
3.875 0.1433
4.125 0.1433
};
\path [draw=black, fill=white, opacity=0.7]
(axis cs:4.75,0.142857142857143)
--(axis cs:5.25,0.142857142857143)
--(axis cs:5.25,0.428571428571428)
--(axis cs:4.75,0.428571428571428)
--(axis cs:4.75,0.142857142857143)
--cycle;
\addplot [black]
table {%
5 0.142857142857143
5 0.142857142857143
};
\addplot [black]
table {%
5 0.428571428571428
5 0.8571
};
\addplot [black]
table {%
4.875 0.142857142857143
5.125 0.142857142857143
};
\addplot [black]
table {%
4.875 0.8571
5.125 0.8571
};
\path [draw=black, fill=white, opacity=0.7]
(axis cs:5.75,0.428571428571428)
--(axis cs:6.25,0.428571428571428)
--(axis cs:6.25,0.531921428571428)
--(axis cs:5.75,0.531921428571428)
--(axis cs:5.75,0.428571428571428)
--cycle;
\addplot [black]
table {%
6 0.428571428571428
6 0.285714285714286
};
\addplot [black]
table {%
6 0.531921428571428
6 0.678571428571428
};
\addplot [black]
table {%
5.875 0.285714285714286
6.125 0.285714285714286
};
\addplot [black]
table {%
5.875 0.678571428571428
6.125 0.678571428571428
};
\addplot [thick, black]
table {%
0.75 0.732321428571428
1.25 0.732321428571428
};
\addplot [thick, black]
table {%
1.75 0.617557142857143
2.25 0.617557142857143
};
\addplot [thick, black]
table {%
2.75 0.377092857142857
3.25 0.377092857142857
};
\addplot [thick, black]
table {%
3.75 0.143035714285714
4.25 0.143035714285714
};
\addplot [thick, black]
table {%
4.75 0.142857142857143
5.25 0.142857142857143
};
\addplot [thick, black]
table {%
5.75 0.466814285714286
6.25 0.466814285714286
};
\end{axis}

\end{tikzpicture}
        \subcaption*{(a) RPDI across State Definitions}
        \label{fig:combined_normalized_price_boxplot_standard}
	\end{minipage}
	\begin{minipage}{0.48\linewidth}
		\centering
\begin{tikzpicture}

\definecolor{darkgray176}{RGB}{176,176,176}
\definecolor{skyblue}{RGB}{135,206,235}
\scriptsize

\begin{axis}[
width=0.9\textwidth, 
height=0.64\textwidth, 
tick align=outside,
tick pos=left,
x grid style={darkgray176},
xmin=0.5, xmax=6.5,
xtick style={color=black},
xtick={1,2,3,4,5,6},
xticklabel style={rotate=30.0,anchor=east},
xticklabels={k1,k2,k3,self\_k1,self\_k2,self\_k3},
y grid style={darkgray176},
ylabel={\(\displaystyle \Delta\)},
ymajorgrids,
ymin=0.0652272448979591, ymax=0.575903367346939,
ytick style={color=black}
]
\path [draw=black, fill=white, opacity=0.7]
(axis cs:0.75,0.408132653061224)
--(axis cs:1.25,0.408132653061224)
--(axis cs:1.25,0.473797704081632)
--(axis cs:0.75,0.473797704081632)
--(axis cs:0.75,0.408132653061224)
--cycle;
\addplot [black]
table {%
1 0.408132653061224
1 0.350676530612245
};
\addplot [black]
table {%
1 0.473797704081632
1 0.55269081632653
};
\addplot [black]
table {%
0.875 0.350676530612245
1.125 0.350676530612245
};
\addplot [black]
table {%
0.875 0.55269081632653
1.125 0.55269081632653
};
\path [draw=black, fill=white, opacity=0.7]
(axis cs:1.75,0.254169897959184)
--(axis cs:2.25,0.254169897959184)
--(axis cs:2.25,0.328622193877551)
--(axis cs:1.75,0.328622193877551)
--(axis cs:1.75,0.254169897959184)
--cycle;
\addplot [black]
table {%
2 0.254169897959184
2 0.162620408163265
};
\addplot [black]
table {%
2 0.328622193877551
2 0.421208163265306
};
\addplot [black]
table {%
1.875 0.162620408163265
2.125 0.162620408163265
};
\addplot [black]
table {%
1.875 0.421208163265306
2.125 0.421208163265306
};
\path [draw=black, fill=white, opacity=0.7]
(axis cs:2.75,0.163541326530612)
--(axis cs:3.25,0.163541326530612)
--(axis cs:3.25,0.221566071428571)
--(axis cs:2.75,0.221566071428571)
--(axis cs:2.75,0.163541326530612)
--cycle;
\addplot [black]
table {%
3 0.163541326530612
3 0.120118367346939
};
\addplot [black]
table {%
3 0.221566071428571
3 0.266326530612245
};
\addplot [black]
table {%
2.875 0.120118367346939
3.125 0.120118367346939
};
\addplot [black]
table {%
2.875 0.266326530612245
3.125 0.266326530612245
};
\path [draw=black, fill=white, opacity=0.7]
(axis cs:3.75,0.132639795918367)
--(axis cs:4.25,0.132639795918367)
--(axis cs:4.25,0.132666326530612)
--(axis cs:3.75,0.132666326530612)
--(axis cs:3.75,0.132639795918367)
--cycle;
\addplot [black]
table {%
4 0.132639795918367
4 0.1326
};
\addplot [black]
table {%
4 0.132666326530612
4 0.132692857142857
};
\addplot [black]
table {%
3.875 0.1326
4.125 0.1326
};
\addplot [black]
table {%
3.875 0.132692857142857
4.125 0.132692857142857
};
\path [draw=black, fill=white, opacity=0.7]
(axis cs:4.75,0.13265306122449)
--(axis cs:5.25,0.13265306122449)
--(axis cs:5.25,0.256442091836735)
--(axis cs:4.75,0.256442091836735)
--(axis cs:4.75,0.13265306122449)
--cycle;
\addplot [black]
table {%
5 0.13265306122449
5 0.0884397959183672
};
\addplot [black]
table {%
5 0.256442091836735
5 0.441159183673469
};
\addplot [black]
table {%
4.875 0.0884397959183672
5.125 0.0884397959183672
};
\addplot [black]
table {%
4.875 0.441159183673469
5.125 0.441159183673469
};
\path [draw=black, fill=white, opacity=0.7]
(axis cs:5.75,0.270917091836735)
--(axis cs:6.25,0.270917091836735)
--(axis cs:6.25,0.353526275510204)
--(axis cs:5.75,0.353526275510204)
--(axis cs:5.75,0.270917091836735)
--cycle;
\addplot [black]
table {%
6 0.270917091836735
6 0.178573469387755
};
\addplot [black]
table {%
6 0.353526275510204
6 0.450765306122449
};
\addplot [black]
table {%
5.875 0.178573469387755
6.125 0.178573469387755
};
\addplot [black]
table {%
5.875 0.450765306122449
6.125 0.450765306122449
};
\addplot [thick, black]
table {%
0.75 0.458953571428571
1.25 0.458953571428571
};
\addplot [thick, black]
table {%
1.75 0.290345408163265
2.25 0.290345408163265
};
\addplot [thick, black]
table {%
2.75 0.192092857142857
3.25 0.192092857142857
};
\addplot [thick, black]
table {%
3.75 0.13265306122449
4.25 0.13265306122449
};
\addplot [thick, black]
table {%
4.75 0.13265306122449
5.25 0.13265306122449
};
\addplot [thick, black]
table {%
5.75 0.318585204081632
6.25 0.318585204081632
};
\end{axis}

\end{tikzpicture}
        \subcaption*{(b) \(\displaystyle \Delta\) across State Definitions}
        \label{fig:combined_delta_boxplot_standard}
	\end{minipage}
        \caption{Standard Bertrand - For states without opponent information (\(self\_k1\)), pricing remains competitive and stabilizes near the Nash price. With complete opponent information (\(k1\), \(k2\), \(k3\)), more state information leads to lower prices. Conversely, for states with only self-information (\(self\_k2\), \(self\_k3\)), longer memory results in higher prices.}
        \label{fig:state_price_profit_diff_boxplot_standard}
\end{figure}

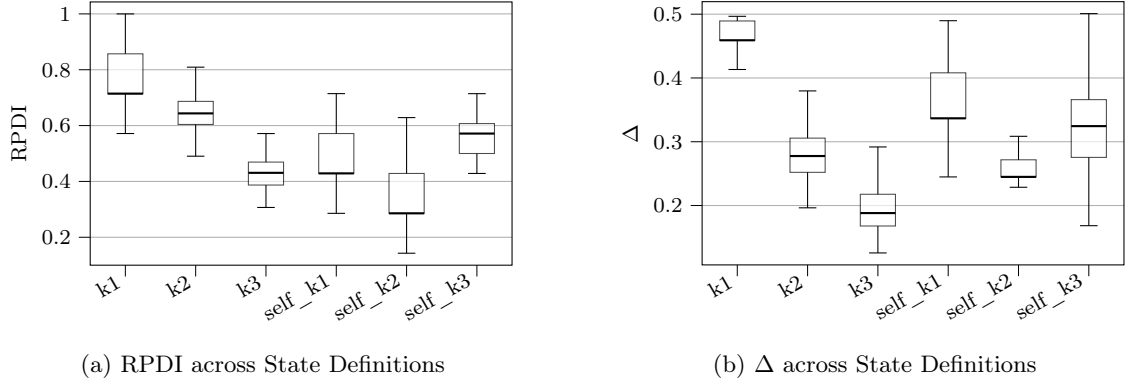
\begin{figure}[htbp]
	\centering
	\begin{minipage}{0.48\linewidth}
		\centering
\begin{tikzpicture}

\definecolor{darkgray176}{RGB}{176,176,176}
\definecolor{skyblue}{RGB}{135,206,235}
\scriptsize

\begin{axis}[
width=0.9\textwidth, 
height=0.64\textwidth, 
tick align=outside,
tick pos=left,
x grid style={darkgray176},
xmin=0.5, xmax=6.5,
xtick style={color=black},
xtick={1,2,3,4,5,6},
xticklabel style={rotate=30.0,anchor=east},
xticklabels={k1,k2,k3,self\_k1,self\_k2,self\_k3},
y grid style={darkgray176},
ylabel={RPDI},
ymajorgrids,
ymin=0.100001428571429, ymax=1.04282714285714,
ytick style={color=black}
]
\path [draw=black, fill=white, opacity=0.7]
(axis cs:0.75,0.714214285714285)
--(axis cs:1.25,0.714214285714285)
--(axis cs:1.25,0.857028571428572)
--(axis cs:0.75,0.857028571428572)
--(axis cs:0.75,0.714214285714285)
--cycle;
\addplot [black]
table {%
1 0.714214285714285
1 0.571542857142857
};
\addplot [black]
table {%
1 0.857028571428572
1 0.999971428571428
};
\addplot [black]
table {%
0.875 0.571542857142857
1.125 0.571542857142857
};
\addplot [black]
table {%
0.875 0.999971428571428
1.125 0.999971428571428
};
\path [draw=black, fill=white, opacity=0.7]
(axis cs:1.75,0.604007142857143)
--(axis cs:2.25,0.604007142857143)
--(axis cs:2.25,0.686682142857143)
--(axis cs:1.75,0.686682142857143)
--(axis cs:1.75,0.604007142857143)
--cycle;
\addplot [black]
table {%
2 0.604007142857143
2 0.490357142857143
};
\addplot [black]
table {%
2 0.686682142857143
2 0.809214285714286
};
\addplot [black]
table {%
1.875 0.490357142857143
2.125 0.490357142857143
};
\addplot [black]
table {%
1.875 0.809214285714286
2.125 0.809214285714286
};
\path [draw=black, fill=white, opacity=0.7]
(axis cs:2.75,0.387057142857143)
--(axis cs:3.25,0.387057142857143)
--(axis cs:3.25,0.469260714285714)
--(axis cs:2.75,0.469260714285714)
--(axis cs:2.75,0.387057142857143)
--cycle;
\addplot [black]
table {%
3 0.387057142857143
3 0.306771428571428
};
\addplot [black]
table {%
3 0.469260714285714
3 0.571428571428571
};
\addplot [black]
table {%
2.875 0.306771428571428
3.125 0.306771428571428
};
\addplot [black]
table {%
2.875 0.571428571428571
3.125 0.571428571428571
};
\path [draw=black, fill=white, opacity=0.7]
(axis cs:3.75,0.428610714285714)
--(axis cs:4.25,0.428610714285714)
--(axis cs:4.25,0.571492857142857)
--(axis cs:3.75,0.571492857142857)
--(axis cs:3.75,0.428610714285714)
--cycle;
\addplot [black]
table {%
4 0.428610714285714
4 0.285685714285714
};
\addplot [black]
table {%
4 0.571492857142857
4 0.714442857142857
};
\addplot [black]
table {%
3.875 0.285685714285714
4.125 0.285685714285714
};
\addplot [black]
table {%
3.875 0.714442857142857
4.125 0.714442857142857
};
\path [draw=black, fill=white, opacity=0.7]
(axis cs:4.75,0.285714285714286)
--(axis cs:5.25,0.285714285714286)
--(axis cs:5.25,0.428571428571428)
--(axis cs:4.75,0.428571428571428)
--(axis cs:4.75,0.285714285714286)
--cycle;
\addplot [black]
table {%
5 0.285714285714286
5 0.142857142857143
};
\addplot [black]
table {%
5 0.428571428571428
5 0.628571428571428
};
\addplot [black]
table {%
4.875 0.142857142857143
5.125 0.142857142857143
};
\addplot [black]
table {%
4.875 0.628571428571428
5.125 0.628571428571428
};
\path [draw=black, fill=white, opacity=0.7]
(axis cs:5.75,0.5)
--(axis cs:6.25,0.5)
--(axis cs:6.25,0.607142857142857)
--(axis cs:5.75,0.607142857142857)
--(axis cs:5.75,0.5)
--cycle;
\addplot [black]
table {%
6 0.5
6 0.428571428571428
};
\addplot [black]
table {%
6 0.607142857142857
6 0.714285714285714
};
\addplot [black]
table {%
5.875 0.428571428571428
6.125 0.428571428571428
};
\addplot [black]
table {%
5.875 0.714285714285714
6.125 0.714285714285714
};
\addplot [thick, black]
table {%
0.75 0.714464285714285
1.25 0.714464285714285
};
\addplot [thick, black]
table {%
1.75 0.643671428571428
2.25 0.643671428571428
};
\addplot [thick, black]
table {%
2.75 0.430757142857143
3.25 0.430757142857143
};
\addplot [thick, black]
table {%
3.75 0.428757142857143
4.25 0.428757142857143
};
\addplot [thick, black]
table {%
4.75 0.285714285714286
5.25 0.285714285714286
};
\addplot [thick, black]
table {%
5.75 0.571428571428571
6.25 0.571428571428571
};
\end{axis}

\end{tikzpicture}
        \subcaption*{(a) RPDI across State Definitions}
        \label{fig:combined_normalized_price_boxplot_edgeworth}
	\end{minipage}
	\begin{minipage}{0.48\linewidth}
		\centering
\begin{tikzpicture}

\definecolor{darkgray176}{RGB}{176,176,176}
\definecolor{skyblue}{RGB}{135,206,235}
\scriptsize

\begin{axis}[
width=0.9\textwidth, 
height=0.64\textwidth, 
tick align=outside,
tick pos=left,
x grid style={darkgray176},
xmin=0.5, xmax=6.5,
xtick style={color=black},
xtick={1,2,3,4,5,6},
xticklabel style={rotate=30.0,anchor=east},
xticklabels={k1,k2,k3,self\_k1,self\_k2,self\_k3},
y grid style={darkgray176},
ylabel={\(\displaystyle \Delta\) },
ymajorgrids,
ymin=0.106888051020408, ymax=0.519637459183673,
ytick style={color=black}
]
\path [draw=black, fill=white, opacity=0.7]
(axis cs:0.75,0.458919132653061)
--(axis cs:1.25,0.458919132653061)
--(axis cs:1.25,0.489352959183673)
--(axis cs:0.75,0.489352959183673)
--(axis cs:0.75,0.458919132653061)
--cycle;
\addplot [black]
table {%
1 0.458919132653061
1 0.413269795918367
};
\addplot [black]
table {%
1 0.489352959183673
1 0.49666306122449
};
\addplot [black]
table {%
0.875 0.413269795918367
1.125 0.413269795918367
};
\addplot [black]
table {%
0.875 0.49666306122449
1.125 0.49666306122449
};
\path [draw=black, fill=white, opacity=0.7]
(axis cs:1.75,0.252027295918367)
--(axis cs:2.25,0.252027295918367)
--(axis cs:2.25,0.30564693877551)
--(axis cs:1.75,0.30564693877551)
--(axis cs:1.75,0.252027295918367)
--cycle;
\addplot [black]
table {%
2 0.252027295918367
2 0.196388775510204
};
\addplot [black]
table {%
2 0.30564693877551
2 0.379698979591837
};
\addplot [black]
table {%
1.875 0.196388775510204
2.125 0.196388775510204
};
\addplot [black]
table {%
1.875 0.379698979591837
2.125 0.379698979591837
};
\path [draw=black, fill=white, opacity=0.7]
(axis cs:2.75,0.167888775510204)
--(axis cs:3.25,0.167888775510204)
--(axis cs:3.25,0.217632040816326)
--(axis cs:2.75,0.217632040816326)
--(axis cs:2.75,0.167888775510204)
--cycle;
\addplot [black]
table {%
3 0.167888775510204
3 0.125649387755102
};
\addplot [black]
table {%
3 0.217632040816326
3 0.291836734693877
};
\addplot [black]
table {%
2.875 0.125649387755102
3.125 0.125649387755102
};
\addplot [black]
table {%
2.875 0.291836734693877
3.125 0.291836734693877
};
\path [draw=black, fill=white, opacity=0.7]
(axis cs:3.75,0.336651887755102)
--(axis cs:4.25,0.336651887755102)
--(axis cs:4.25,0.408058163265306)
--(axis cs:3.75,0.408058163265306)
--(axis cs:3.75,0.336651887755102)
--cycle;
\addplot [black]
table {%
4 0.336651887755102
4 0.244824489795918
};
\addplot [black]
table {%
4 0.408058163265306
4 0.489756734693877
};
\addplot [black]
table {%
3.875 0.244824489795918
4.125 0.244824489795918
};
\addplot [black]
table {%
3.875 0.489756734693877
4.125 0.489756734693877
};
\path [draw=black, fill=white, opacity=0.7]
(axis cs:4.75,0.244897959183673)
--(axis cs:5.25,0.244897959183673)
--(axis cs:5.25,0.271683673469388)
--(axis cs:4.75,0.271683673469388)
--(axis cs:4.75,0.244897959183673)
--cycle;
\addplot [black]
table {%
5 0.244897959183673
5 0.228617142857143
};
\addplot [black]
table {%
5 0.271683673469388
5 0.308531428571428
};
\addplot [black]
table {%
4.875 0.228617142857143
5.125 0.228617142857143
};
\addplot [black]
table {%
4.875 0.308531428571428
5.125 0.308531428571428
};
\path [draw=black, fill=white, opacity=0.7]
(axis cs:5.75,0.275461683673469)
--(axis cs:6.25,0.275461683673469)
--(axis cs:6.25,0.366017551020408)
--(axis cs:5.75,0.366017551020408)
--(axis cs:5.75,0.275461683673469)
--cycle;
\addplot [black]
table {%
6 0.275461683673469
6 0.168367346938775
};
\addplot [black]
table {%
6 0.366017551020408
6 0.500876122448979
};
\addplot [black]
table {%
5.875 0.168367346938775
6.125 0.168367346938775
};
\addplot [black]
table {%
5.875 0.500876122448979
6.125 0.500876122448979
};
\addplot [thick, black]
table {%
0.75 0.459071224489796
1.25 0.459071224489796
};
\addplot [thick, black]
table {%
1.75 0.277594183673469
2.25 0.277594183673469
};
\addplot [thick, black]
table {%
2.75 0.18814306122449
3.25 0.18814306122449
};
\addplot [thick, black]
table {%
3.75 0.336695510204081
4.25 0.336695510204081
};
\addplot [thick, black]
table {%
4.75 0.244897959183673
5.25 0.244897959183673
};
\addplot [thick, black]
table {%
5.75 0.324489795918367
6.25 0.324489795918367
};
\end{axis}

\end{tikzpicture}
        \subcaption*{(b) \(\displaystyle \Delta\) across State Definitions}
        \label{fig:combined_delta_boxplot_edgeworth}
	\end{minipage}
        \caption{Edgeworth Bertrand: With complete opponent information (\(k1\), \(k2\), \(k3\)), more state information consistently leads to lower pricing.}
        \label{fig:state_price_profit_diff_boxplot_edgeworth}
\end{figure}

\end{appendices}

\newpage
\onehalfspacing
\begin{appendices}
	\normalsize
\section{Pseudocodes of Algorithms Selection}
\label{appendix: pseudocodes_drl}

This section introduces the \gls{acr:rl} algorithms used in our experiments, including \gls{acr:tql}, \gls{acr:dqn}, and \gls{acr:ppo}. We briefly overview their algorithmic steps, highlighting their advantages and applicability in complex decision-making. 

\subsection{\gls{acr:tql}}

\gls{acr:tql} is a value-based algorithm, designed to address \gls{acr:mdp} problems with explicit rewards. Introduced by \citep{watkins1989learning}, it's a model-free, asynchronous learning method. The fundamental concept involves learning a Q-function to choose the optimal action, with the relevant pseudocode provided in Algorithm~\ref{alg:qlearning}. Here, $Q$, as in $Q(s, a)$, represents the expected reward at a particular moment in state $s$ when taking action $a$. Specifically, the \gls{acr:tql} algorithm uses a Q-table to store each state and action's Q-value (expected reward). This table is initially a matrix $Q_0$ of size $|S|\times|A|$ and is updated at each time step as the agent selects an action, then observes the feedback from the environment (i.e., the reward and the next state). The Q-values are updated using the Bellman equation 
\begin{equation}
    Q_{t+1}(s, a) = (1-\alpha) Q_t(s, a) + \alpha ( r_t + \gamma \max_{a\in A} Q_t(s^{\prime}, a)),
\end{equation}

where $\alpha$ is the learning rate. This process is repeated until the Q-table converges.


\begin{algorithm}
\caption{Tabular Q-learning}
\label{alg:qlearning}
\begin{algorithmic}[1]
\State Initialize $Q(s, a)$, for all $s \in \mathcal{S}$, $a \in \mathcal{A}$, arbitrarily
\State Initialize $S_1$ arbitrarily
\For{$t=1$ to $T$}
    \State With probability $\epsilon_t$ play a random action $A_t$
    \State Otherwise play $A_t = \arg\max_a Q(S_t, a)$
    \State Observe $R_{t+1}, S_{t+1}$
    \State $Q(S_t, A_t) \gets Q(S_t, A_t) + \alpha [R_{t+1} + \gamma \max_a Q(S_{t+1}, a)-Q(S_t, A_t)]$
\EndFor
\end{algorithmic}
\end{algorithm}

\gls{acr:tql} requires trying all actions in all states to balance exploration and exploitation. Exploration involves selecting suboptimal actions to acquire new information, while exploitation leverages existing knowledge. A common approach is the \(\epsilon\)-greedy strategy, where an action is chosen randomly with probability \(\epsilon_t\) and greedily with probability \(1 - \epsilon_t\) to maximize \(Q_t(s_t, a)\). The exploration rate \(\epsilon_t = \exp(-\beta t)\) decreases over time, allowing the agent to initially explore more and gradually shift towards exploitation as it gains knowledge. This balance prevents premature convergence to suboptimal policies while improving learning efficiency by focusing on high-reward actions.

\gls{acr:tql} adapts to unknown environments without prior knowledge, making it suitable for handling uncertainty in fields like economics and finance. However, it has several limitations. Updating the Q-table requires extensive trials, leading to high computational costs in large state spaces. Convergence to the optimal policy can be slow, impacting efficiency in practical applications. Additionally, \gls{acr:tql} is prone to overfitting in dynamic environments, where historical data may become obsolete. 

\subsection{\gls{acr:dqn}}

\gls{acr:dqn} integrates \gls{acr:tql} with \gls{acr:dnn} to approximate the optimal Q-function \(Q^{*} (s, a)\), enabling it to handle complex, high-dimensional state-action spaces. By leveraging \gls{acr:dnn}s, \gls{acr:dqn} generalizes learning across similar state-action pairs, improving efficiency and mitigating the curse of dimensionality in \gls{acr:tql}, making it suitable for large-scale problems.


\begin{algorithm}
\caption{Deep Q-Network (DQN)}
\label{alg:dqn}
\begin{algorithmic}[1]
\State \textbf{Initialize:} local network $Q(s, a, \theta)$ with arbitrary weights $\theta$
\State Initialize target network $\hat{Q}(s, a, \hat{\theta})$ with weights $\hat{\theta} = \theta$
\State Initialize average reward estimate $\Bar{R}$ arbitrarily
\State Initialize $S_1$ arbitrarily
\For{$t=1$ to $T$} 
    \State With probability $\epsilon_t$, play a random action $A_{t}$
    \State Otherwise, play $A_{t} = \arg\max\limits_{a} Q(S_{t}, a, \theta)$
    \State Observe $R_{t+1}, S_{t+1}$
    \State Store transition $(S_t, A_t, R_{t+1}, S_{t+1})$ in $B$
    \State Sample random minibatch of transitions $(S_j , A_j , R_{j+1}, S_{j+1})$ from $B$
    \State Set target value:
    \[
    Y_j = R_{j+1} - \Bar{R} + \max\limits_a \hat{Q}(S_{j+1}, a, \hat{\theta})
    \]
    \State Perform a gradient descent step on 
    \[
    \left[Y_j - Q(S_j, A_j, \theta)\right]^2
    \]
    with respect to $\theta$
    \State Update average reward:
    \[
    \Bar{R} \gets \Bar{R} + \lambda \left[R_{t+1} - \Bar{R} + \max\limits_a \hat{Q}(S_{t+1}, a, \hat{\theta}) - \hat{Q}(S_t, A_t, \hat{\theta})\right]
    \]
    \If{$t \mod C == 0$}
        \State Update $\hat{Q} \gets Q$
    \EndIf
\EndFor
\end{algorithmic}
\end{algorithm}

In \gls{acr:dqn}, the \gls{acr:dnn} is represented as \( f(s) = W_2\sigma(W_1s + v_1) + v_2 \), where \( W_1, W_2, v_1, v_2 \) are network parameters, and \( \sigma(\cdot) \) is a nonlinear activation function, such as ReLU. This structure enables \gls{acr:dqn} to approximate complex functions and process large input data. However, \gls{acr:dqn} requires substantial data and computing resources and can be unstable during training. To address this, it incorporates experience replay and a target network. Experience replay stores and randomly samples past experiences, reducing correlation and improving learning stability. The target network updates weights periodically, further enhancing stability.

The pseudocode for DQN is provided in Algorithm~\ref{alg:dqn}. The gradient of the loss function \( L \) with respect to network weights \( \theta \) is:  

\begin{equation}
    \nabla_{\theta}L(\theta) = \mathbb{E}_{s,a,r,s'}\left[ \left( r + \gamma \max_{a'} Q(s', a'; \theta^-) - Q(s, a; \theta) \right) \nabla_{\theta}Q(s, a; \theta) \right]
\end{equation}  

A mini-batch of samples (typically 16 or 32) approximates this gradient, which is then used to update parameters via:  

\begin{equation}
     \theta_{\text{new}} = \theta_{\text{old}} - \alpha \nabla_{\theta} J(\theta_{\text{old}})
\end{equation}  

where \( \alpha \) is the learning rate. The target network, updated every \( C \) iterations, mitigates instability by providing fixed reference values.  Despite high computational demands and stability challenges, \gls{acr:dqn} effectively integrates deep learning with reinforcement learning, making it a powerful tool for complex decision-making. Given its advanced nature relative to \gls{acr:tql}, some researchers believe that the use of \gls{acr:dqn} will accelerate the emergence of collusive behavior.

\subsection{\gls{acr:ppo}}

\gls{acr:ppo} is a policy-based reinforcement learning algorithm that directly optimizes the policy function to learn an optimal strategy. The policy function $\pi(a|s)$ defines the probability distribution over actions given a state $s$. \gls{acr:ppo} primarily improves training stability by constraining policy updates while maintaining sample efficiency.


\begin{algorithm}
\caption{PPO-Clip}
\label{alg:ppo}
\begin{algorithmic}[1]
\State \textbf{Input:} initial policy parameters $\theta_0$, initial value function parameter $\phi_0$
\For{$k=0,1,2,\ldots$}
    \State Collect set of trajectories $\mathcal{D}_k=\{\tau_i\}$ by running policy $\pi_k=\pi(\theta_k)$ in the environment
    \State Compute rewards-to-go $\hat{R}_t$
    \State Compute advantage estimates $\hat{A}_t$ based on current value function $V_{\phi_k}$
    \State Update policy by maximizing the PPO-Clip objective:
    \[
    \theta_{k+1} = \arg\max_{\theta} \frac{1}{|\mathcal{D}_k|T}\sum_{\tau \in \mathcal{D}_k}\sum_{t=0}^{T} \min\left(\frac{\pi_{\theta}(a_t|s_t)}{\pi_{\theta_k}(a_t|s_t)}A^{\pi_{\theta_k}}(s_t,a_t), g(\epsilon,A^{\pi_{\theta_k}}(s_t,a_t))\right)
    \]
    \State Fit value function by regression:
    \[
    \phi_{k+1} = \arg\min_{\phi} \frac{1}{|\mathcal{D}_k|T}\sum_{\tau \in \mathcal{D}_k}\sum_{t=0}^{T} (V_{\phi}(s_t)-\hat{R}_t)^{2}
    \]
\EndFor
\end{algorithmic}
\end{algorithm}

\gls{acr:ppo} optimizes the policy by clipping the objective function to mitigate instability from excessively large updates:  
\begin{equation}
L^{CLIP}(\theta) = \hat{\mathbb{E}}_t \left[ \min(r_t(\theta) \hat{A}_t, \text {clip} (r_t(\theta), 1 - \epsilon, 1 + \epsilon) \hat{A}_t) \right]
\end{equation}  
where \( r_t(\theta) = \frac{\pi_{\theta}(a_t | s_t)}{\pi_{\theta_k}(a_t | s_t)} \) is the probability ratio between the new and old policies, \( \hat{A}_t \) is the advantage estimate, and \( \epsilon \) constrains policy updates to prevent performance degradation from abrupt changes. The clipping function restricts \( r_t(\theta) \) within \([1-\epsilon, 1+\epsilon]\), capping the objective when the ratio exceeds this range, thereby stabilizing training.

Compared to \gls{acr:dqn}, \gls{acr:ppo} is more stable as it avoids overestimation of Q-values and is naturally suited for continuous action spaces, whereas \gls{acr:dqn} primarily handles discrete action spaces. However, \gls{acr:ppo} is sensitive to hyperparameters, particularly the clipping range \( \epsilon \) and the learning rate, both of which significantly affect training performance.

\end{appendices}

\end{document}